# DeepFoldit - A Deep Reinforcement Learning Neural Network Folding Proteins

**Dimitra Panou[1], Martin Reczko[2]**

[1]University of Athens, Department of Informatics and Telecommunications

[2]Biomedical Sciences Research Center “Alexander Fleming”

## ABSTRACT

Despite considerable progress, *ab initio* protein structure prediction remains suboptimal. A crowdsourcing approach is the online puzzle video game Foldit [1], that provided several useful results that matched or even outperformed algorithmically computed solutions [2]. Using Foldit, the WeFold [3] crowd had several successful participations in the Critical Assessment of Techniques for Protein Structure Prediction. Based on the recent Foldit standalone version [4], we trained a deep reinforcement neural network called DeepFoldit to improve the score assigned to an unfolded protein, using the Q-learning method [5] with experience replay. This paper is focused on model improvement through hyperparameter tuning. We examined various implementations by examining different model architectures and changing hyperparameter values to improve the accuracy of the model. The new model’s hyper-parameters also improved its ability to generalize. Initial results, from the latest implementation, show that given a set of small unfolded training proteins, DeepFoldit learns action sequences that improve the score both on the training set and on novel test proteins. Our approach combines the intuitive user interface of Foldit with the efficiency of deep reinforcement learning.

**KEYWORDS**: ab initio protein structure prediction, Reinforcement Learning, Deep Learning, Convolution Neural Networks, Q-learning

## 1. ALGORITHMIC BACKGROUND

Machine learning (ML) is the study of algorithms and statistical models used by computer systems to accomplish a given task without using explicit guidelines, relying on inferences derived from patterns. ML is a field of artificial intelligence. The goal of machine learning algorithms is to understand the structure of data and build representative models of them, by constructing knowledge representations and inference mechanisms that captured the underlying distribution. The models should have the capability to extrapolate novel data.

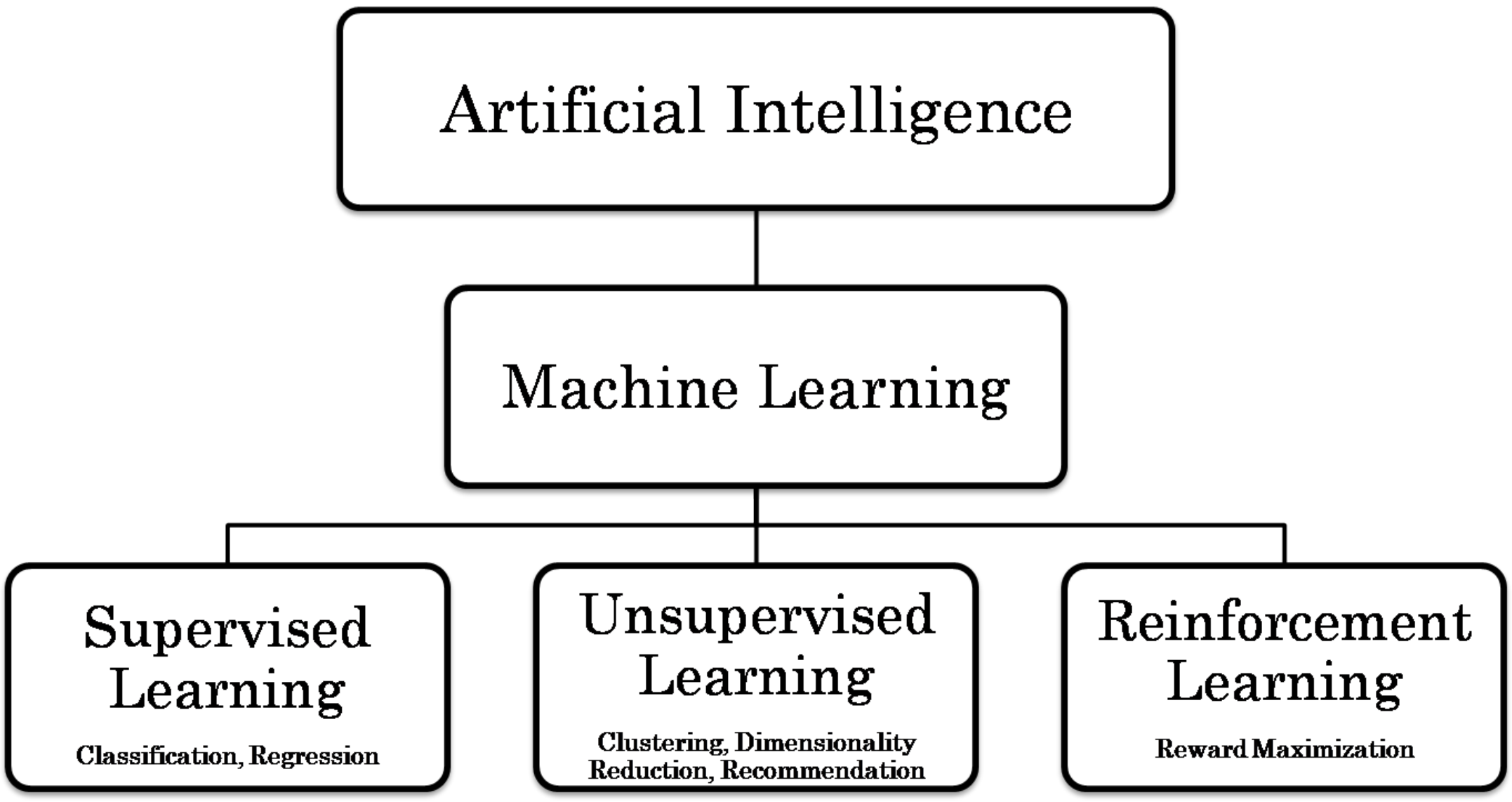

**Figure 1: Types of Machine learning techniques**

Machine learning can be categorized based on the system of the learning process, or based on the output data types. Depending on the nature of the learning "signal" or "reaction" accessible to a learning scheme, machine learning implementations are categorized into three major categories, which are supervised, unsupervised and reinforcement learning. Supervised learning includes feedback indicating the exact outcome of a forecast, whereas unsupervised learning does not require any labeling of the data: the algorithm attempts to categorize information based on its hidden structure. Reinforcement learning is similar to supervised learning because it receives feedback, but not necessarily for each state or input and only in the form of penalties. Reinforcement learning will be described further in the next section. Machine learning is a continuously developing and very promising field.

## 1.1 ARTIFICIAL NEURAL NETWORKS (NN)

Artificial Neural Networks (ANN), are mathematical models inspired by biological neural networks that are used mostly as machine learning algorithms and have lately gained a lot of attention thanks to the availability of Big Data and fast computing. ANNs try to mimic biological neural networks. Their theory has been developed ever since the 70s, however, wrong assumptions about their capabilities and the computational cost to implement them led people to design alternative algorithms for automated model building. In recent years, there has been a rapid increase in interest for the ANNs, thanks to many breakthroughs in computer vision, speech recognition and text processing in the context of deep learning.

In supervised learning there are plenty algorithms such as, , linear classifiers [6], Bayesian classifiers [7], K-nearest neighbors (KNNs) [8], Hidden Markov model (HMM) [9], and decision trees [10]. On the other hand, unsupervised methods that are popular include Autoencoders [11], expectation maximization [12], self-organizing maps [13], k-means [14], fuzzy [15], and density-based clustering [16]. Finally, reinforcement learning (RL) problems are modeled using the Markov decision process (MDP), which will analyzed further in section 1.3.2 and dynamic programming [17]. RL algorithms have been successfully combined with a number of deep Neural Network architectures, including Convolutional Neural Networks (CNN) [18, 19], Recurrent Neural Networks (RNN) [18] such as LSTM [20] and GRU [21], Autoencoders [22] and Deep Belief Networks (DBN) [23].

Convolutional Neural Networks (CNN) are multi-layer neural networks that are primarily used to analyze images for image classification, segmentation and object identification. They are based on reducing input images to their key features and classify them using a combination of them. Recurrent Neural Networks (RNN) are used for sequential input such as text, audio data and videos for classification and analysis. RNNs work by evaluating sections of the input stream, using weighted temporal memory and feedback loops. There are algorithms that use hybrid models, a combination of CNN and RNN referred to as CRNN, to increase their effectiveness. Both networks can classify images, texts and video [24-27].

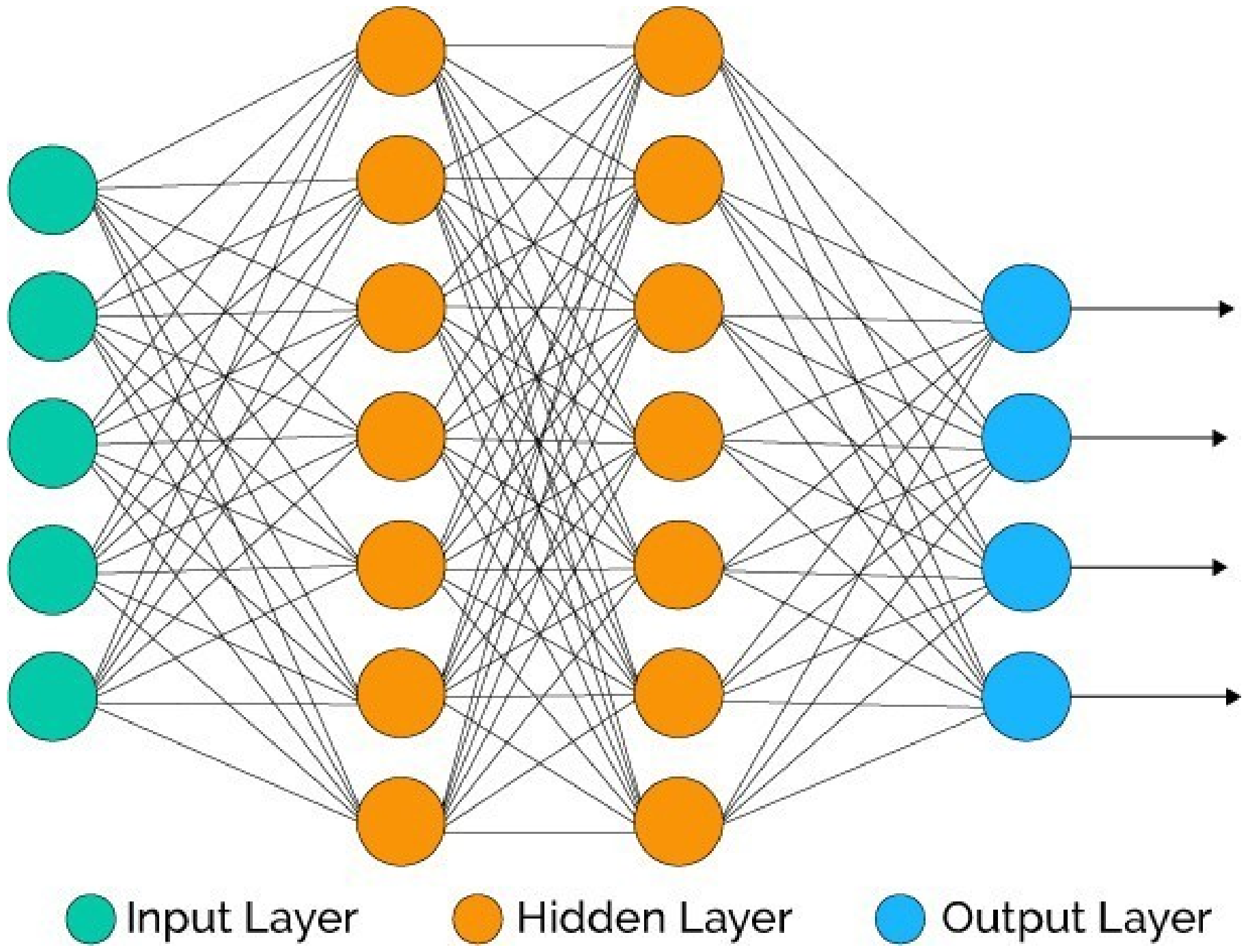

**Figure 2: Artificial Neural Network with 2 hidden layers [28]**

Generally, an ANN is an information processing system that is based on simple processing elements, the neurons. The central idea is to create a machine that would be able to simulate how biological brains processes information through transmitting signals between its neurons. Different signals representing information traverse different paths through the network. Similar to biological networks, artificial neurons are organized in layers. Each layer consists of nodes (neurons), weighted connections between these nodes (weights) that are updated during the training or learning process and an activation function that defines the output value of each node depending on the input. Every time the network is called to answer a question about the input, neurons are activated or deactivated, producing outputs that are fed to the neurons of the next layer.

### 1.1.1 Computational models for neurons

ANNs have long history that starts with the first efforts to comprehend how biological brains operate and the structure of intelligence. The structural component of the brain is the neuron, which is made of three basic components: Dendrites, the cell body and axons. Each neuron is connected to other neurons through unique links called synapses. It is estimated that the human brain comprises of 86 billion neurons and approximately $10^{14}$ to $10^{15}$ synapses [29]. Each neuron receives input signals from its dendrites and generates output signals along its axon, which connects with other neurons through synapses with their dendrites (Figure 3).

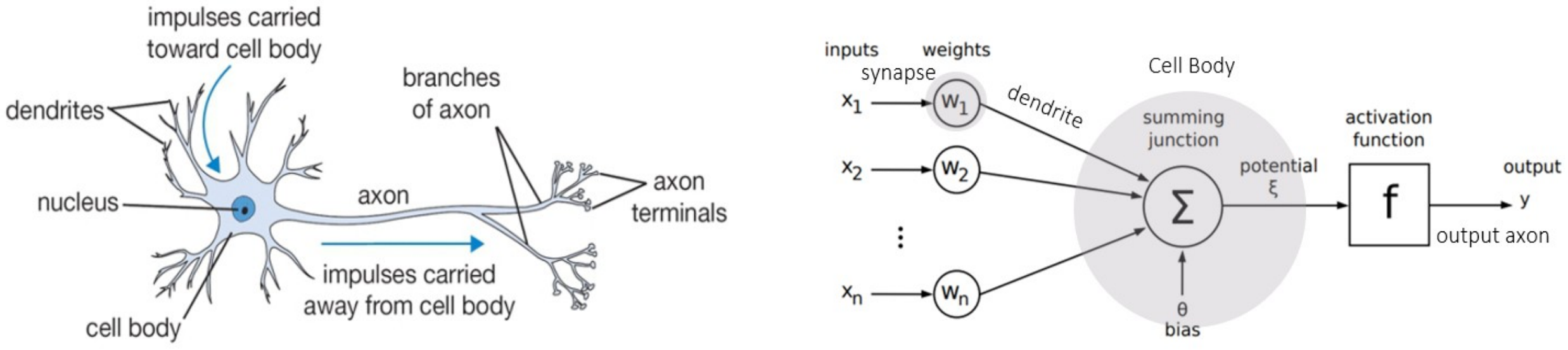

**Figure 3: Biological Neuron (left), artificial neuron model (right). [30]**

Using the same logic, an artificial neuron or perceptron is the basic unit of an artificial neural network. In the non-spiking computational model (Figure 3) of a neuron, the signals moving along its axon ($x_1$) interact multiplicatively ($w_1 x_1$) with the dendrites of another neuron based on the synaptic weight ($w_1$). Synaptic weights can be adapted during the training. The neuron integrates all inputs and passes it through an activation function which determines if the neuron should be activated or not and generates an output activation. This activation is the normalized output of each neuron. In supervised training, the objective is to determine the weights that can generate some desired output activations for a subset of the neurons.

### 1.1.2 Single Neuron as Classifier

The perceptron is the smallest and simplest modle for a neuron that exists for a neural net, proposed by F. Rossenblatt [31]. A perceptron can solve a simple binary classification problem by giving as output signals “yes” or “no”, 0 or 1. Modeled by the idea of how biological brains work, a single layer perceptron may be a very simple learning machine. Mathematically this is how one perceptron neuron works:

Considering $x_1, x_2, \ldots, x_n$ the input values, we pass x values from an operation $g$ where

$$g(x_1, x_2, \ldots, x_n) = \sum_i^n w_i x_i + bias$$

$$y = f(g(x)) = \begin{cases} 0, iff(g(x)) < \theta \\ 1, iff(g(x)) \geq \theta \end{cases}$$

Where $g()$is a function, which takes a vector of inputs $\bar{x}$, performs an aggregation (weighted sum) and passes the aggregated value through an activation function. If the final sum is greater than a certain threshold called bias, the neuron is activated. The bias is used to shift the activation function.

### 1.1.3 Multilayer Perceptron as Classifier

A multilayer perceptron (MLP) is a deep artificial neural network that consists of multiple layers of perceptrons between the input layer and the output layer. The output of each neuron in one layer is usually connected to every neuron in the next layer. The input layer is a feature vector that needs to be classified and the activations of the output layer are decoded into class assignments for the input features to make decisions about the input. The weights are determined using a training set of feature vectors with known class labels. The MLP is also called *feed-forward neural network*, because the information is processed successively, layer by layer, from the input layer to the output layer. An MLP must contain at least one hidden layer. The motivation behind designing multilayer networks is to support solving more complex tasks by adding a hierarchy of internal representations in each layer.

### 1.1.4 Activation Function

One critical step in building a neural network is the selection of an activation function. The activation function is a mathematical equation defined for each neuron in the network, and determines its degree of activation. Activation functions typically have an output range between 1 and 0 or between -1 and 1. This function is the feature that gives the network the ability to compute and represent arbitrarily complex functions (non-linear). Most activation functions are nonlinear functions, as the linear activations in adjacent layers can be combined into a single layer.

Typical activation functions are: binary step, linear or nonlinear, and differentiable non-linearities. In the binary step function, if the input value is above a certain threshold the neuron is activated and sends its activation to the next layer. A linear function has the form of $f(x)=cx$, as it takes inputs multiplied by the weights for each neuron and creates an output proportional to the input. With linear activation functions, all adjacent layers of the net become one, because no matter how many layers exist in the network, the last layer will always be a linear combination of the first. Non-linear activation functions manage to create complex mappings between the network's input and outputs, which is ideal for learning data such as images, audio and videos with multiple dimensions.

Differentiable non-linear activation functions allows backpropagation [32] and multiple hidden layers. In **,** some of the most popular activation functions are shown.

**a) Sigmoid**

It is also referred as logistic activation function. This function has a Real number as input and normalizes it in the range 0 to 1. Converts large negative numbers to 0, while large positive numbers provide an approximate output of 1. This function only works with 2-class classification A generalized version of sigmoid activation function that is used in multiclass classification is the SoftMax. The most important problem with sigmoid function is known as vanishing gradients (the gradient of weight vanishes or goes down to zero) [33] and is occurred in backpropagation-trained networks. For very high or very low values there is almost no change to the prediction causing a vanishing gradient problem. The problem lies in the fact that the derivative of the sigmoid function is close to zero both for very large positive and negative inputs, which results in no alteration in the weights of some neurons. Another issue we face while using sigmoid activation function is that the output is always positive (accumulated towards the positive side), so it's not a zero centered function. Sigmoid functions are mostly used in output layers and in classification problems.

**b) Hyperbolic Tangent Function (Tanh)**

Tanh function is quite similar to the sigmoid with the difference that it is a zero centered function. Zero-centered activation functions have a mean activation value around zero. It receives a real number as input and outputs a number in the range of -1 to 1. It is considered a better function than the sigmoid since a zero centered function is less dependent on further normalization measures. However, it also suffers from the vanishing of gradients problem.

**c) Rectified Linear Unit (ReLU)**

This function assigns the input value to a value in the range 0 to x, where x is a positive number. For negative input values the output is 0 while for positive inputs the output is x. This function is widely used today as it overcomes part of the vanishing gradient problem, for positive input values, and it is easy to implement.

**Table 1: Most known activation functions plots obtained from [34]**

| Name | Plot | Equation | Derivative |
|---|---|---|---|
| **Identity** | 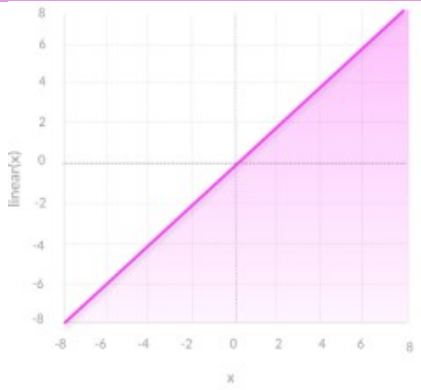 | $f(x)=x$ | $f'(x)=1$ |
| **Binary Step** | 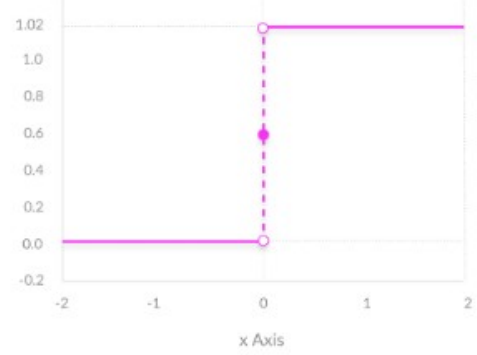 | $f(x)=\begin{cases}0\ for\ x<0\\1\ for\ x\geq 0\end{cases}$ | $f'(x)=\begin{cases}0\ for\ x\neq 0\\\infty\ for\ x=0\end{cases}$ |
| **Sigmoid/Logistic** | 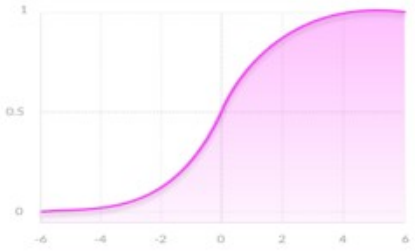 | $f(x)=\frac{1}{1+e^{-x}}$ | $f'(x)=f(x)(1-f(x))$ |
| **Hyperbolic/ Tanh** | 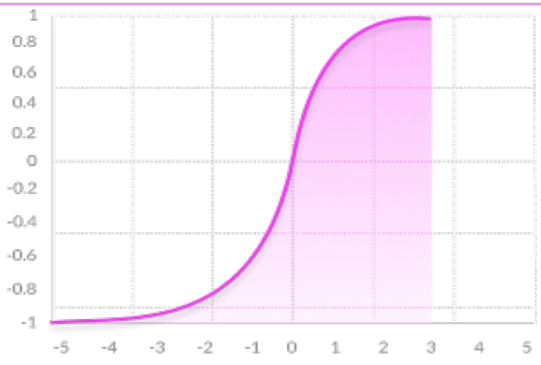 | $f(x)=\frac{2}{1+e^{-2x}}-1$ | $f'(x)=1-f(x)^2$ |
| **ArcTan** | 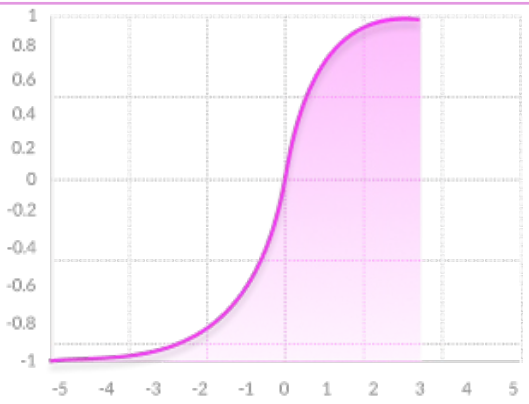 | $f(x)=\tan^{-1}(x)$ | $f'(x)=\frac{1}{x^2+1}$ |
| **ReLU** | 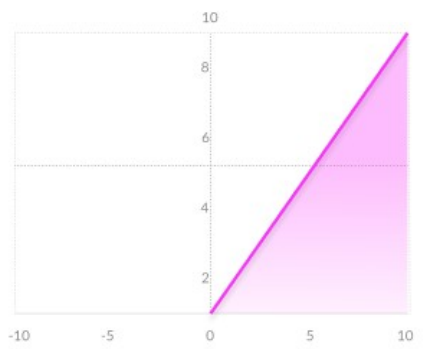 | $f(x)=\begin{cases}0\ for\ x<0\\1\ for\ x\geq 0\end{cases}$ | $f'(x)=\begin{cases}0\ for\ x<0\\1\ for\ x\geq 0\end{cases}$ |
| **PReLU or Leaky ReLU** | 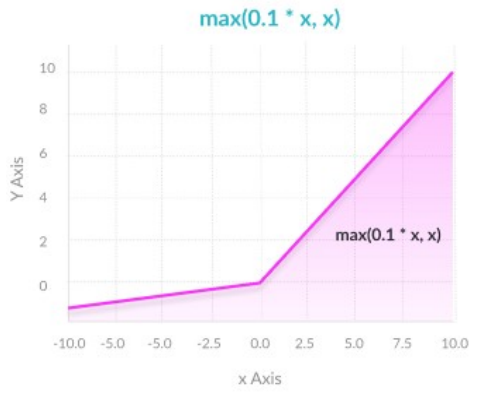  | $f(x)=\begin{cases}ax\ for\ x<0\\x\ for\ x\geq 0\end{cases}$ | $f'(x)=\begin{cases}a\ for\ x<0\\1\ for\ x\geq 0\end{cases}$ |

### 1.1.5 Training

The learning process is the key concept of a neural network. During the learning process, the network is searching for its optimal parameters which can solve the given problem and before the learning process starts, we have to initialize our parameters. Usually the initial values are chosen randomly, but there are some heuristic algorithms [24-27], which speed up the identification of the optimal parameters. There are two data sets, the training, and the test set. The training set is fed to the neural network during the training process. The **learning phase** is an iterative process where the outputs that were produced form every input of the training set are analyzed and the network is repeatedly being adjusted to produce better results. In the simplest setting, the network is considered to be trained after reaching a predefined target performance on the training data. There exist different metrics for assessing the network performance, with the most common being the mean squared error. If after learning the error rate is still high, usually some hyperparameters are changed and the training is repeated.

Depending on the type of problem, different types of training are used. Specifically, in supervised learning the training process is focused on the identification of the scoring function which is inferred from the labeled training data. The data consists of training examples with known classes. These form pairs of inputs and the desired outputs. In order to avoid a common problem known as **overfitting** (when the network's predictive performance is improving on the training set but it is worsening on the unseen test data), we split the dataset into two subsets. One is for the actual training (training set), and the other is to control how the training is going (validation set). The main reason to use the **validation set** is that it shows the error rates on data independent of the **training set**. The objective of training in supervised methods is to identify the model that best expresses the input. This process is evaluated using a loss function. Typically, we seek to minimize the error. As such, the objective function is often referred to as a cost function or a loss function. During the training procedure, the network will update the weights of each layer by comparing the result with the desired output value.

In unsupervised learning, the training set does not contain expected outputs for the given inputs. The objective of these methods is to form clusters, groups of data that show similarities, common behavior according to specific features. The features are known, so these algorithms aim to find those clusters. There are also methods that combine supervised and unsupervised learning methods [35].

#### 1.1.5.1 Learning Rate

**Hyperparameters** are system-external configurations whose values cannot be estimated from data. In other words, are the parameters that are not learned by the model and must be chosen by the user before the training process. These pre-fixed values help to estimate model parameters and setting up the right values can affect the accuracy of the model.

Learning rate is a hyperparameter used in training that controls the stepsize used to change adaptable parameters. The learning rate has range between 0.0 and 1.0 and controls the stepsize for the gradient decent. If the learning rate is low, the model requires more training epochs to update the weights but may better follow actual gradient, while in high values it has rapid changes at the risk of introducing oscillations. A higher learning rate has a higher risk of falling in a sub-optimal solution of the loss function. This is the reason why the right value selection for the learning rate is important in the process of creating a model.

#### 1.1.5.2 Cost Function

Before training starts, all the model's parameters are usually initialized randomly. During the training, some parameters are updated and our main goal is to find the optimal values for them. Defined by an evaluation method. A cost function is an evaluation of the performance of a model, a measure of "how good" a network is. With respect to the given training samples and the expected output a cost function produces a single number that represents the performance of the network. A cost or loss function specifies how to calculate the error between prediction and the label of a given training example. This error is backpropagated during training in order to update the learnable parameters (weights). Broadly, loss functions can be classified into two major categories, depending on the type of learning task, to Regression losses and Classification losses. In classification losses, the predicted output belongs to a set of finite categorical values, for example the problem of categorizing hand-written digits into one of the classes corresponding to the 0–9 digits. Cross Entropy loss function belongs to Classifications losses. Cross entropy is a simple and effective method, which works with Gaussian distributions, repeatedly updating the mean and variance of a distribution over candidate parameters. In Regression losses, on the other hand, the predicted output is a continuous value. Popular loss functions that belong to this category is Mean Square Error (MSE), Mean Absolute Error (MAE), Mean Bias Error (MBE). Since the cost function represents the loss-error, our main concern is to minimize it at the end of the training. To achieve this, we use an optimization function called Gradient Descent. A penalty term usually related to the size of the weights may be added to the loss function or the gradient update equation to prevent overfitting and/or to make the network more robust against noisy input data.

#### 1.1.5.3 Gradient Descent

In practice, the most commonly used procedure is the stochastic gradient descent (SGD). This consists of providing the input vector a few examples, computing the outputs and the errors, computing the average gradient for those examples, and adjusting the weights accordingly. This process is repeated for many small sets of examples from the training set, called batches, until the average of the objective function stops decreasing. It is called stochastic because each small set of examples gives a noisy estimate of the average gradient of the entirety of the examples.

#### 1.1.5.4 Backpropagation

Backpropagation is an algorithm frequently used to train neural networks.

**Updating the weights during backpropagation**

There are three ways to update the weights during the training. The first way is to calculate the optimal weights and update them after the presentation of each sample (instance) of the training set (online method). This method is really simple, although it can be sensitive to outliers and time consuming for large datasets.

The second way for weight updating is by using batches. The training samples are divided into batches and then training is performed iteratively on each batch. Backpropagation is calculated on all the samples that belong to the same batch. This method is more accurate and less sensitive to outliers.

Finally, the third approach is between the other two and is the random selection of small batches from the training data, and then run forward pass and backpropagation on each batch, iteratively. This [36, 37] prevents a biased selection of samples in each batch, which can lead training into a local optimum.

### 1.1.6 Hyperparameters

There are several parameters we use to define and train an ANN. Some of them are internal and adapted automatically from samples during the training process, and others have fixed values and are chosen before the training starts.

The internal parameters of the network are learnable during the training process. They are used to make predictions in a production model and are referred to as the model's parameters. Parameters such as weights and bias are those internal parameters that are changing during the training and need to be initialized before the training starts.

The external parameters are set by the designer of the ANN and called hyperparameters. Changes in hyperparameters can have an impact on the performance of the network. Tuning of hyperparameters helps the network provide accurate predictions. Common hyperparameters are related to the network's structure are the number of hidden layers, the *dropout rate* [38] (specifies the probability at which outputs of the layer are dropped out or retained.), the activation function and the weight initialization methods. Others that are involved in the training algorithm are the learning rate, the number of epochs, the optimizer algorithm and the momentum. There is no predefined way to choose values for a hyperparameter, only by trials and comparing the results. For some parameters there exist suggested values that have been discovered empirically. For example, most implementations suggest the minibatch size, which is the number of training samples that belong to a batch that is used for weight update during backpropagation, to be a power of two that fits the memory requirements of a CPU/GPU such as 32, 64, 128, 256. Hyperparameters optimization will be analyzed further in the final chapter.

### 1.1.7 Model Evaluation

The final part of the training process is the evaluation of the constructed model. This is an important step, since it allows us to understand the effectiveness of the model we deployed while solving the problem. At this point, error estimation is required and the most popular method is splitting the data set into three parts, specifically the training, validation and testing data sets. Training the model is accomplished by using the training set, composed of labeled data that allow the model to learn the connections from input to output. Using the validation set, we pinpoint when to terminate the model. Finally, we test our model to the unbiased data of the test set to estimate how well it behaves. This process aims to eliminate overfitting. Overfitting is a common problem in ML and happens when a model overly adapts to a specific data set and cannot generalize. Essentially the model is adapted to the noise from the training set and it does not perform well with new unbiased data. There are many methods to avoid over-fitting. One such method is cross-validating our results. The most common version is k-fold cross-validation, in which we split the training set into k sets and proceed to use k-1 parts for training and one part for validation followed by iterating k times while alternating validation parts for each iteration. For big data sets, a standard choice for the k value is 10. This algorithm is effective but can be proved computationally complex since we iterate the training process k times. The performance of a model for classification can be accurately determined using a set of metrics like accuracy (ratio of the number of accurate predictions to the number of total predictions), precision (ratio between true positives to the number of positives), recall and the F-score (weighted mean of the precision and recall).

$$Accuracy = \frac{TP + TN}{P + N}$$

$$Recall = \frac{TP}{TP + FN}$$

$$Precision = \frac{TP}{TP + FP}$$

$$F_1 = \frac{Precision \cdot Recall}{Precision + Recall} = \frac{2\,TP}{2\,TP + FP + FN}$$

## 1.2 CONVOLUTIONAL NEURAL NETWORKS (CNN)

Convolutional neural networks (CNN) are multi-layer feed-forward neural networks that are specifically designed for image processing, classification, clustering and feature recognition, and signal processing. Convolutional networks can also perform optical character recognition (OCR) to digitize text and make natural-language processing possible on hand-written documents, where the images are symbols to be transcribed. The architecture of CNN is  inspired by the findings of P.H Hubel and T. N. Wiesel in 1959 [39, 40], in trying to explain how mammals perceive the world around them in a hierarchical way. Their paper was a study of signal processing in a cat's visual cortex [41]. Using a layered architecture of neurons of the cat's brain, inspired engineers to develop a similar mechanism for computer vision. CNNs are usually applied for data types that can be presented as multi-dimensional matrices. An image can be stored as a matrix with its height, width and color channels as its dimensions are height, width, 3 for RGB images. Deep learning has greatly enhanced the state of the art for many problems faced by the machine learning and AI community especially in image recognition and object identification and CNNs are responsible for this improvement and one of the main reasons why deep learning is famous nowadays. The big success of AlexNet opened the path for 2D image recognition in 2012 [19].

### 1.2.1 CNN layer architecture

A convolutional network is architecturally split into layers. Each layer is intended to fulfill a different purpose and to learn different levels of abstraction. The network's layers close to the input layer perform feature extraction, and consist of convolutional and subsampling layers. The layers closer to the output usually perform classification based on the extracted features detected in the lower layers. Neurons are grouped, creating layers, depending on the different levels of abstraction they are learning. The first layer of the network is the input layer, all other layers are hidden layers except the last which is the output layer. The number of hidden layers denote the depth of the network. The layers of the network perform different tasks based on their connection structure and their activation functions. Examples of these types of layers are: the **fully connected layers**, which take into consideration all neurons from the previous layer; **the pooling layers**, that perform a down sampling operation (for example, max-pooling takes the maximum value from the inputs) or the **convolutional layers** that compute the output of neurons that are connected to a local region of the previous layer and share these connections to cover the complete previous layer. The layer-types of a CNN will be further analyzed in the next sections. The networks that have at least one convolutional layer are known as Convolutional Neural Networks (CNN) or ConvNets.

A typical CNN for image processing operates in three stages. The first is a *convolution*, in which the image is "scanned" a few pixels at a time, and a *feature map* is created with probabilities, so that each feature belongs to the required class (in a simple classification example). The second stage is *pooling* (also called down sampling), which reduces the dimensionality of each feature while maintaining its most important information. The pooling stage creates a "summary" of the most important features in the image. The third phase is the fully connected layer which ends up to the output.

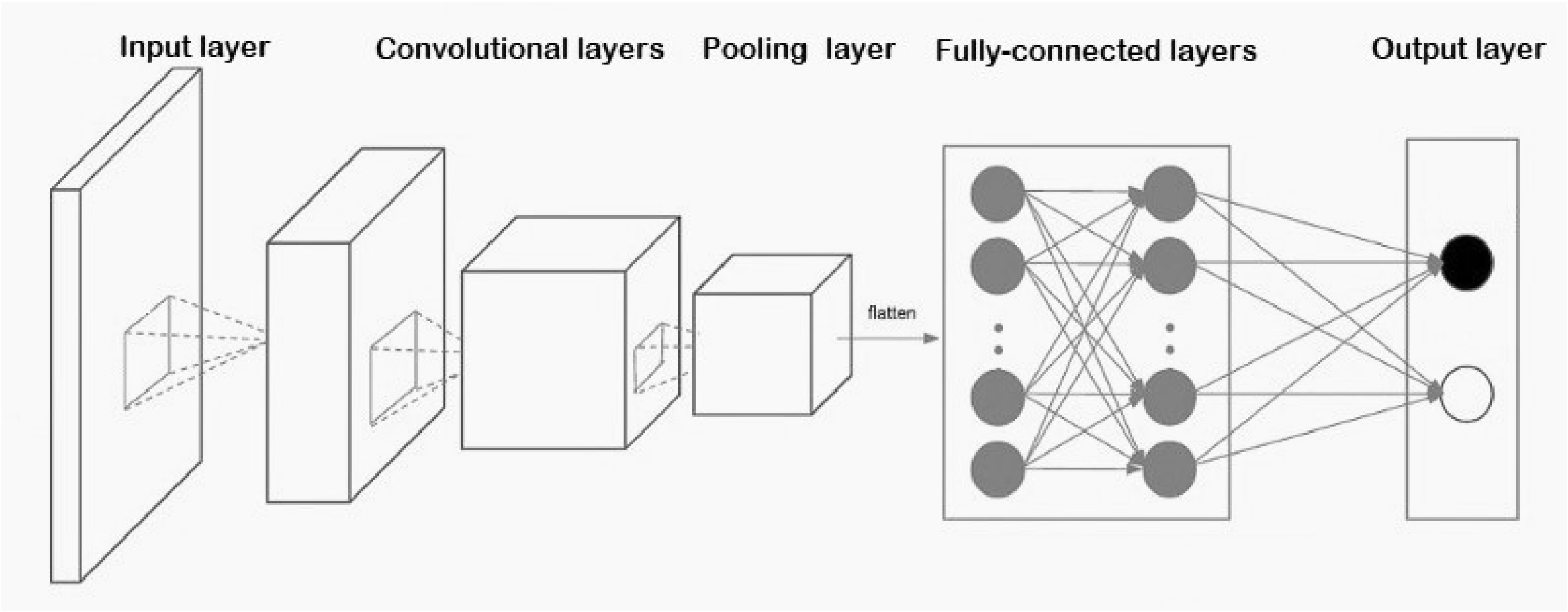

**Figure 4: A possible structure of a CNN which is used to classify objects on different images, source [42]**

### 1.2.1.1 Convolution layers

A digital image is a 2D array of pixels, where each element in position $(x, y)$ in the array is the position of the pixel, starting from left to right, and the value is the pixel's value. For colored images (RGB), pixels are characterized by three values, one for each color channel. So, an RGB image with height M and width N is stored as a MxNx3 matrix.

The Convolutional layer makes use of a set of learnable filters.

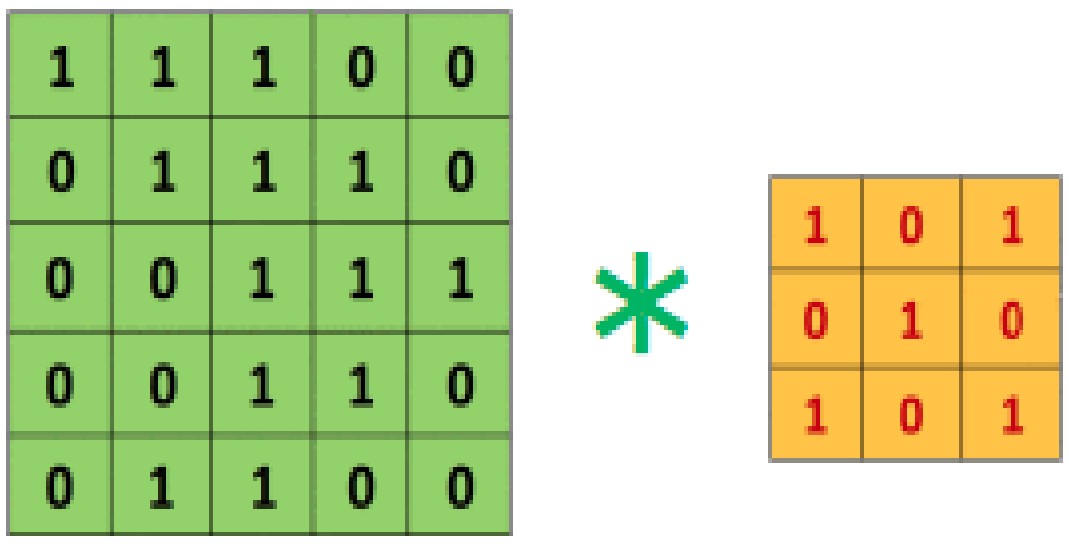

**Figure 5: 5x5 Input image with 3x3 filter, source: [43, 44]**

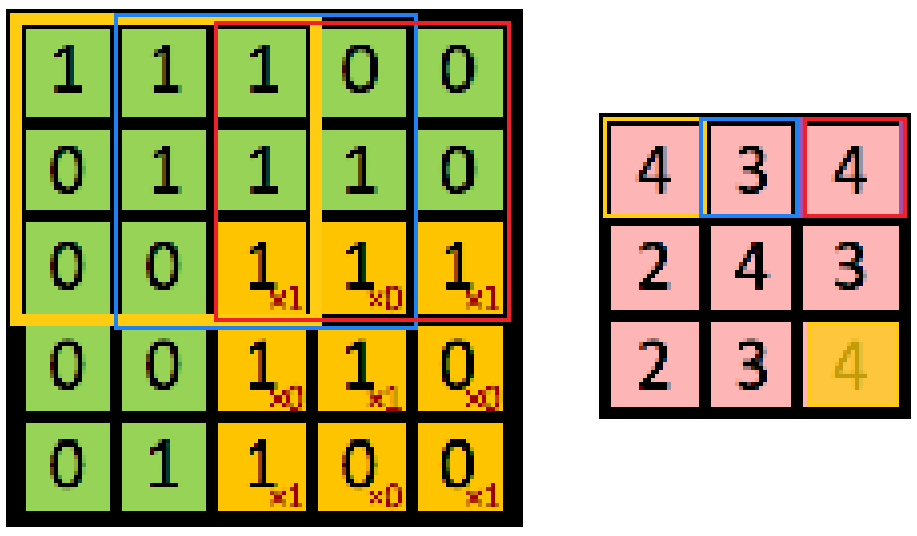

**Figure 6: 3x3 output feature, source: [44]**

An example of a filter that is convolved with an image array is shown in Figure 5. With pink color in Figure 6 is the result of the convolution with the yellow filer. The input is a 5x5 matrix and the filter of size 3x3 is applied with stride of 1. Therefore, the output volume size has spatial size:

$$output\,\frac{5-3+0}{1}+1=3$$

The filter slides one pixel at a time until it reaches the final and is applied on a sub-matrix with dimensions 3x3 of the input array, having as first pixel the current pixel.

$$pixel_{ij}=\sum_{i}^{N}\sum_{j}^{N} input_{pixelij}\cdot filter_{iJ}$$

So, the last element of the output array (pink) marked with yellow is computed:

$$1\cdot 1+1\cdot 0+1\cdot 1+0\cdot 0+1\cdot 1+1\cdot 0+0\cdot 1+0\cdot 0+1\cdot 1=1+1+1+1=4$$

A filter is used to detect specific features or patterns present in the input image. It is usually expressed as a matrix (MxMx3), with smaller dimensions than the input size, but with the same depth. This filter is convolved (slided) across the width and height of the input file, moving horizontally, starting from the upper left pixel, and a dot product is calculated to produce an activation map. The number of pixels the filter shifts over the input volume is called *stride*. When the stride is 2 then the filter moves 2 pixels at a time.

The parameters of the convolutional layer are mainly the set of its filters or weights that are learned through the training process. Every filter is small spatially but it extends in the entire depth of the input volume. For example, we have an input image of size 160x160x3. The receptive field will also have depth of size 3, the same depth as the input for example 15x15x3. During the forward pass, the filter slides over the width and the height of the input and the responses of the filter at every spatial position is given by the activation map. Each filter has one activation map as a result (Figure 6) and all the activation maps along the depth of the input, produce the output volume. If we symbolize the input size as *IS*, and the receptive field or filter size as *RFS*, the number of zero padding (zeros around the input on border) as *P* and the stride as *S*, then the output size is given by:

$$\frac{IS-RFS+2P}{S}+1$$

in this example, the output of the first layer will be 30 activation maps with size:

$IS=160$

$RFS=15$

$P=0$

$S=1$

$$output \frac{160-15+0}{1}+1=146$$

Every neuron of the convolutional layer will have weights to a [15x15x3] region in the input volume, having 15x15x3 equal to 675 total weights (+ 1 bias) parameters. That makes 15x15x30 connections equal to 6750.

So, the output size will be 146x146x30. This is the input for the second convolutional layer and 10 receptive fields of 20x20x30 (same depth as the input) are applied with 146x146x30 = 639480 neurons and 12000 connections. The output of the second layer would produce 10 activation maps of size

$$output \frac{146-20+0}{1}+1=127$$

So, the output would be 127x127x10. In Figure 7 we can see the results of the convolution of an 7x7x3 input with 2 filters 3x3, stride = 2 and zero padding = 1.

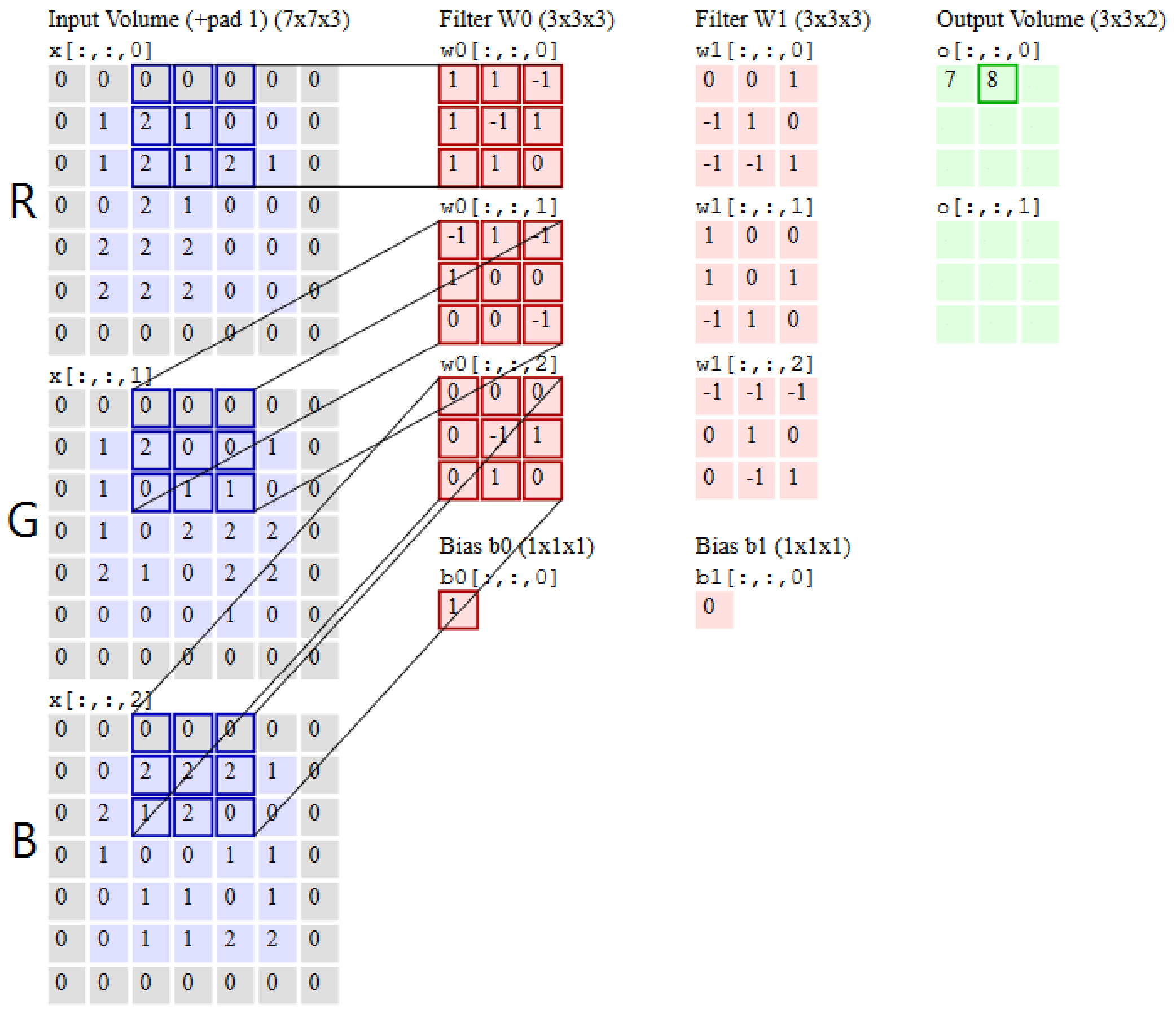

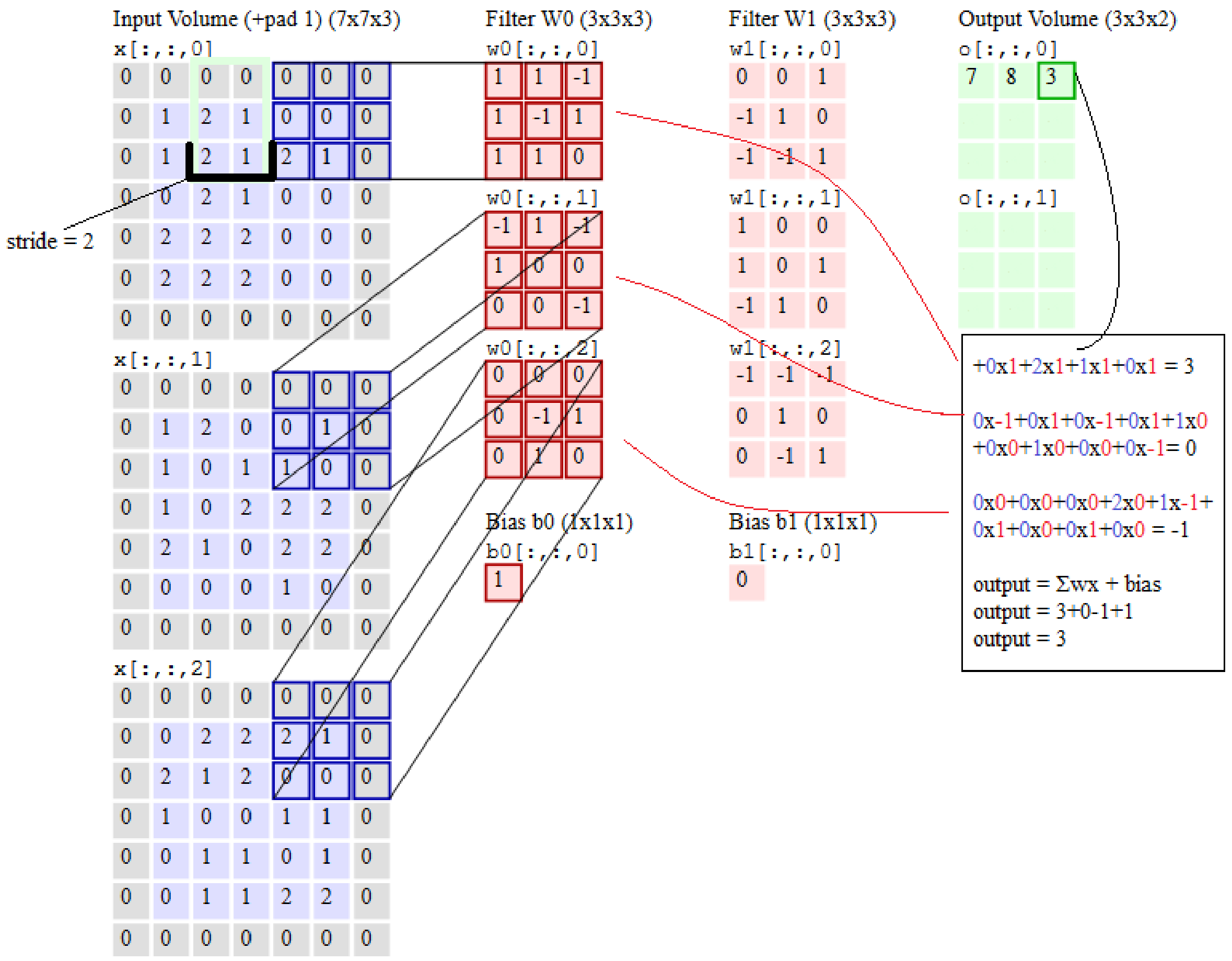

**Figure 7: Convolution of an RGB image with 2 filters and the computation of the activation maps.**

#### 1.2.1.2 Activation function layers

The activation function layer is the layer where the activation function is applied. The ReLU function is the most widely used activation function in CNNs today. One of its greatest advantages, over other activation functions, is that it does not suffer from the vanishing gradient problem [45]. In practice, ReLU converges six times faster than tanh and sigmoid activation functions.

A disadvantage ReLU possesses is that it is saturated at the negative region, meaning that the gradient in that region is zero. With the gradient equal to zero, during backpropagation all the weights will not be updated. To fix this, we use a handy tool, Leaky ReLU. Also, ReLU functions are not zero-centered. This means that for them to get to their optimal point, they have to use a zig-zag path which may be longer.

#### 1.2.1.3 Pooling layers

A Pooling Layer is very important for the operation of the CNN. The addition of a pooling layer after a convolutional layer is a common step that is used to reduce the spatial size of the representation, the amount of data and the number of required moves. It is usually placed between two convolutional layers. This step may be repeated one or more times in a given model. With pooling layers in CNNs, a down sampling feature map is created that summarizes the presence of different features in patches. The common pooling methods are max pooling, average pooling and sum pooling. Max pooling summarizes the maximum presence of a feature, while the average pooling the average presence of a feature. A Max-Pooling Layer slides a window of a given size $k$ over the input matrix with a given stride $s$ and gets the max/average/summarized value in the scanned submatrix.

An example of average and max pooling operation is shown in Figure 8. Here as input we have a 4x4, the window size or kernel is 2 ($k=2$) and the stride is 2 ($s=2$). The window is applied on the input array and slides with stride 2 until it reaches the final pixel. For max pooling, the output is the maximum value of the first sub-array (green), which is 21 (between 8 ,12, 19, 21). Then we slide the window 2 pixels right, so the first pixel of the new window (orange) is now 8. The second output is the maximum between 7, 8, 9, 12, which is 12. We follow the same procedure with the average.

The average in the first window (green) is $\frac{21+8+12+19}{4}=15$ and in orange window $\frac{8+12+9+7}{4}=9$.

Pooling is capable except for the reduction of data, to increase receptive fields.

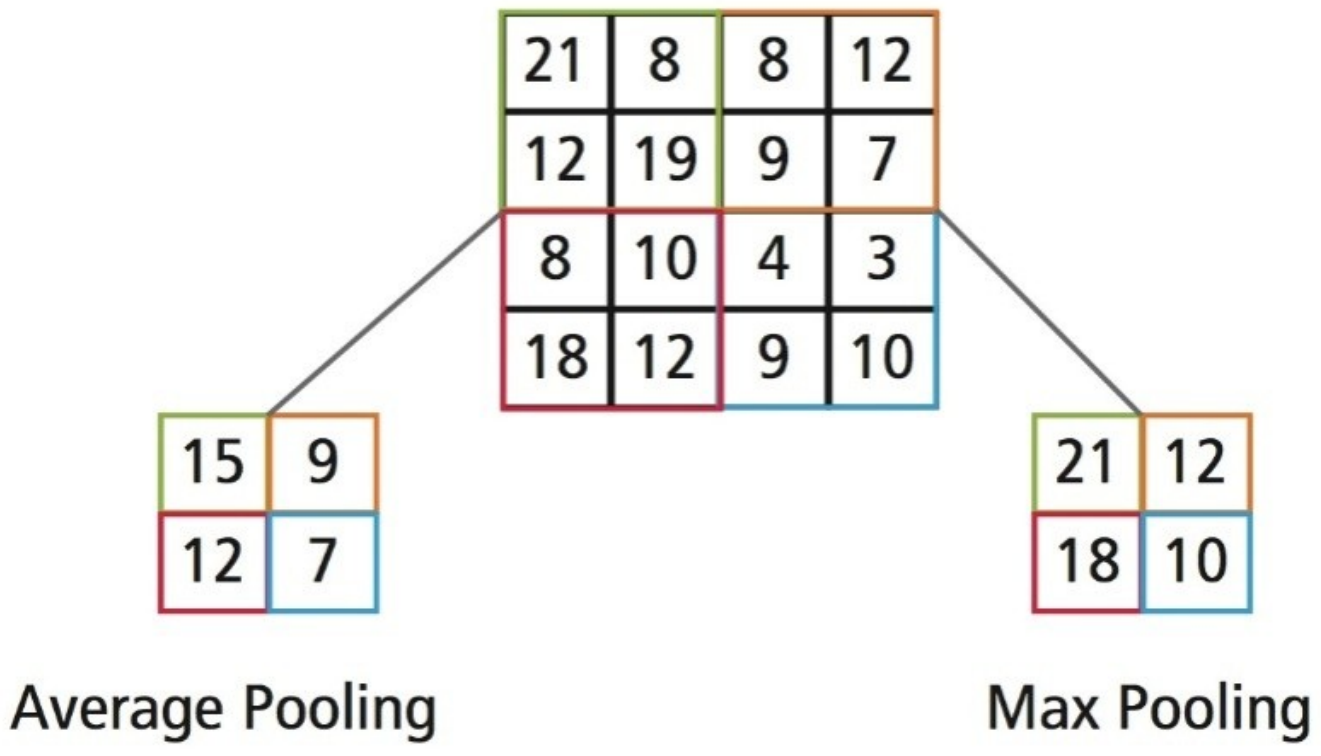

**Figure 8: Average pooling and max pooling with stride 2**

#### 1.2.1.4 Fully connected layers

A layer where each neuron receives input from each neuron in the previous layer is called fully connected. The ouput layer usually is fully connected.

**Dropout**

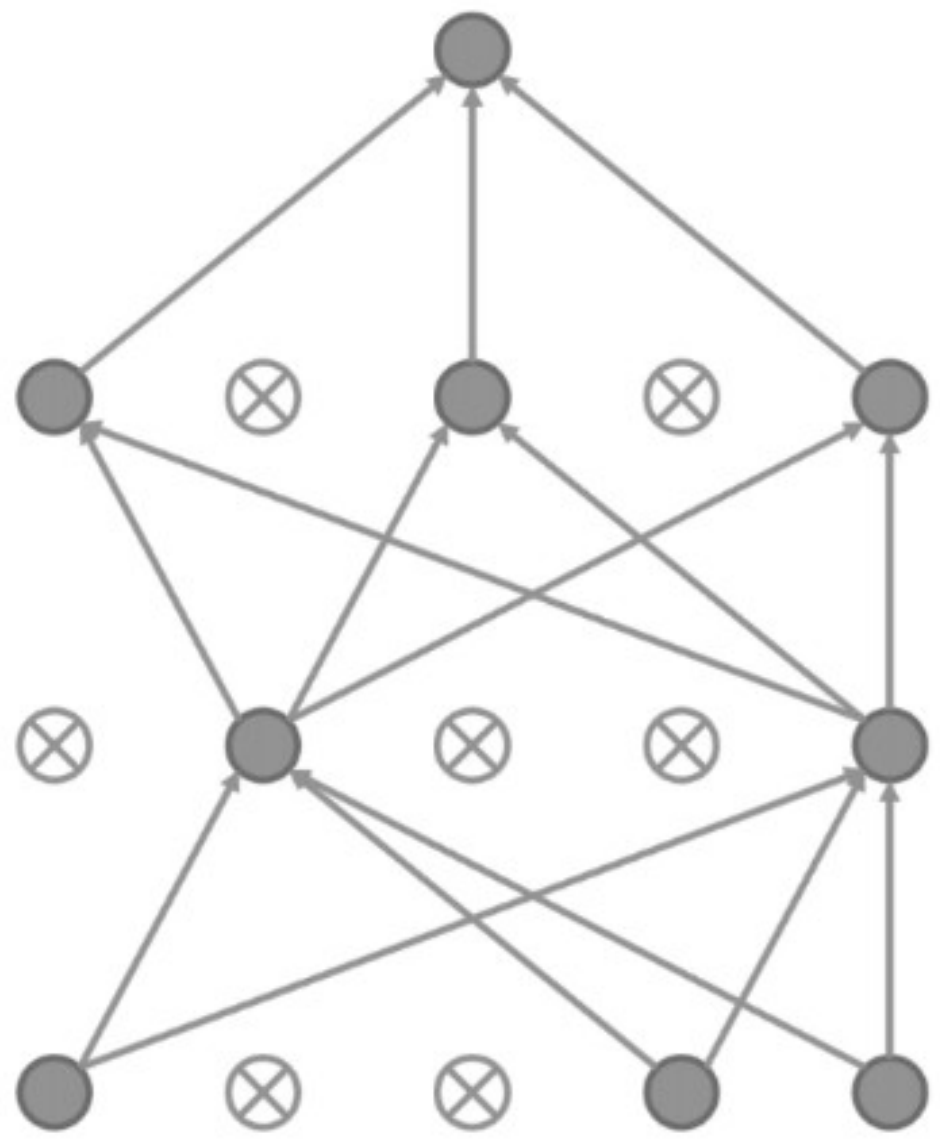

Figure 9: Dropout active

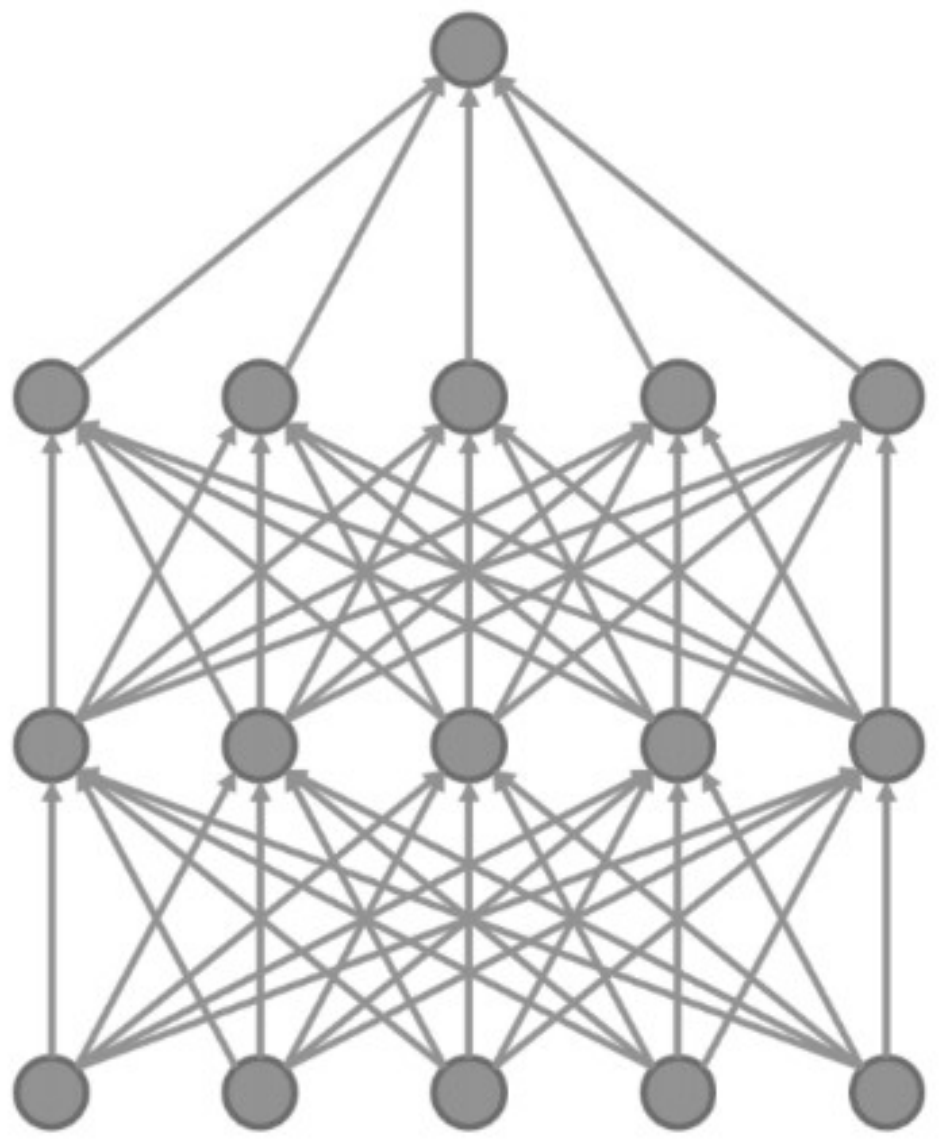

Figure 10: Dropout inactive

Dropout [38] is a regularization technique to train sub-networks by dropping non-output units from the original network randomly. This increases the generalization ability of the network and prevents overfitting. In this technique the output for each hidden neuron is set to zero (turned off) with probability around 0.5, called the dropout rate, so the network is forced to learn robust representations for the data. The units which are ignored in this way do not contribute to the forward pass and do not participate in back-propagation (Figure 9). Dropout usually increases the number of iterations to converge.

## 1.3 REINFORCEMENT LEARNING

### 1.3.1 Introduction

Researchers from many scientific fields have started using deep neural nets to model a wide range of new tasks including how to learn intelligent behavior in complex dynamic environments. In most machine learning applications people use supervised learning. This means that you give an input to your neural network model knowing the output your model should produce and therefore you can compute gradients and the backpropagation algorithm to train the network to produce your desired outputs. If we want to train a network to play a game, what we should do in a supervised setting is to have a human gamer play this game for a couple of hours and create a data set where we log all of the frames that the human saw on the screen and the actions that she/he performed in response to those frames. Then, we can feed these input frames through a feed-forward neural network. The output of this net can be mapped to the actions of the game. By training on the data set of the human game-play using backpropagation, we can actually train that neural network to replicate the actions of the human gamer. There are two significant disadvantages of this approach. First, the construction of the data set is tedious, and second, if we train our neural network model to imitate the actions of the human player, by definition our agent can never surpass the human player. So, if we want to surpass the best human performance we can't use supervised learning. In reinforcement learning the mechanism is quite similar with supervised models, but we don't know the target label.

Reinforcement learning is a set of machine learning algorithms based on trial and error in an environment. This technique lets an AI agent learn to complete an objective in an environment using time delayed labels, or what we call rewards, as a signal.

Reinforcement learning (RL) is the branch of machine learning that is concerned with making sequences of decisions. It considers an agent placed in an environment. The agent is trying to achieve progress through its actions toward a desired goal. RL algorithms, under the right conditions, can achieve superhuman performance. The environment penalizes or rewards the agents, according to the choice of action they make.

### 1.3.2 Markov Decision Process (MDP)

The agent starts at a state and follows its internal policy. At each step, it records the rewards obtained and saves the history of all visited states until reaching a terminal state. This sequence of states, from the starting state until reaching the terminal, we call an episode. This sequence, together with the transitioning rules, forms a Markov decision process.

$$Episode \rightarrow s_0, a_0, r_1, s_1, a_1, r_2, \ldots s_{n-1}, a_{n-1}, r_n$$

where $s_i$ is the state in $i_{th}$ timestep,$a_i, r_{i+1}$ are the action and the reward after performing the action $a_i$.

Every time our agent steps into a state, it is as if we are picking a value from the state set $S$ for the random variable $s$. For each state $s_i$ of each episode, we can calculate the returned reward and store it in a list. Repeating this process for a large number of times, the expected Q -value is guaranteed to converge to true utility.

### 1.3.3 Definitions of RL

Reinforcement learning is defined using the concepts of agents, environments, states, actions and rewards. In Figure 11 we can see all the elements of a Markov decision process (MDP).

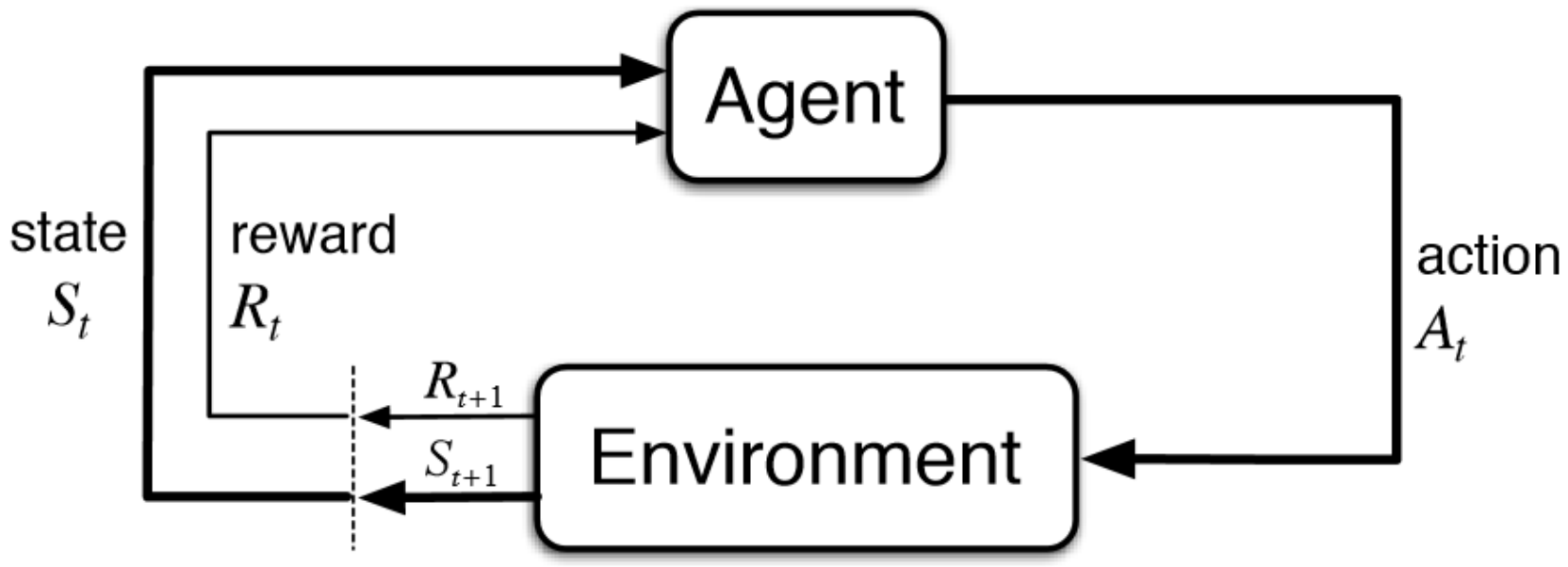

**Figure 11: The agent–environment interaction in a Markov decision process.**

**At time $t$, the agent receives state $S_t$ from the environment. The agent uses its policy to choose an action $A_t$. Once the action is executed, the environment transitions a step, providing the next state $S_{t+1}$ as well as feedback in the form of a reward $R_{t+1}$. The agent uses its knowledge of state transitions, of the form ($S_t, A_t, S_{t+1}, R_{t+1}$), in order to learn and improve its policy.**

- Agent: An **agent** is somebody or something that takes actions. Here when we refer to agents, we mean the algorithm, the running program.
- Environment: The environment is either a simulation of the world through which the agent moves or the real world. It takes the agent's current state and action as input, and returns the agent's potential reward and the next state as output.
- Action ($A$): A is the set of all possible moves an agent can make in the environment. Possible moves we can meet in some video games are up, down, left and right. So, every time an agent performs a move, it chooses among this list of possible actions.
- State ($S$): A **state** is a current, immediate, situation returned by the environment. A specific place and moment, an instantaneous configuration that puts the agent in relation to other significant things.
- Reward ($R$): A **reward** is the feedback send back from the environment to evaluate the last action. Rewards can be immediate or delayed. Using rewards, we effectively measure the success or failure of an agent's actions. Negative rewards are penaltied
- Policy ($\pi$): A **policy** is the strategy an agent employs to determine the next action based on the current state.

- Discount factor ($\gamma$): The parameter $\gamma$ is a number between 0 and 1. This quantifies the importance between immediate and future rewards. The further into the future the reward is, the less we take it into consideration. For example, if $\gamma$ is 0.8, and there's a reward of 10, after 3 time steps the present value of that reward is $0.8^3$ x 10. A discount factor of 1 would make future rewards worth just as much as immediate rewards.
- Value ($V$): $V\pi(s)$ is defined as the expected long-term return of the current state under policy π.
- Q-value or action-value ($Q$): **Q-value** is similar to Value, except that it takes an extra parameter, the current action $a$. $Q\pi(s,a)$ refers to the long-term return of the current state $s$, taking the action $a$ under policy $\pi$. $Q$ maps state-action pairs to rewards.
- Trajectory ($T$): Transitions between states**.** A sequence of states and actions that influence those states.

### 1.3.4 MDP Policies

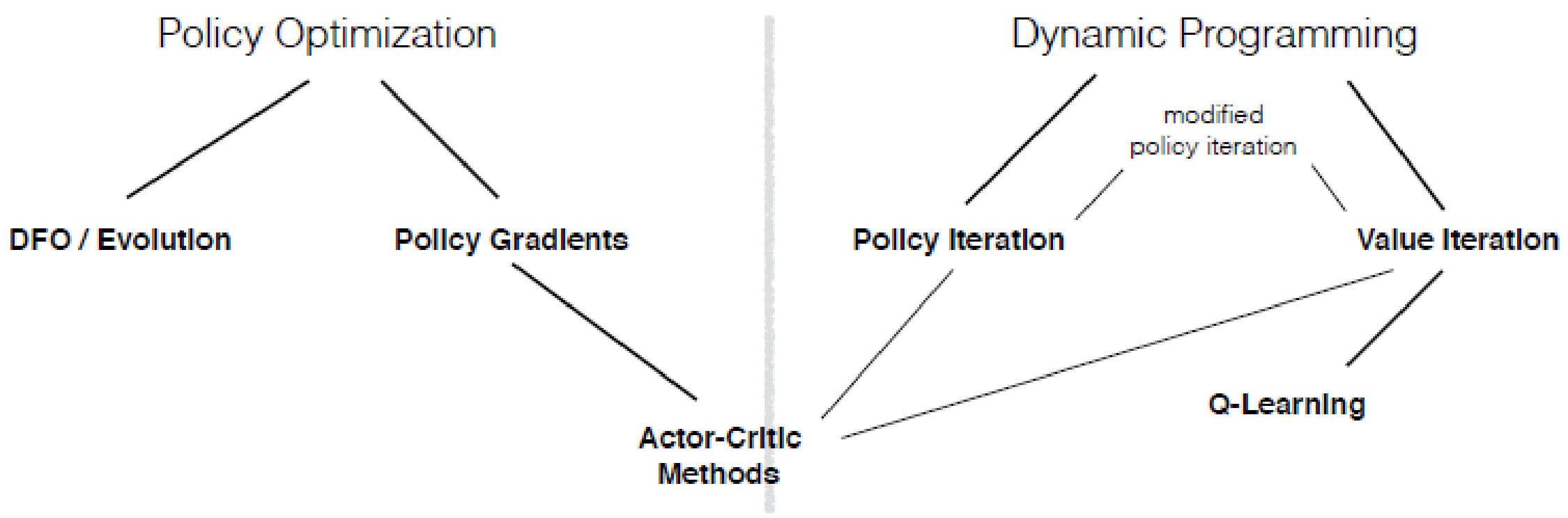

**Figure 12: Taxonomy of model-free RL algorithms, (source [17])**

There are two paths for computing optimal policies for an AI system, the policy iteration and the value iteration. In policy iteration, we study how an AI agent chooses its actions. We modeled our environment using a Markov decision process and we use a transition model to describe the probability of moving from one state to the other. This is the most common way to formalize a reinforcement learning problem.

When an agent provides an action to the environment as a function of the state and reward, this causes the environment to update and to provide a new state and reward to the agent in a feedback loop. So far, we have assumed that the agent knows what all the elements of the Markov decision process are. We can just compute the solution (the result after taking performing an action) to this decision-making problem, before actually executing this action in the environment.

Both value iteration and policy iteration algorithms are examples of planning algorithms but there is a big difference between planning algorithms and reinforcement learning algorithms. What makes a problem a reinforcement learning problem is that the agent does not know all the elements of the Markov decision process, so it would not be able to plan a solution. The agent has to try taking actions in the environment, observing what happens until somehow it finds a good policy and builds a model. The model doesn't know how the world would change in response to its actions, the transition function or what immediate reward it will receive. The pressing is how to find a good policy. There are two approaches that answer this question. The first approach for the agent to learn a model is through observations of the environment. Then by using all these observations the agent can plan a solution. If an agent is currently in a state $s$, takes an action $a$ and then observes the environment's transition to the next state with a reward, all that information can be used to improve the estimate of transition function $t$ and reward function $r$. Once the agent has a model for the environment, it can use a planning algorithm like policy iteration or value iteration with its learned model to find a policy. Reinforcement learning solutions that follow this framework are called model-based algorithms. This is when an agent exploits a previously learned model to accomplish a task in hand. But, it turns out that the agent doesn't have to learn a model of the environment to find a good policy. Sometimes, our agent can simply rely on trial and error experience from action selection. This is called model-free learning. In model-free reinforcement learning, the first thing we miss is the transition model and the second the reward function, which gives the agent the reward associated to a particular state. There are two approaches here, a passive and an active one. In passive approach we have a policy which the agent can use to move in the environment. In state $s$**,** the agent always produces an action $a$ given by a policy $\pi(\alpha)$. The goal here, is for the agent to learn the utility function. This is the case for Monte Carlo prediction. But it is also possible to estimate the optimal policy while moving in the environment and in this case, we are in the active approach.

### 1.3.5 Discounted Future Reward

To consider also long-term rewards we have to define the reward function in a way that the current reward is a linear function of the current and the future rewards. For an episode the total future reward from time $t$ until the end of the episode is

$$R_t = r_t + r_{t+1} + r_{t+2} + \cdots + r_n$$

The environment is stochastic and we need to focus on closer in time rewards. For that reason, it is common to use discounted future reward instead:

$$R_t = r_t + \gamma r_{t+1} + \gamma^2 r_{t+2} + \cdots + \gamma^{n-1} r_n$$

$$R_t = r_t + \gamma R_{t+1}$$

If we set the discount factor $\gamma = 0$, then the strategy will be short-sighted relying only on the immediate rewards.

### 1.3.6 Q-learning

If we want our agent to always choose an action that maximizes the discounted future reward, we want to use some form of TD (Temporal Difference) learning. We can define a function that represents the maximum discounted feature reward when we perform an action $a$ in state $s$ and continue optimally from that point on. This function gives the best possible score/reward at the end of the game after performing the action $a$. It is called **Q-function** because it represents the quality of a certain action in a given state [49]. We want to select the action that has as a result, the highest score at the end of the game. Once we built the Q function, things can become really simple, because all we have to do is to select the action with the highest Q value. Using the Q function we can estimate the score at the end of the game knowing just the current state and action and not knowing actions and rewards coming after that. The main idea is that we can iteratively approximate the Q-function using the Bellman equation:

$$Q(s_t, a_t) \leftarrow (1-\alpha) \cdot \underbrace{Q(s_t, a_t)}_{old\ value} + \underbrace{\alpha}_{learning\ rate} \cdot \overbrace{\left( \underbrace{r_t}_{reward} + \underbrace{\gamma}_{discount\ factor} \cdot \max_{a_{t+1}} Q \underbrace{(s_{t+1}, \alpha)}_{estimate\ of\ optimal\ future\ value} \right)}^{learned\ value}$$

where ***a*** is the learning rate, and ***γ*** is the discount factor described in 1.3.5 section.

The $maxQ(s_{t+1}, \alpha)$ that we use to update $Q(s_t, a_t)$ is an approximation, which in the early stages of learning is wrong but is getting through iterations until it reaches the true Q-value [50]. For $\alpha = 1$, the update is the same as the Bellman equation. In the simplest case, Q is implemented as a state-action table, where the states are the rows of the table and actions the columns (Figure 13). Starting running the algorithm, the Q table is initialized randomly. Then, the agent starts to interact with the environment and upon each iteration, the agent will observe the reward of its action and the state transmission.

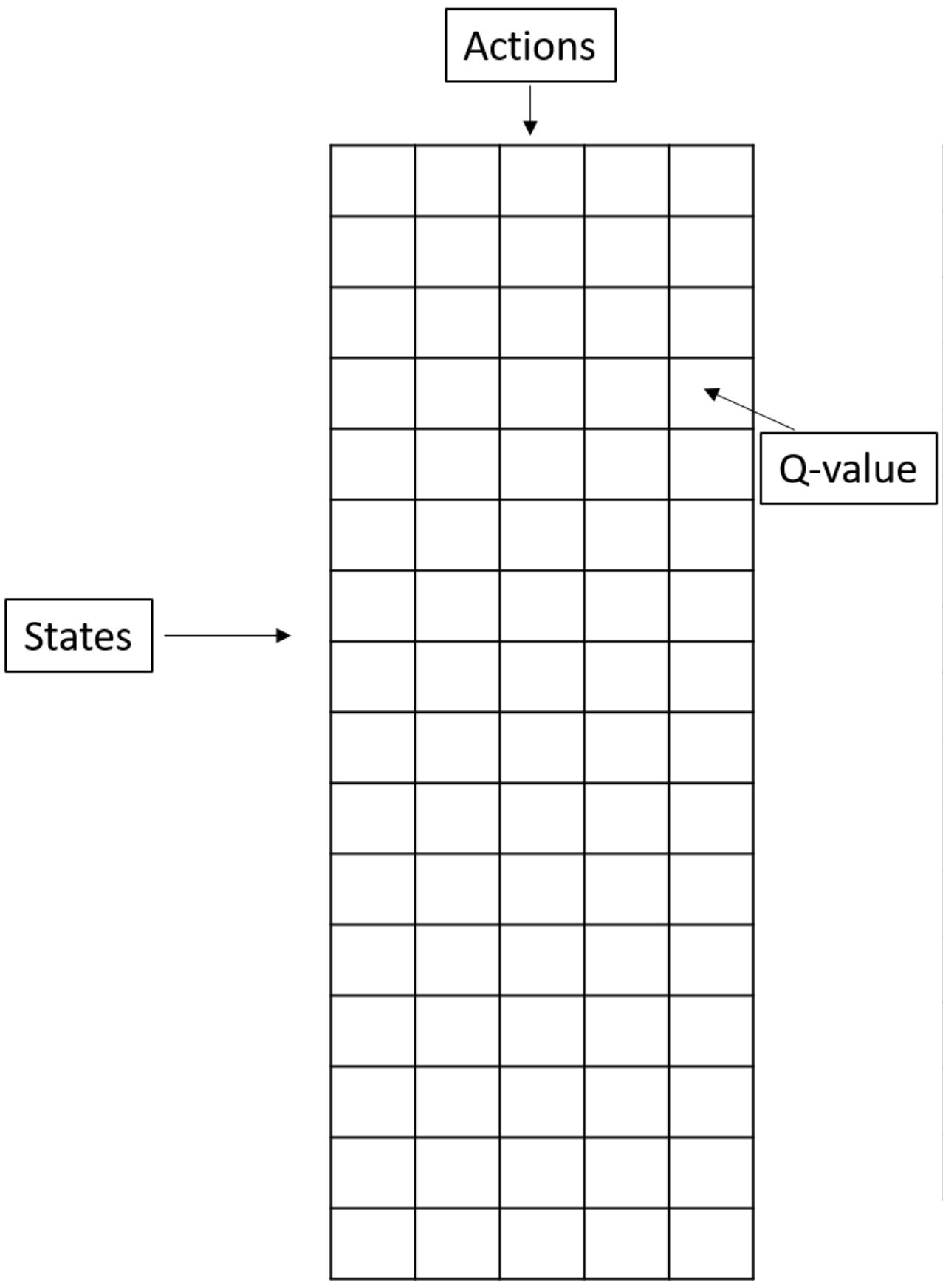

| | Up | Down | Left | Right | Shift |
|---|---|---|---|---|---|
| 1 | -500 | -100 | 200 | -100 | -100 |
| 2 | 0 | 0 | 0 | 500 | 0 |
| 3 | 0 | 0 | 100 | -100 | -100 |
| 4 | 0 | 0 | -100 | 100 | 0 |
| 5 | 100 | -900 | 0 | 0 | 0 |
| 6 | 0 | 0 | -300 | 500 | 500 |
| 7 | 10 | 200 | -100 | 0 | 0 |
| 8 | -500 | -200 | -100 | 20 | 0 |
| 9 | 10 | -200 | 100 | 0 | 0 |
| 10 | 0 | 0 | 0 | 500 | 0 |
| | 100 | -900 | 0 | 0 | 0 |
| | 0 | -900 | 0 | 500 | 0 |
| | 0 | 0 | 0 | 500 | 0 |
| | -200 | 0 | 0 | 500 | 0 |
| | 0 | -300 | -300 | -300 | 0 |
| 200 | 100 | -900 | 0 | 500 | 600 |

**Figure 13: Q-matrix**

The Q-learning algorithm needs to learn what actions can maximize the reward, and which actions need to be avoided. The algorithm (Algorithm 1) works as follows at a general level [51]:

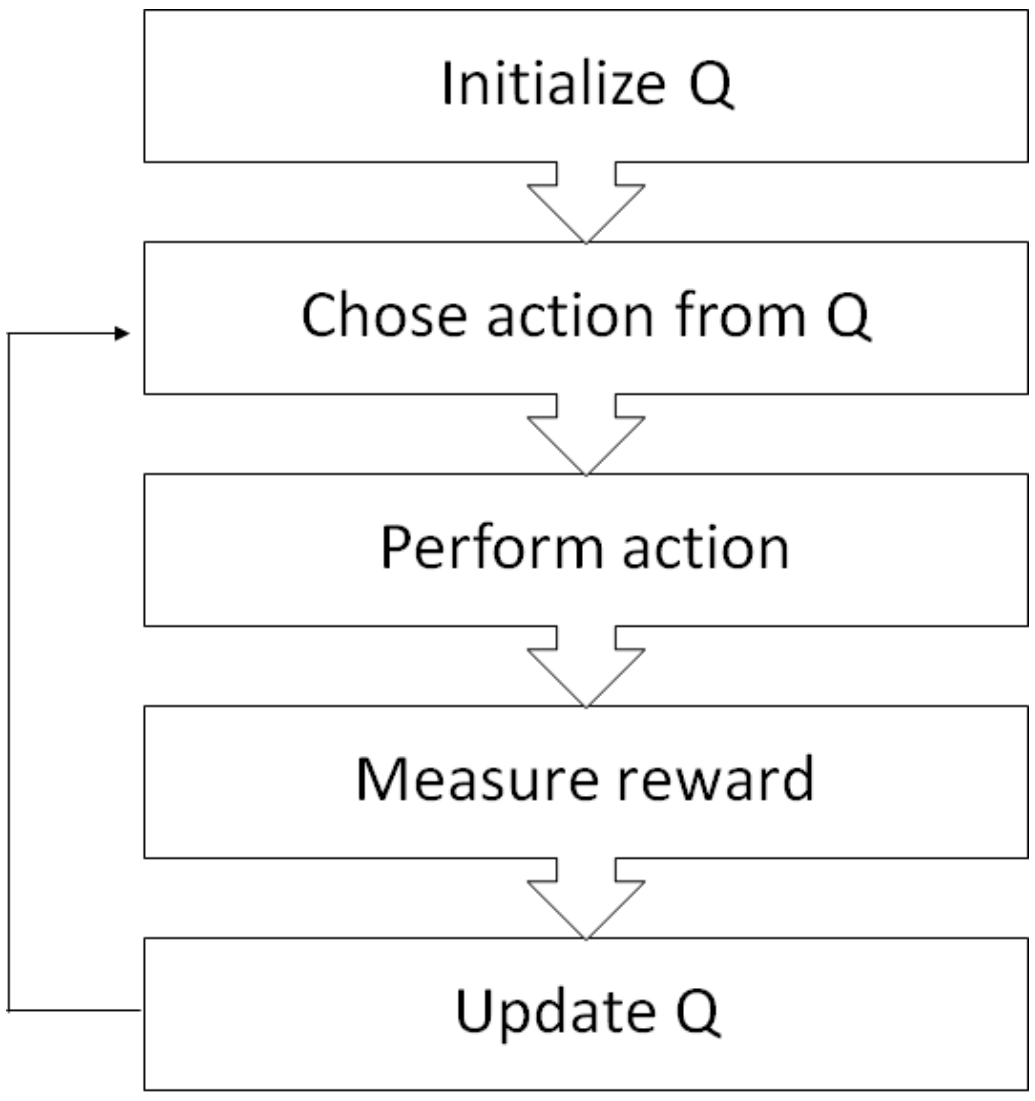

The game starts and the action is selected randomly. Then, the system receives the current state $s$ and performs an action randomly or on the basis of its neural network, depending on $s$. Often, it selects random behavior during the first iterations, in order to maximize exploration. Later, the system is increasingly relying on its neural network and collects the reward when the AI agent conducts the action. Receiving the new state $s_{t+1}$, it updates its Q-value as stated above with the Bellman equation. It also stores the original state, the action, the state reached after the action has been performed.

Q learning is a direct TD-method which learns the function $Q(a,s)$ in order to map state-action pairs to the expected return. In the traditional Q-learning algorithm the collected experiences through trial and error are used to adjust the network once and they are never reused again. Apart from the state-action table, the Q function can be implemented using a neural network where its final nodes represent actions. There is a technique called experience replay [52] that speeds up reinforcement learning by combining RL with teaching from experiences.

**Algorithm 1**

---

$Q$-learning: Learn function $Q : \mathcal{X} \times \mathcal{A} \rightarrow \mathbb{R}$

---

**Require:**
  Sates $\mathcal{X} = \{1, \ldots, n_x\}$
  Actions $\mathcal{A} = \{1, \ldots, n_a\}$, $\quad A : \mathcal{X} \Rightarrow \mathcal{A}$
  Reward function $R : \mathcal{X} \times \mathcal{A} \rightarrow \mathbb{R}$
  Black-box (probabilistic) transition function $T : \mathcal{X} \times \mathcal{A} \rightarrow \mathcal{X}$
  Learning rate $\alpha \in [0, 1]$, typically $\alpha = 0.1$
  Discounting factor $\gamma \in [0, 1]$
  **procedure** QLEARNING($\mathcal{X}$, $A$, $R$, $T$, $\alpha$, $\gamma$)
    Initialize $Q : \mathcal{X} \times \mathcal{A} \rightarrow \mathbb{R}$ arbitrarily
    **while** $Q$ is not converged **do**
      Start in state $s \in \mathcal{X}$
      **while** $s$ is not terminal **do**
        Calculate $\pi$ according to Q and exploration strategy (e.g. $\pi(x) \leftarrow \arg\max_a Q(x, a)$)
        $a \leftarrow \pi(s)$
        $r \leftarrow R(s, a)$ ▷ Receive the reward
        $s' \leftarrow T(s, a)$ ▷ Receive the new state
        $Q(s', a) \leftarrow (1 - \alpha) \cdot Q(s, a) + \alpha \cdot (r + \gamma \cdot \max_{a'} Q(s', a'))$
        $s \leftarrow s'$
    **return** $Q$

---

In conclusion, Reinforcement learning is a technique that lets an AI agent learn how to complete an objective in an environment, using time delayed labels as a signal. We can formally call all this representation a Markov decision process, which relates states, actions and rewards for an agent. Two fundamental ways of solving MDP problems are either value iteration or policy iteration algorithms.

## 1.4 DEEP LEARNING

Deep learning is a new and developing field of machine learning whose main objective is to bring machine learning closer to artificial intelligence. Deep learning models and algorithms learn representations of data with multiple levels of abstraction, realized by computational models that consist of multiple processing layers. The word "deep", refers to the number of layers of the neural network. In many cases, deep learning models are based on CNNs.

When processing images, lower layers comprise local combinations of features for example dots or edges. The edges form motifs, then motifs combine into parts, and parts form objects. Similar hierarchies exist in speech and text from sounds to phones, phonemes, syllables, words, and sentences. Deep learning can model the data structure in large datasets using the backpropagation algorithm on the raw data.

Deep Learning was introduced to the ML community in the 60s [53-55] . There are two main reasons why Deep learning has become popular today. The first is related to the fact that the computational complexity needed to train deep networks with thousands of neurons is very large and it would be extremely difficult or even impossible for a single-threaded computer or a computer cluster of the past to complete it. Nowadays, training such networks is much faster by using graphic cards (GPUs) that are able to achieve data parallel processing at high speed. The second reason was the lack of data for the training. Deep learning's specialty is that it requires big data as the network needs millions of instances to be trained and work properly. Deep networks are popular due to its ability to deal with large amounts of data, and today there are countless datasets available online from databases that can be used.

Lately, deep learning applications have gained much attention in the field of computational biology and bioinformatics. DNNs aid important studies, such as the activity of drug molecules [56, 57], the effects of mutations in non-coding DNA, gene expression and diseases [58-62]. Splice junctions can be identified easily using as input DNA sequences [63], sequence analysis [64], predict enhancers and regulatory regions [65-68], identify potential long non-coding RNA [69, 70], predict DNA methylation state [71, 72], single cell sequencing analysis [72-75], X-Ray classification, prediction of protein-protein interactions in PPI network, discovery of biomarkers [76], and RNA-protein binding sites prediction [66, 77] are some of the applications in bioinformatics.

### 1.4.1 Deep Reinforcement learning

Deep reinforcement learning belongs to the category of standard reinforcement learning where a deep neural network approximates a policy or a value function. The state is given as the input and the Q-value of all possible actions is generated as the output. To achieve high level of accuracy, it is required to train a deep neural network with a huge amount of data and a lot of real/simulated interactions with the environment.

### 1.4.2 Deep Q-Learning with experience replay

#### 1.4.2.1 Deep Q-Networks (DQN)

Neural networks are exceptionally good at learning good features for highly structured data. Q-function can be represented with a neural network, which will take every time $t$ the state $s$ and the action $a$ as input and will output the corresponding Q-value. A deep Q-network (DQN) is a combination of reinforcement learning with deep neural networks, which acts as the approximate function to represent the Q-value in Q-learning. Deep Q-learning is an introduction to deep reinforcement learning. In fact, Deep Q-learning is using a deep neural network function approximator, called the Q-network and leverages the advanced deep learning to learn policies from a high dimensional sensory input. In addition, it uses a discrete and finite set of actions A. The agent uses epsilon-greedy policy see section 1.4.2.2 to select actions and probabilistically choose between the action with the highest Q value and a random action. The core idea is for the neural network to learn a non-linear hierarchy of features-feature representations that gives accurate Q-value estimates. The neural network has a separate output unit for each possible action, which gives estimated Q-value for that action given the input state and is trained using mini-batch stochastic gradient updates and experience replay.

#### 1.4.2.2 Epsilon-greedy Policy

Q-learning is an online action-value function learning with an exploration policy. In epsilon-greedy policy the Q function is able to maximize Q at a given state $s_{t+1}$. More specifically in the epsilon-greedy policy, the agent either follows the greedy strategy with probability 1-$\varepsilon$, or a random action with probability $\varepsilon$. At each time step, the agent selects an action to take place. If ε has a higher value than a randomly generated number $p$, $0 \leq p \leq 1$, the AI agent picks a random action from the action space. Otherwise, the action is chosen according to the $Q(s, a)$.

#### 1.4.2.3 Exploration vs. exploitation

One of the main challenges in reinforcement learning, is to find a balance between exploration and exploitation [17, 78, 79]. An RL agent must always discover new actions that hasn't selected before (explore), but at the same time prefer the ones that have proved their effectiveness in the past. The agent has to exploit what has already been experienced in order to obtain more rewards, but it also has to explore in order to make better action selections in the future. The key to the dilemma is to find the point where the agent will stop using its time and resources to explore new and more efficient solutions and start capitalizing its already known methods (exploit). The exploration–exploitation dilemma has been studied by mathematicians for many decades. One simple strategy is to use 80% of its time for exploitation and the other 20% for exploring new actions.

#### 1.4.2.4 Experience Replay

We have to use random sample experiences instead of sequential experiences because, sequential experiences are highly correlated with each other. More specifically, each state is a sequence of actions and observations (states), a tuple

$s_t = s_1, a_1, s_2, \ldots, a_{t-1}, s_{t-1}$

We store the agent's experiences at each time step t, as $e_t = (s_t, a_t, r_t, s_{t+1})$

in a dataset $D = e_1, \ldots, e_n$ and pool over many episodes into a replay memory. Random sampling of experiences, breaks this temporal correlation of behavior and distributes it over many of its previous states. In practice, we only store the last $N$experience tuples in the replay memory and sample uniformly from $D$ when performing updates. Then we sample a random mini-batch of experience tuples uniformly at random from $D$. The following pseudo-algorithm (Algorithm 2) [80] implements the Deep-Q Learning with Experience Replay.

**Algorithm 2: Deep Q-learning with Experience Replay [80]**

Initialize replay memory $\mathcal{D}$ to capacity $N$
Initialize action-value function $Q$ with random weights
**for** episode $= 1, M$ **do**
  Initialise sequence $s_1 = \{x_1\}$ and preprocessed sequenced $\phi_1 = \phi(s_1)$
  **for** $t = 1, T$ **do**
    With probability $\epsilon$ select a random action $a_t$
    otherwise select $a_t = \max_a Q^*(\phi(s_t), a; \theta)$
    Execute action $a_t$ in emulator and observe reward $r_t$ and image $x_{t+1}$
    Set $s_{t+1} = s_t, a_t, x_{t+1}$ and preprocess $\phi_{t+1} = \phi(s_{t+1})$
    Store transition $(\phi_t, a_t, r_t, \phi_{t+1})$ in $\mathcal{D}$
    Sample random minibatch of transitions $(\phi_j, a_j, r_j, \phi_{j+1})$ from $\mathcal{D}$
    Set $y_j = \begin{cases} r_j & \text{for terminal } \phi_{j+1} \\ r_j + \gamma \max_{a'} Q(\phi_{j+1}, a'; \theta) & \text{for non-terminal } \phi_{j+1} \end{cases}$
    Perform a gradient descent step on $(y_j - Q(\phi_j, a_j; \theta))^2$
  **end for**
**end for**

### 1.4.3 DRL Applications

In this subsection we will discuss about the applications of deep reinforcement learning that have been released in recent years and we'll further analyze the most important achievements in the history of DRL in electronic games. Deep Reinforcement learning has various application domains including but not necessarily limited to computer vision [81-83] field, which is dealing with how computers gain understanding from digital images or videos, computer systems [84-86], robotics [87-89] and games [80].

Also, application of machine learning in healthcare focused mostly on diagnosing has yielded many impressive results. In 2017 Google DeepMind launched the DeepMind Health [90] to develop effective healthcare technologies. Advice on selecting treating methods can indirectly help at treating people where there are many available treatment options, figuring out the best treatment policy to use for a particular patient is challenging for human decision makers. n RL literature this is referred to as “Off-Policy Evaluation”. Many RL algorithms such as Q-learning can, “in theory,” learn the optimal policy effectively in the off-policy context. In a normal RL context to evaluate a policy we would simply have the agent make decisions then compute the average reward based on the outcome. However, as mentioned above, this is not possible due to ethical and logistical reasons. There are several biomedical applications of DRL, such as new designed molecules can be optimized using DDQN (Double Deep Q-Network) [91] and  classification of skin cancer [92],. In 2017 [93] one of the first articles to discuss the application of DRL to healthcare problems was published featuring a Double-Deep Q Network working on identifying a policy for sepsis treatment. Another similar work for sepsis treatment was published later [94] focusing on the slightly different approach of looking only at glycemic control. In cancer treatment, another study [95] with a model consisted of actions in the form of quantities of doses for given durations and an agent equipped with a Q learning algorithm rewarding reductions in tumor diameter, attempted to propose the best chemotherapy treatment. Furthermore, a paper using supervised RL in conjunction with 3 RNNs utilized the full MIMIC-III dataset to provide a treatment plan among 1000 possible medications [96]. Last but not least a study on RL on Graft Versus Host Disease (GVHD) [97] stands out for challenging a dynamic state-action space (diseases-treatments). Another interesting, trained a CNN to classify open source images of suspect lesions as melanoma or atypical nevi and its results were outperformed 136 of 157 dermatologists and physicians.

All these methods utilize a range of neural network architectures, including CNN, multilayer perceptrons, restricted Boltzmann machines and RNN.

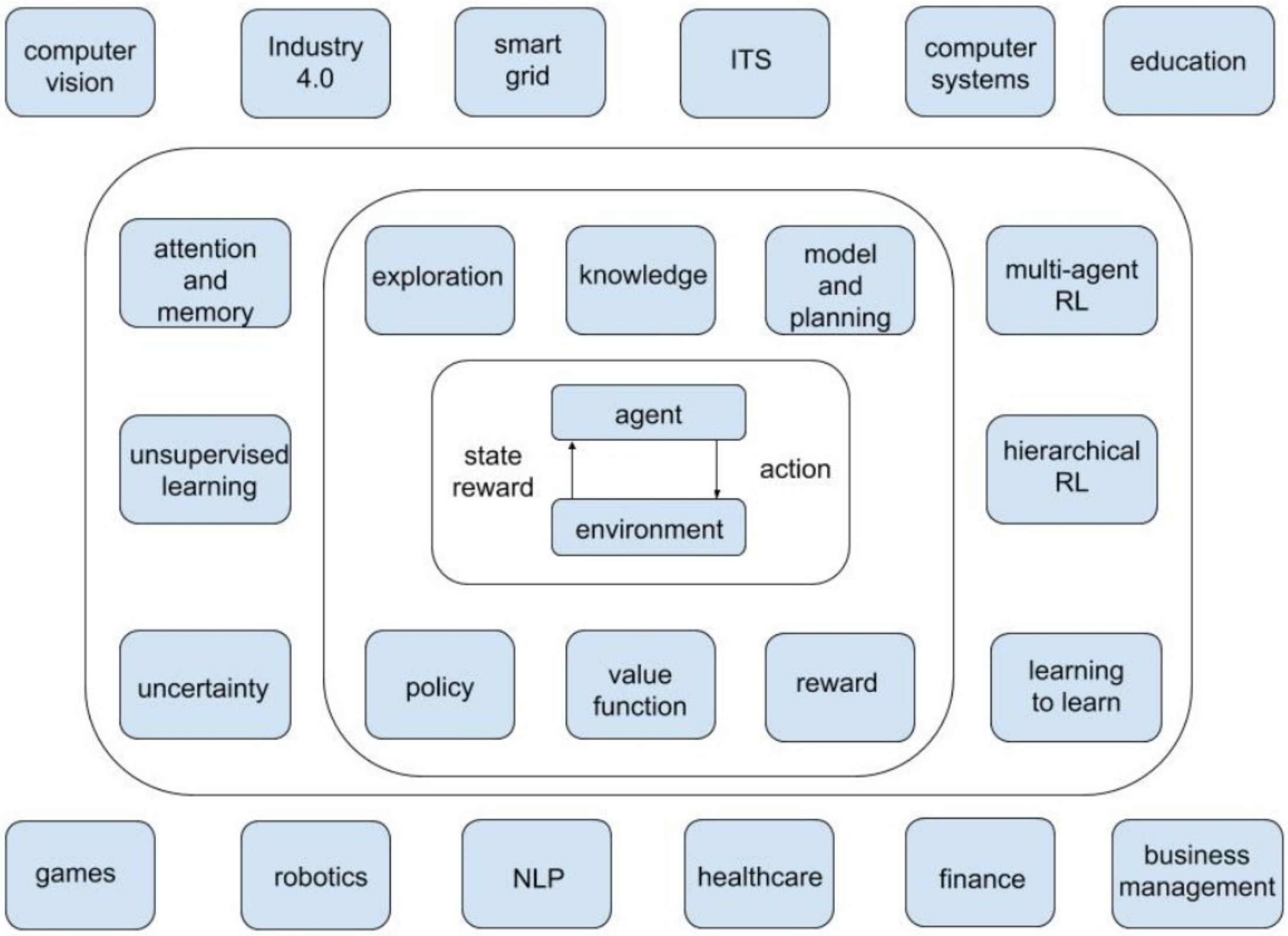

**Figure 14: Applications of DRL, source [98]**

Games and especially board games like backgammon and chess, are very popular testbeds for DRL algorithms and research companies like DeepMind use games to test their algorithms. The reason is that games support the concept of an environment where the agent has to explore it and best interact with it by performing actions. Most games have enough diversity in simulation environments to be an important first step towards AI. Methods like DQN are especially successful for video games, where one can learn using video frames and the instant reward.

The primary work in the field started with the development of agents, appropriate to interact with humans through conversation as a tech support helpdesk. The big change on the field started in 2015 by Google [80] when Google trained an AI agent that could discuss morality, express opinions, and answer general facts-based questions. At the same time, DeepMind [80, 99] developed an agent that surpassed human-level performance at 49 classic Atari 2600 games, receiving only the pixels and game score as inputs. This research filled the gap between high-dimensional sensory inputs and actions, developing the first artificial agent capable to succeed in a variety of challenging tasks.

Soon after, in 2016, DeepMind released a new game play method called A3C [100]. Go is a Chinese strategy game that has $10^{170}$ possible board positions and 46 million players all over the world. This game was still dominated by humans for two decades after machines first conquered chess (Deep Blue) [101]. AlphaGo [47] managed to defeat one of the best human players, using a combination of supervised, reinforcement learning and traditional heuristic search algorithm. In addition, in March 2017, OpenAI [102] created agents that invented their own language to cooperate and achieve more effectively their goal. Facebook also, has reportedly trained agents to negotiate and even lie [103]. Recently OpenAI reached another milestone by defeating, at world-competitive levels in 1-on-1 matches, the world's top professionals of the online multiplayer game Dota 2 [104], using Proximal Policy Optimization (PPO) algorithm [105].

In January of 2017 [106], a superhuman AI for Texas hold'em poker was presented. The algorithm was designed by N. Brown and T. Sandholm, of Carnegie Mellon University and Facebook. The name of the agent was Libratus and it was able to defeat 10.000 hands of a multiplayer poker with 6 human players. Among all the achievements this was considered the most impressive for two reasons. The first is that poker is a multiplayer game and the most referred research until then was only on 1-to-1 matches. The second is that Texas hold'em has an extra difficulty in performing the best action than chess and GO, because most of the elements are not visible. Therefore, it is not possible to predict the necessary information. Other examples of DRL agents playing games are Ms. Pac-Man [107, 108], Project Malmo [109, 110], Brood War API (BWAPI) [111], StarCraft II [112] ,Quake III Arena and Montezuma's revenge [113].

Many implementation tools exists for AI applications, including libraries and toolkits such as TensorFlow, PyTorch, OpenAI Gym [114], which are integrated with many game engines (Unity, Unreal Engine). There are many libraries that allow researchers to rapidly build controllable environments for their experiments. One famous library for developing game applications is Pygame [115]. Pygame is free and Open Source and has more than 4845 games so far. On the other hand, there are frameworks with interesting learning environments along with the necessary APIs to interact with them. Their extensive use consolidated them as benchmarks for game AI applications. Some of the most famous examples are:

- Arcade Learning Environment [116, 117]: An object-oriented environment that offers more than 50 different Atari video games to develop AI agents on [78, 80, 118]. It is mostly used for General Video Game Artificial Intelligence (GVGAI) applications.

- VizDoom [119]: A reinforcement learning environment based on “Doom” game. The learning process is focused on raw visual data, it is thus suited for deep reinforcement learning applications.
- TORCS: An AI research platform for car racing agents in a 3D environment, primarily focused on visual reinforcement learning to test DLR algorithms with continuous actions. It offers built-in data structures for neural networks applications [120].
- Project Malmo [109, 110] from Microsoft, is an AI research and experimentation platform built on top of Minecraft.
- Twitter torch-twrl: an open-source framework for RL development [121].

**Table 2: DRL Frameworks & SDKs**

| **Deep Learning Frameworks** | | | |
|---|---|---|---|
| **a. Frameworks** | | | |
| Tensorflow | https://www.tensorflow.org | CNTK (Microsoft) | https://github.com/Microsoft/CNTK |
| Caffe | http://caffe.berkeleyvision.org | MatConvNet | http://www.vlfeat.org/matconvnet/ |
| KERAS | https://keras.io | MINERVA | https://github.com/dmlc/minerva |
| Theano | http://deeplearning.net/software/theano | MXNET | https://github.com/dmlc/mxnet |
| Torch | http://torch.ch | OpenDeep | http://www.opendeep.org/ |
| PyTorch | http://pytorch.org | PuRine | https://github.com/purine/purine2 |
| Lasagne | https://lasagne.readthedocs.io/en/latest | PyLerarn2 | http://deeplearning.net/software/pylearn2 |
| DL4J (DeepLearning4J) | https://deeplearning4j.org | TensorLayer | https://github.com/zsdonghao/tensorlayer |
| DIGITS | https://developer.nvidia.com/digits | LBANN | https://github.com/LLNL/lbann |
| Tensorforce | https://github.com/hill-a/stable-baselines | Stable Baselines | https://github.com/hill-a/stable-baselines |
| OpenAI Baselines | https://github.com/openai/baselines | TF Agents | https://github.com/tensorflow/agents |
| **b. SDKs** | | | |
| cuDNN | https://developer.nvidia.com/cudnn | cuBLAS | https://developer.nvidia.com/cublas |
| TensorRT | https://developer.nvidia.com/tensorr | cuSPARSE | http://docs.nvidia.com/cuda/cusparse/ |
| DeepStream SDK | https://developer.nvidia.com/deepstream-sdk | NCCL | https://devblogs.nvidia.com/parallelforall/fast-multi-gpu-collectives-nccl/ |

## 2. BIOLOGICAL BACKGROUND

Part 1.2 is an introduction to the main problem we examine in this thesis, the protein folding problem. At the end of this section we suggest the novel approach Deep Foldit, which is analyzed further in Chapter 3. In the first section we describe the problem we face and the current scientific work around it. Section 2.2.2 describes a gamification approach to the protein folding problem, its rules and a modified version called **Foldit Standalone**.

## 2.1 PROTEIN FOLDING

Protein structure prediction (PSP) is a developing and very important subfield of bioinformatics. Based on a protein's amino acid sequence, PSP targets to infer the protein three-dimensional folded structure, which in turn, determines to a great extent the protein's biological function. Advances in PSP can be of invaluable importance to the study of many proteins, the functionality of which is still to be determined. The more information we know about how a certain protein folds, the better we can design new protein structures to combat diseases that are related to proteins. PSP and protein folding are considered as NP-complete problems in computational theory [122] and are among the most difficult in terms of computational requirements. The complexity is due to the protein's size, that determines the size of the conformational space reflects all the positions and orientations a protein can take. Among many possible folds we have to find the one structure the protein folds into, which is usually the one with the minimal free energy.

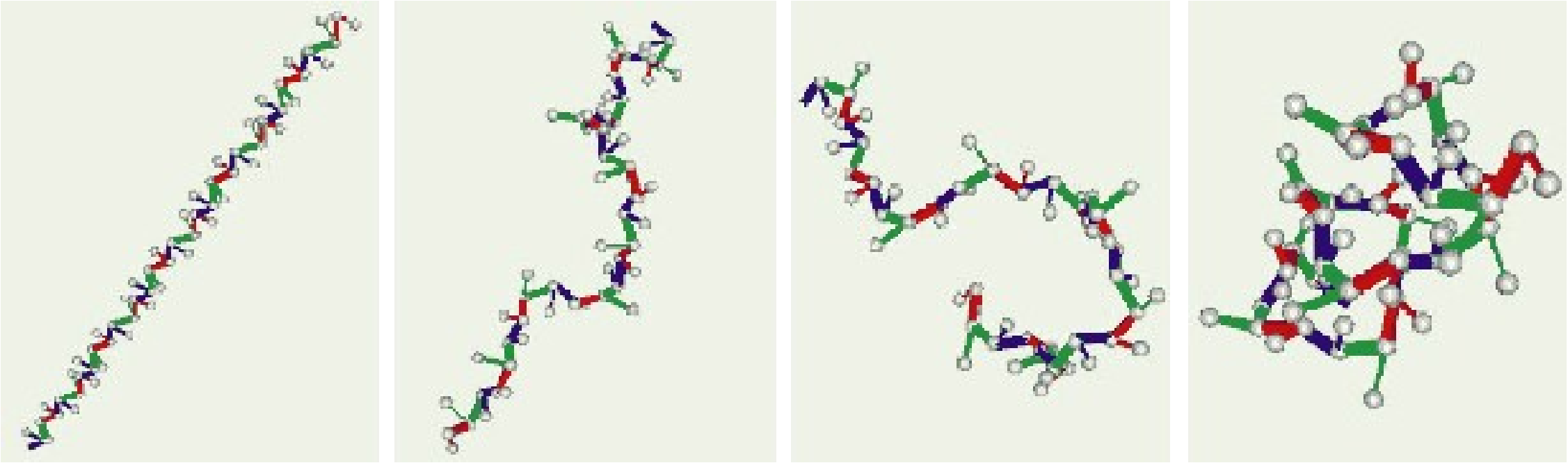

**Figure 15: Stages of a folding protein**

Every protein can fold up into a very specific shape, the same shape every time. Some proteins form their 3D structure in space by themselves and some others need extra help from chaperone proteins. These unique shapes proteins have is the most stable state a protein can adopt, the point where the total free energy is minimized. To determine which of all the possible structures is the best one is regarded as one of the hardest problems in biology today and current methods are expensive and very slow. There has been an important research focus on the protein structure prediction computationally, mainly through distributed large-scale computers. Unfortunately, these projects have shown promising but limited success [123].

Proteins are long sequences of amino acids (each amino acid is one of out of the 20 amino acids), linked together into contiguous chains, the polymers. A protein proceeds through three main structural classes to obtain its final form, primary, secondary and tertiary. Primary is the simplest form and essentially the linear amino acid sequence. Secondary structure includes local folded structures, helices, sheets, and coils that are formed when the sequences of primary structures tend to arrange themselves into regular conformations due to interactions between atoms of the backbone. Tertiary or 3D structure is the overall three-dimensional structure of a polypeptide when secondary structure elements are packed against each other in the most stable configuration [124]. The tertiary structure is formed primarily due to interactions between the R groups, which are the side chains, (hydrogen bonding, ionic bonding, dipole-dipole interactions, and London dispersion forces) of the amino acids that make up the protein. In some proteins, there is another final level called quaternary structure. Quaternary structure defines the arrangement of multiple polypeptide chains, that are grouped together.

Although there has been a rapid growth in the sequencing of considerable genomes in the last decades, in the post-genomics there is a huge gap between the proteins that have been identified experimentally in Protein Data Bank (PDB) [125] and proteins with unknown structures, whose sequences are identified and stored in databases [126]. The Human Genome project has produced a huge amount of protein sequences by large-scale DNA sequencing, but the identification of their 3D shape, through experimental methods such as X-ray crystallography, electron microscopy or nuclear magnetic resonance (NMR), is hugely, computationally expensive, time-consuming and lagging far behind the current output of protein sequences. Only a small portion of the protein sequences that have been discovered , have had their 3D structures experimentally uncovered , which is less than 1% [127].

The need for computational methods rather than laboratory techniques alone to predict protein structure becomes inevitable. In the past ten years several computational methodologies and algorithms have been employed to predict the three-dimensional protein structure and can be separated into three major categories, namely, homology modeling, threading (fold recognition) and ab initio modeling methods. Homology modeling was founded by Greer on 1981 [128] and it was the first semi-automated program. The basic idea is to predict an unknown protein by comparing and utilizing the available information with known homologous sequences [129]. The homological approach is very successful because there are experimental data for at least one member of every protein family, which can be used as a template for modeling.

Threading algorithms are used to find a good homologous protein to use it as a template structure. In threading, the number of folds is limited and that's why homologous proteins have similar structures. The threading method is based on the distance between the optimal alignment score and the mean alignment scored and is called the Z-score. This is obtained by random shuffling of the target sequence. With this method, we create a database for standard protein structures and a minimization scoring function that is used to find the optimal alignment between the target and the standard protein.

If the target protein has a homolog, the task is relatively easy and high-resolution models can be built by copying the solved structure framework. However, this procedure does not answer the question of how and why a protein adopts its particular structure. If structure homologs do not exist or exist but it is difficult to be recognized, models must be built from scratch. This procedure called ab initio modeling, and is essential in finding a universal solution to the predictive problem of the protein structure [130]. Ab initio modeling conducts a conformational search under the guidance of a designed energy function using thermodynamic laws and molecular energy parameters. Ab initio protein folding is considered a global optimization problem, where the goal is to find optimal positions for the atoms. Ab initio modeling can also guide us to realize the physicochemical principle of the way a protein folds through its natural code and why it adopts this particular shape. Only these methods can obtain novel protein folds. However, the conformational space is really big, even for small molecules, making ab initio modeling a difficult problem to solve, and is thus restricted to small proteins (less than 100 residues).

Some tools and algorithms that are worth mentioning are Rosetta [[131] (generate a substantial number of protein models due to the typically large number of local minima using Monte Carlo method), CHARMM [132], AMBER [133] and GROMOS [134] (molecular dynamics simulations by solving Newton's equations).

## 2.2 FOLDIT

Despite considerable progress, ab initio protein structure prediction remains unsolved. Protein folding is computationally a very difficult task due to the large numbers of protein solutions that need to be tested. A crowdsourcing approach to this is the online puzzle video game Foldit [1]. Foldit attempts to predict the structure of a protein by taking advantage of humans' puzzle-solving intuitions and having people play competitively to fold the best proteins. It has provided several useful results that matched or outperformed algorithmically computed solutions. Foldit manages to pool creative solutions of protein folding from people around the world, while at the same time being amusing as a game.

In May 2008, Foldit was developed at the University of Washington, Center for Game Science, in collaboration with the UW Department of Biochemistry, Northeastern University, Vanderbilt University, University of California and University of Massachusetts. The idea was to build a game that is fun and approachable but also motivates people to play it. Foldit started as a 3D Tetris but designed for proteins. In Tetris, the only rule is to fit all the blocks together and fill as much empty spaces as possible to remove the lines. It resembles protein folding in the sense that we try to remove, the empty space from the interior of the protein and pack everything as tightly as possible. So, instead of the different types of blocks that Tetris has, Foldit has amino acids.

A protein is presented to Foldit players, which they can fold using a host of provided tools. The game evaluates how good every move a player performed was, by returning them a score (positive or negative) as also a total score of the fold. There is also a ranking for every player, that is calculated in comparison with other online players' scores.

In addition, the game records structures, moves and strategies of its players and uses all these gathered data to improve the game in every aspect (more qualified results, how many introductory levels the game has etc). Foldit often releases updates with new features, aiming to endear the game to its players. Once a week, Foldit publishes new sets of puzzles keeping its audience connected.

Except from the puzzles that are just protein structure prediction, which are already existing problems, Foldit can be a really creative game, as it allows users to build their own molecules similar to MineCraft [135]. Players are free to design and build a protein from scratch combining amino acids, or use tools to add, replace and move amino acids changing the protein structure with no constraints. This creative site is more attractive to players instead of just trying to fold an existing protein. Foldit attempts to predict the structure of a protein by having people play competitively to fold the best proteins.

With over 460,000 players, Foldit produced some predictions that outperform the best known computational methods [1] and showed that non experts can work together and develop new strategies and algorithms that differ from traditional software solutions. Foldit players managed to design an enzyme that catalyzes the Diels-Alder reaction and was about 20 times more efficient in catalyzing the reaction than the one the scientists had started with [136]. Foldit has contributed to many applications. In May of 2008 it was used as the first computer game that could predict the 3D crystal structure of a protein with human ability [3]. In 2011 player of Foldit helped to decipher an accurate crystal 3D structure of the Mason-Pfizer monkey virus (M-PMV) retroviral protease. The puzzle was solved in 10 days and the solution that was investigated then used to find the structure with the method of molecular replacement [137]. The strategies and mechanics of the best players of Foldit can be formalized and structured and contribute in this way to the creation of algorithms. These algorithms can be used by computers or further modified and expanded by other users [138].Foldit's rules are simple. Every week, players are presented a problem to solve and their goal is to reach the best score. Every time a player performs a move, the game returns the score of the folding. Proteins fold at the lowest energy states, although in the original game the score is computed by multiplying the energy with -100, so higher scores are better. A player acts using simple mouse moves, dragging parts of the protein and observing which moves increase the score and which return penalties. Then, he constructs strategies that can lead him to better folds. In the introductory levels he learns some already existing techniques and adopts them for other puzzles. In Figure 16 we can see the Foldit online environment.

Players can interact with the protein in a variety of ways:

- pull on some parts directly,
- place bands and indirectly pull whole chains,
- freeze pieces in place and block movement to some parts.

and some more complex automatic moves (optimizations) from Rosetta package [131] that will computationally improve the protein:

- “Shake” (combinatorial sidechain rotamer packing),
- Global “wiggle” (gradient-based minimization),
- Local “wiggle” (gradient-based minimization with loop closure) and
- “Rebuild” (fragment insertion)
- “Tweak” (helix rotation or shift)

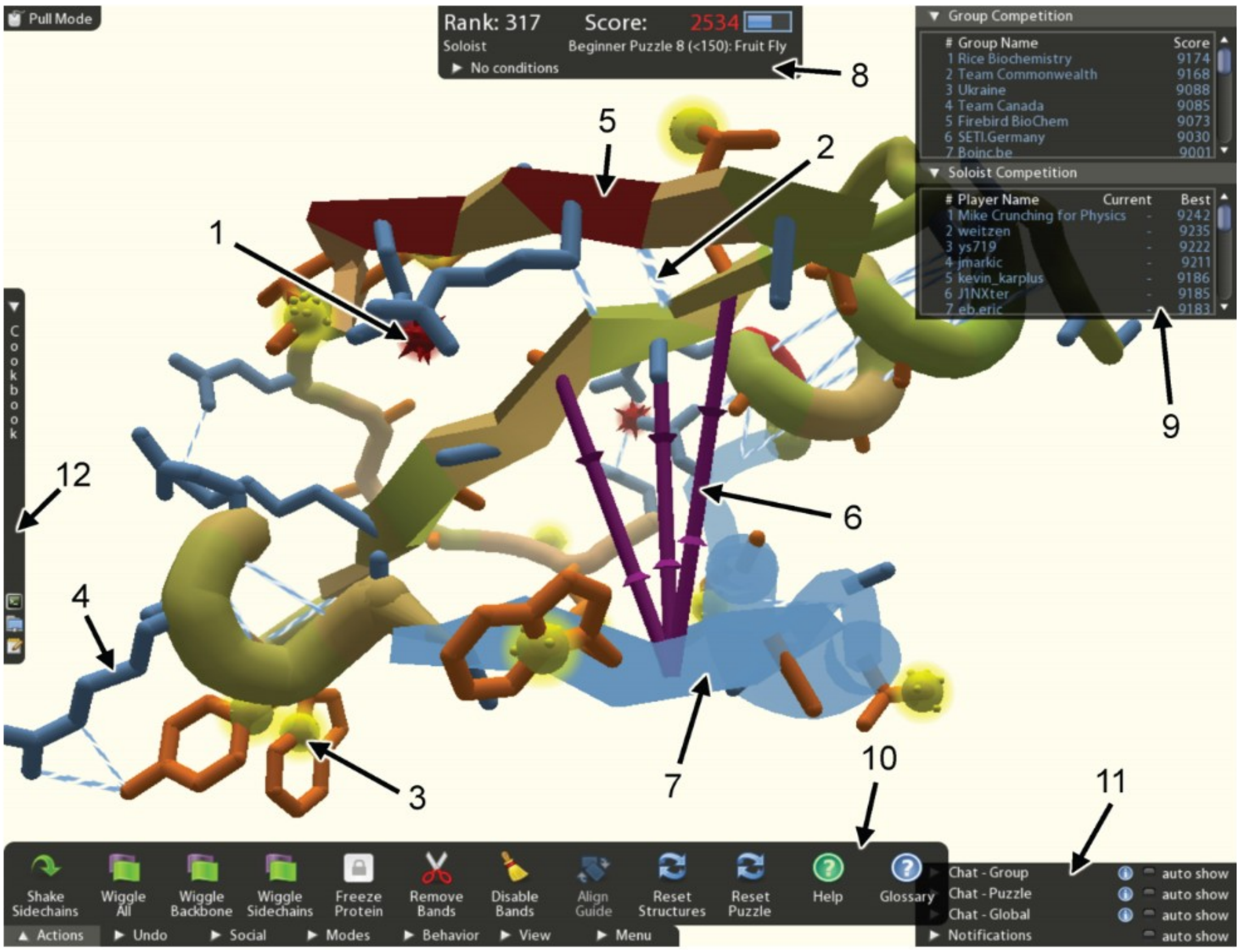

**Figure 16: Figure from [1]. Foldit guide**

**1) a hydrogen bond, 2) an exposed hydrophobic sidechain (yellow), 3) a hydrophilic sidechain, 4) segment of the backbone that is red due to high residue energy, 5) players can make modifications including bands, 6) add constraints, 7) freeze parts, 8) score, 9) leaderboard for players and groups 10) toolbar, 11) chat for interacting with other players, 12) cookbook**

Foldit uses an achievement system that gives extra points to players for sharing solutions, complete the introductory levels and solve many puzzles. In this way it attracts players to share solutions and strategies they have developed, with other players online, as well as to form teams, leading faster to better protein structure results. Players can form groups, work together and share their solutions. The score of the player’s solution is updated in real-time. There is also a leaderboard where the best scores are displayed and an extra leaderboard for the teams.

### 2.2.1 Cookbook and Foldit recipes

A recipe is an automated move that runs multiple simple game actions at the same time and is written by Foldit players to make things more interesting. A recipe is practically a written strategy or a part of a player’s strategy. Players can expose their strategies by sharing their recipes within the WeFold [3] community or along with their teammates. Most known Foldit recipes are shake, wiggle, rebuild and remix. The Foldit website has a section for recipes [139], contributed by Foldit players over the years.

The Cookbook [140] is a tool of Foldit, created to store all the recipes and to provide an interface to write new ones. This GUI uses Lua scripting language [141] which has custom functions that enable players to execute game moves and query game state. Cookbook is one of the most important features of the game.

### 2.2.2 Foldit Standalone

In order to open Foldit for biochemistry applications by experts, the designers decided to develop another separate version, a stand-alone desktop application for protein structure manipulation, called Foldit Standalone [4]. Derived from the user interface of the puzzle game Foldit, the Standalone version misses all the game competitive features, while adapting more advanced ones, centered to biochemists, as well as many options for visualization.

First of all, users are able to load their own molecules and easily reshape them in real time and save them, using the powerful Rosetta molecular modeling package [142]. Rosetta algorithm, is one of the most useful and successful methods that are able to accomplish prediction, design and analysis on a diverse set of bio-molecular systems. Rosetta is most known for its energy function, parameterized from small-molecule and X-ray crystal structure data, used to approximate the energy associated with each molecular conformation. This method is based on the Monte Carlo technique, achieved accuracy between 3 and 6 $A^o$ [143] and has been used in a variety of computational modeling applications. Foldit is an interface for Rosetta tool as it uses the same mechanism for the energy computation as Rosetta does, but rather than having a computer doing this (finding the best conformation), humans do it.

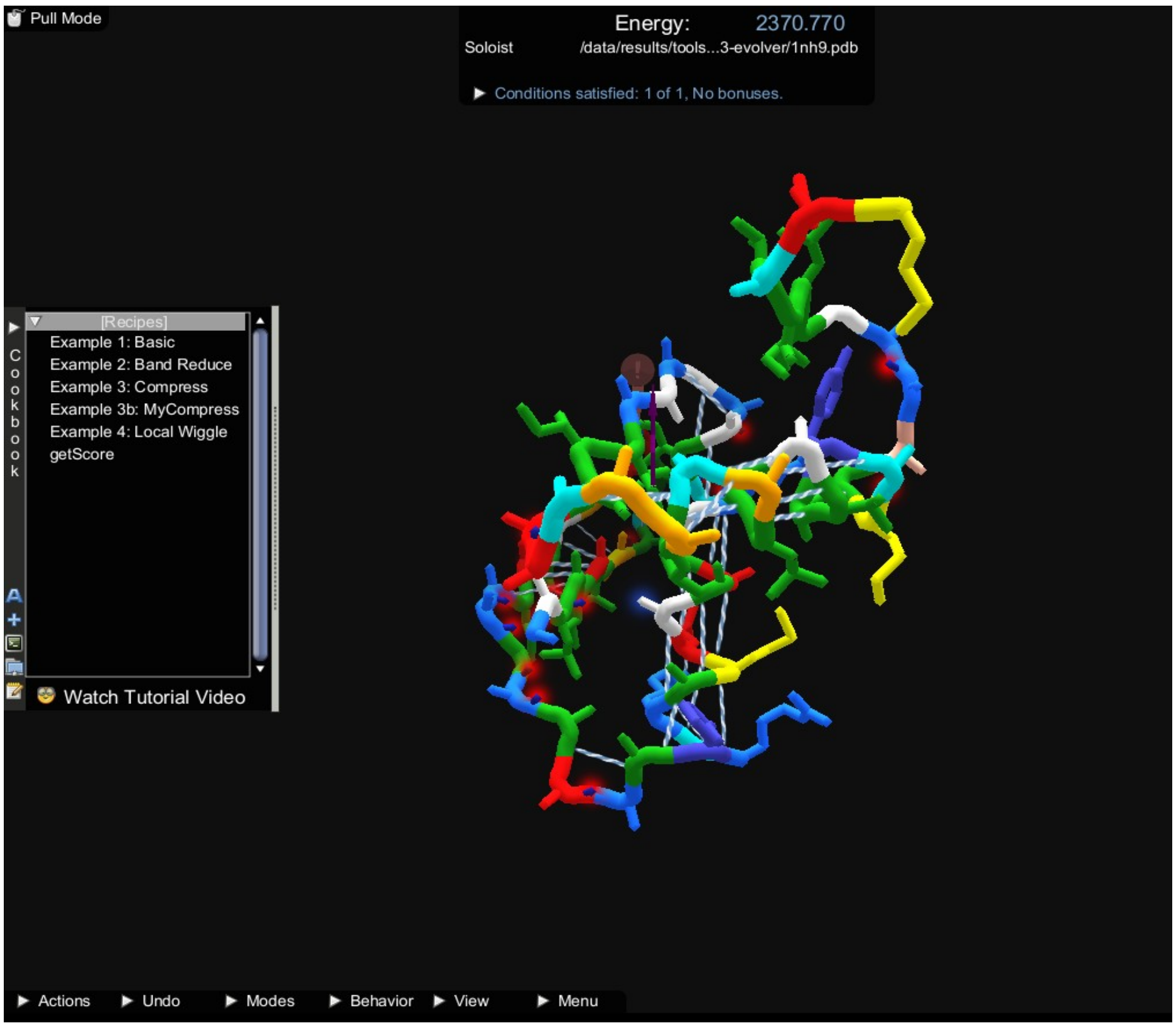

**Figure 17: Foldit Standalone interface**

Foldit Standalone supports a variety of visualization options, including electron density and contact map, different geometries such as lines, spheres and cartoon, different colorings such as energy-based, CPK and rainbow. It also enables to hide/show elements (hydrogens, sidechains). Among other features, the Standalone version of Foldit contains sequence alignment tools for template modeling, rigid body transformation controls and an embedded Lua interpreter.

## 3. IMPLEMENTATION

Using deep reinforcement learning we are trying to discern an optimal strategy for the puzzle game Foldit. Our future and main goal is to use Foldit to predict 3D protein structures from their amino-acid sequence. This calls for the design of an agent that needs to learn an optimal policy in order to solve our input “puzzles”, in other words, proteins with no homologues. The agent is built using a CNN architecture based on the Flappy bird implementation [144] and the algorithm Q learning with experience replay [52, 145]. The AI agent starts with no knowledge of the environment, only some possible moves. Exploring the problem space, blindly at first by executing random moves, the AI agent quickly develops an intuition of how the game is played through the rewards and the penalties it receives after each move.

Using the Rosetta algorithm [4, 131] as a black box, the best so far for computing the energy of a molecule, we feedforward our network with the current difference in energy. The rewards here are how much the energy is minimized (score maximized) and the penalties are the growth (score reduction). As more episodes pass, the agent starts realizing more ways to handle different proteins and prefers actions that maximize its payoff. Once the training is complete we examine the average rewards each training model had. Using this as an indicator of its accuracy we experiment with different parameters in order to optimize the number of total rewards and choose the best model for extended training and testing. The agent receives game screenshots as input in the form of a pixel array. The image will be fed into a convolutional neural network which will give a decision about which action is best to executed. Then, the network will be trained millions of times via an algorithm called Q-learning, to maximize the future expected reward.

This was an overview of the Implementation chapter. The first section is about the input of the DQN, the different dataset and the image preprocessing. Next, is the Q-learning algorithm and the differences with the classical Q-learning method and the set-up environment and final section is about the DQN architecture and the required libraries for the implementation.

## 3.1 Data

Our dataset consists of 40 proteins from the PDB database obtained by the X-ray Diffraction method. These proteins have no sequence homologues with known structure, as we choose the ones that had less than 30% similarity and contain less than 100 residues. Deep Foldit has as main its goal to be able to fold the input proteins in a fixed number of moves. Initially, we set this number at 200 iterations, and we chose molecules with small number of residues in order to fold them within 200 moves. Also, with small proteins we reduce complexity and the network will learn faster.

The original set of 40 molecules was split into 20 proteins for training and 20 for testing.

**Table 3: List of proteins in each dataset**

| Training Set | Test Set |
|---|---|
| 1aho | 2e3i |
| 1eoe | 2hdz |
| 1h75 | 2igd |
| 1hyp | 3e4h |
| 1lpl | 3e21 |
| 1nh9 | 3kzd |
| 1t2i | 3rjp |
| 1tud | 3uci |
| 1ulr | 3zhi |
| 1wkx | 3zzp |
| 1xak | 4cvd |
| 1yu5 | 4hcs |
| 1zeq | 4hti |
| 2f15 | 4pti |
| 2fht | 4zai |
| 2nls | 4zc3 |
| 2och | 5gua |
| 2pko | 5nod |
| 3e4h | 6atn |
| 6atn | 6av8 |

The structures in each set were modified as follows: For deepFoldit-Soloist, linear extended conformations were created by processing the extracted amino acid sequences using the LEaP program of Amber tool, a suite of biomolecular simulation programs [133]. Tleap is a function that builds extended protein chains given only a sequence. However, several warnings and breaks occurred over the protein sequence. It was observed that the amino acid HIE was not recognized by the Foldit program, thus breaking the chains at the sites where HIE was, replaced by HIS. For deepFoldit-Evolver, the native structures were first optimized until convergence by the energy minimization function of FolditStandalone, global “wiggle”. Then 100 random moves were performed to denature the rest of the proteins, selected from our action set creating the deepFoldit-Evolver dataset. This dataset was used only to observe the improvement of the score in the first trials and justify whether the implemented algorithm and the interaction between the constructed actions and Foldit environment is working satisfyingly. As the Soloist dataset is consisted only of linear conformations, the changes in the score and the protein structure are difficult and not easy to be observed, because the constructed actions are very specific (fixed mouse positions).

For data visualization we kept the default Cartoon display and for color the option AAColor, which displays a unique color for each amino acid (see Figure 18). Other displays were also tested, but this one seemed more appropriate for image processing after checking the pre-processing results.

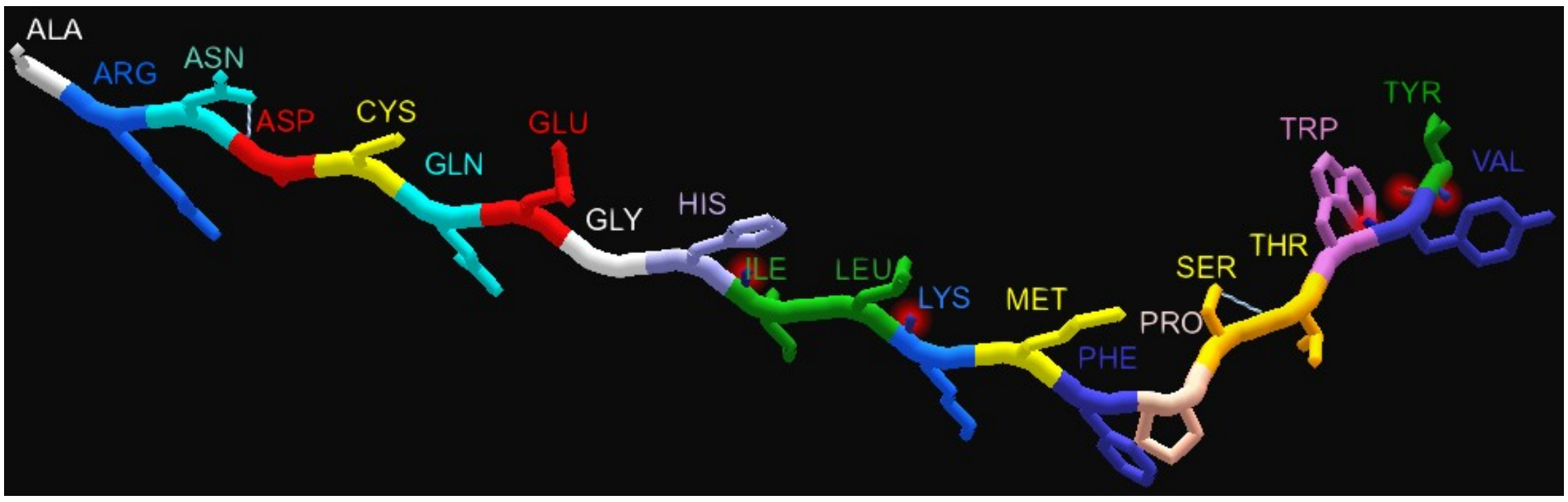

**Figure 18: Colors used in Foldit for amino acids**

## 3.2 Applying Deep Reinforcement Learning to Foldit

### 3.2.1 Preprocessing

Working directly with raw Foldit Standalone frames, which are 1024x1024 pixel images with a 256-colour palette, can be computationally demanding affecting runtime and memory requirements. We apply a basic preprocessing step by cropping the original image, forming an 600x600 window, aimed at reducing the input dimensionality and dealing with some artifacts of Foldit (menu bar, cookbook). To further reduce the dimensionality and the computational cost, we down-size (rescale) the input to 160x160 pixels (square images are more suitable for processing). In some runs we applied a grayscale filter as an extra step to reduce the three-color channels to one. However, this didn't have promising results and we continued with the original idea for the network to identify as features the 20 amino acids using the shape and the color as identifiers.

Foldit uses different colors for each amino acid and colors could help the network identify easier the features. The final input representation is obtained by applying log polar transformation, to obtain invariance with respect to rotations and scaling [146].

The log-polar coordinates (or logarithmic polar coordinates) is a coordinate system in two dimensions that is performed by remapping points from the 2D Cartesian coordinate system ($x$, $y$) to the 2D log-polar coordinate system ($\rho$, $\theta$) according to the transformation below:

$$\rho = \log\left(\sqrt{(x - x_c)^2 + (y - y_c)^2}\right)$$

$$\theta = atan2(y - y_c, x - x_c)$$

where $\rho$ is the logarithm of the distance of a random point $(x, y)$ from the origin $(x_c, y_c)$ and $\theta$ is the angle of the line through the random point and the center. By using this transformation, both rotations and the scaling between two images can be represented by translations.

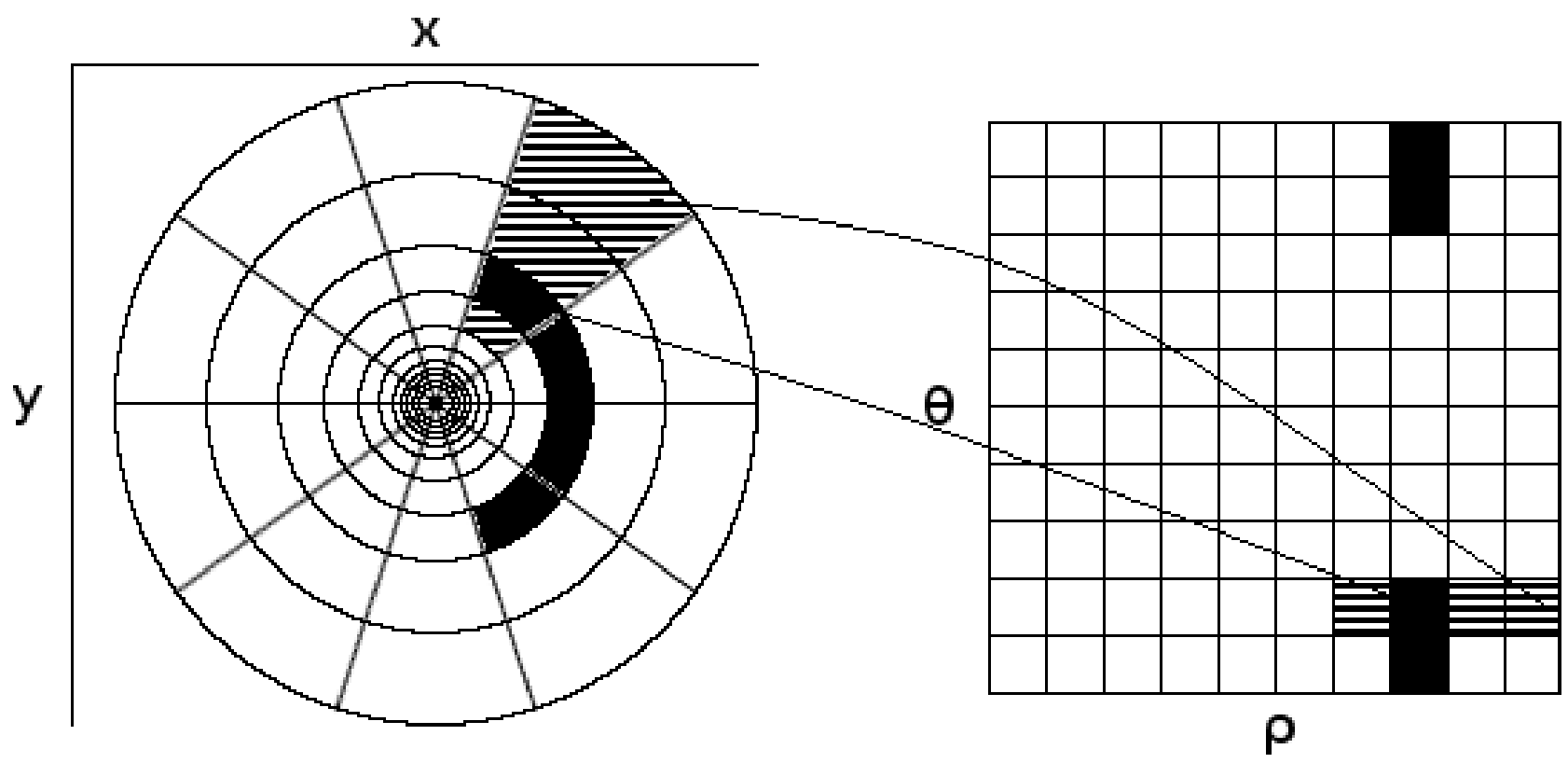

**Figure 19: Log-polar transform, source [147]**

Using this log-polar remapping, the objects occupying the central high-resolution portion of the visual field become dominant over coarsely sampled background elements in the image periphery. This focuses the attention in the visual center field where the target is supposed to be. Rotation and scaling differences between two images can be converted to translation differences of the polar-transformed images. In Figure 20 we can see the transformed input using log polar transform.

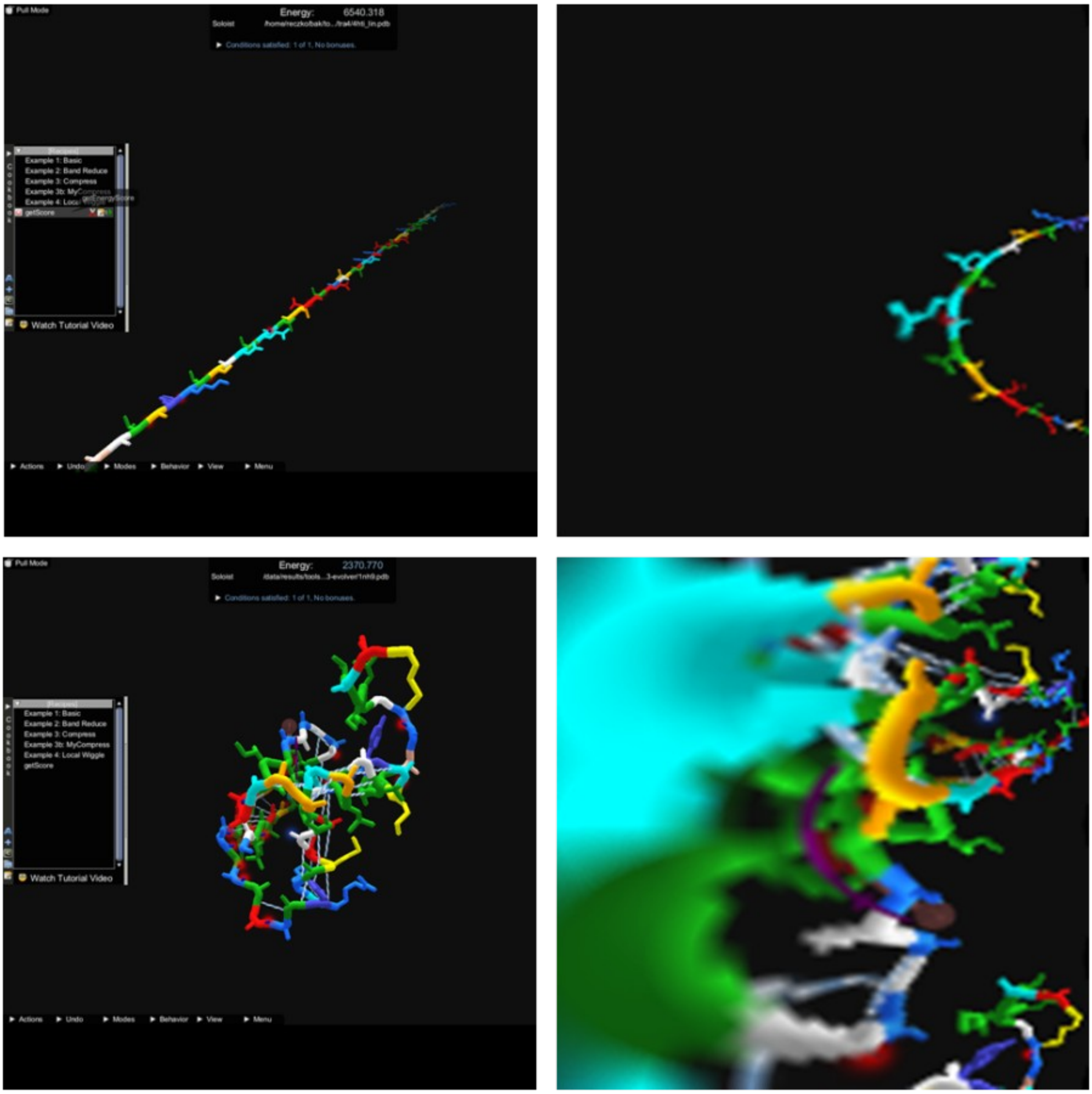

**Figure 20: top left: Linear protein conformation input (start for DeepFoldit-Soloist); top right: log-polar transformation of the left image; bottom left: A suboptimal protein fold (start for DeepFoldit-Evolver); bottom-right: log-polar transformation of the image to the left.**

### 3.2.2 Actions

An action is a command that you can give in the game in the hope of reaching a certain state and reward. Most of Atari games are implemented in python and there are libraries such as pygame, which provide an interaction with the game and its actions. Foldit does not have a library that allows users to interact with it and poses extra difficulty for move specification.

As mentioned in Section 2.2.2, Foldit has many automatic moves through the shortcut keys and it has also plenty tested recipes from WeFold community on cookbook. These recipes contain automatic moves, which correspond to multiple actions using mousepad and keys, and it’s natural to bring major changes in the protein's structure. One really famous and commonly used Foldit move is the "wiggle" recipe, which is activated using the key "w". Wiggle applies changes in the molecule, in order to achieve a conformation with less energy. the more time this movement is used in the game, the more changes are made to the structure until the energy reaches a minimum. However, it does not seem to be responsive to linear molecules that are used as inputs, since it only modifies the local structure. As a result, the molecule can’t fold and take a 3D structure.

From the Foldit moves, DeepFoldit skipped the following:

**Table 4: List of Foldit automatic moves**

| Band | Rebuild |
|---|---|
| Freeze | Secondary structure |
| Global/ local wiggle | Tweak |
| Shake | |

All these automated actions have great impact on score-energy and, during the training phase, the neural network tries to learn those which affect the score the most and seeks to use them more than others. Choosing only actions like "freeze", "Wiggle" etc it is impossible to get the optimal structure. Most of the times, by using those operations, we can find some local minimums but not the optimal structure. The shortcut keys can improve the initial structure's state, but staying focused on applying small moves that don’t affect the backbone’s structure much. Although Foldit has lots of shortcut keys for main moves (mouse moves), we have to construct new actions. The only way to interact with the game screen, is by using xdotool, a simulation of keyboard input and mouse activity. In Table 5: DeepFoldit action set we can see our scripted actions that interact with Foldit Standalone environment.

**Table 5: DeepFoldit action set**

| Up | Up Less | Shift Up |
|---|---|---|
| **Down** | Down Less | Down Up |
| **Right** | Right Less | Shift Right |
| **Left** | Left Less | Shift Left |
| **Zoom in** | Zoom out | Home |

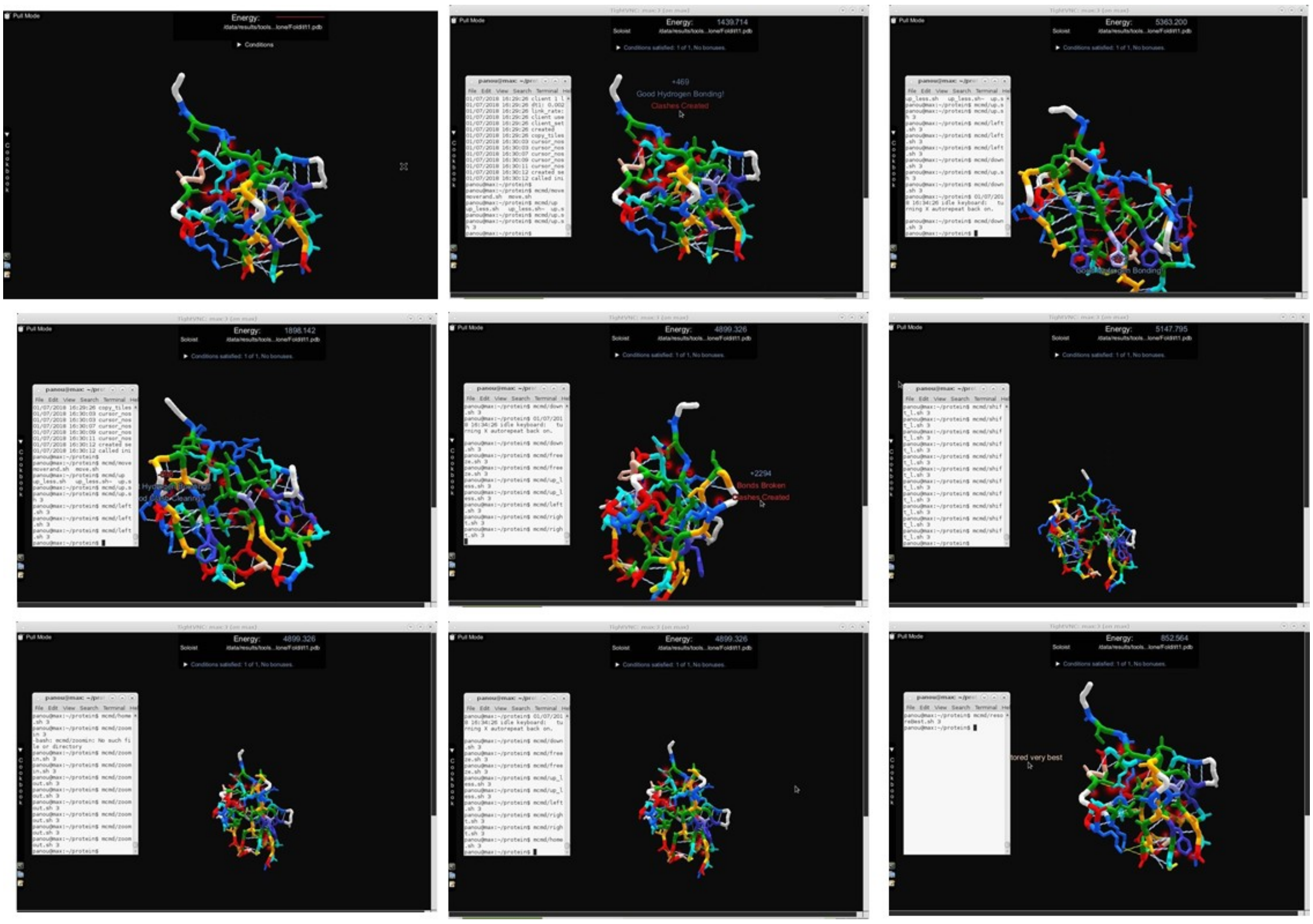

**Figure 21: Original image, b) Action up with energy penalty, c) Action down, d) Action left with reward, e) Action right with energy penalty, f) Action shift to the left (Rotation), g) Action zoom out, h) Action Home (place protein to the center), i) Restore best score**

### 3.2.3 Algorithm

The goal of an RL algorithm is to maximize the total game score. We use the Bellman equation as an iterative update:

$$Q(s_t, a_t) = r + \gamma \max_{a_{t+1}} Q(s_{t+1}, a_{t+1})$$

where, $s_{t+1}$ and $a_{t+1}$ are the state and action at the next timestep, r is the reward, γ the discount factor and $Q(s,a)$ is the Q-value for $(s,a)$at timestep $t$.

A sequence of observations $(s_t, a_t, r_t, s_{t+1})$, becomes an input point to the learning algorithm. The Q-function should fit these inputs to the model, in a way that it will prefer the ones that maximize the total reward. For the update of the weights we use a loss function and its gradient:

$$L = \sum_{s_t, a_t, r_t, s_{t+1}}^{\square} \left( Q(s_t, a_t; \theta) - \left( r_t + \gamma \max_{a_{t+1}} Q(s_{t+1}, a_{t+1}) \right) \right)^2$$

$$\nabla_\theta L = \sum_{s_t, a_t, r_t, s_{t+1}}^{\square} -2$$

where $\theta^-$are non-updated weights for the Q-value function. We apply stochastic gradient descent and backpropagation on the above loss function to update the weights $\theta$. The pipeline for the entire DQN algorithm for training is presented in Algorithm 3. The agent selects and executes actions according to an e-greedy policy based on $Q$. The reward should essentially be the score of the game. It starts out as the energy the lined molecule has and every time we apply a move the game returns a positive score if it’s a reward or negative if it is a penalty. As mentioned previously in this section, the experiences are stored in a replay memory and at regular intervals, a random mini-batch of experiences are sampled from the memory queue and used to perform a gradient descent on the DQN parameters. Then we update the exploration probability as well as the target network parameters $\theta$. The reward $r_t$ is:

$$r_t = \tanh(0.1 \cdot \Delta S) \qquad (1)$$

where $\Delta S$ is the Foldit score difference before and after an action. The usual Q-learning algorithm with experience replay uses replay memory that contains all state-action pairs, irrespective of the reward received. As rewards (or penalties) occur very sparsely, we split the replay memory into two queues that contain the more informative states with non-zero reward (queue $D_2$) and the non-reward states (queue $D$). Training is accelerated by randomly selecting the mini-batch so that it contains equal shares of each queue.

**Algorithm 3: DeepFoldit Algorithm of Q-learning**

---

$Q$-learning: Learn DNN $Q$

---

**Require:**
  Initialize replay memories $D, D_2$ to capacity $N$
  Initialize DNN $Q$ with random weights $\theta$
  **for** episode = 1, M **do**
    Initialize arbitrary first sequence of frames for initial state
    **for** t = 1, T **do**
      With probability $\epsilon$ select a random action $a_t$, otherwise select $a_t = max_a Q(s_t, a)$
      Execute action $a_t$ and observe reward $r_t$ and state $s_{t+1}$
      **if** $r_t <> 0$ **then**
        Store state transition $(s_t, a_t, r_t, s_{t+1}, term)$ in $D_2$
      **else**
        Store state transition $(s_t, a_t, r_t, s_{t+1}, term)$ in $D$
      Sample random mini batch $(s_i, a_i, r_i, s_{i+1}, term)$ from $D$ and $D_2$
      Set $y_i = Q(s_i)$
      **if** $term = true$ **then**
        Set $y_{i,a_i} = r_i$
      **else**
        Set $y_{i,a_i} = r_i + \gamma max_a Q(s_{i+1}, a)$
    Train $Q$ on $(s_i, y_i)$

---

## 3.3 Architecture

The input to the neural network is a game screenshot, specifically a 160x160x3 RGB image, sampled from Foldit-standalone, followed by 3-5 rectifier hidden layers. Two or three of them are convolutional layers, for feature extraction and one or two fully connected with ReLU layer for prediction. The first layer of the model is a convolutional layer of 30 filters of size 15x15x3 with stride 1 and with zero padding, followed by a nonlinear rectifier. In some implementations the first layer is convolving one 1x1 filter with stride 1 (for adaptive RGB to gray conversion). The second layer is another convolutional layer with of 10 filters of size 20x20 with stride 1, followed by a nonlinear rectifier unit. The final hidden layer with 200 fully-connected rectifier units is connected to the output layer. The output layer is a fully-connected linear layer with a single output for each valid action. The number of valid actions is 15. In some implementations, between the fully connected layer of 200 rectifier units, there is another one fully connected layer of 30 rectifier units. Figure 22 depicts our model’s architecture. With (*) we mark the extra layers that are used in some implementations. For a shorter and better representation, the following abbreviations are used for the different types of layers and their configurations:

**Table 6: Acronyms of the components of the DQNN**

| Code | Meaning |
|---|---|
| **ISi** | Input Size Layer i |
| **RFSi** | Receptive field size layer i |
| **CLDi** | Convolutional layer i depth |
| **Si** | Convolutional layer stride i |
| **ND** | Dense size |

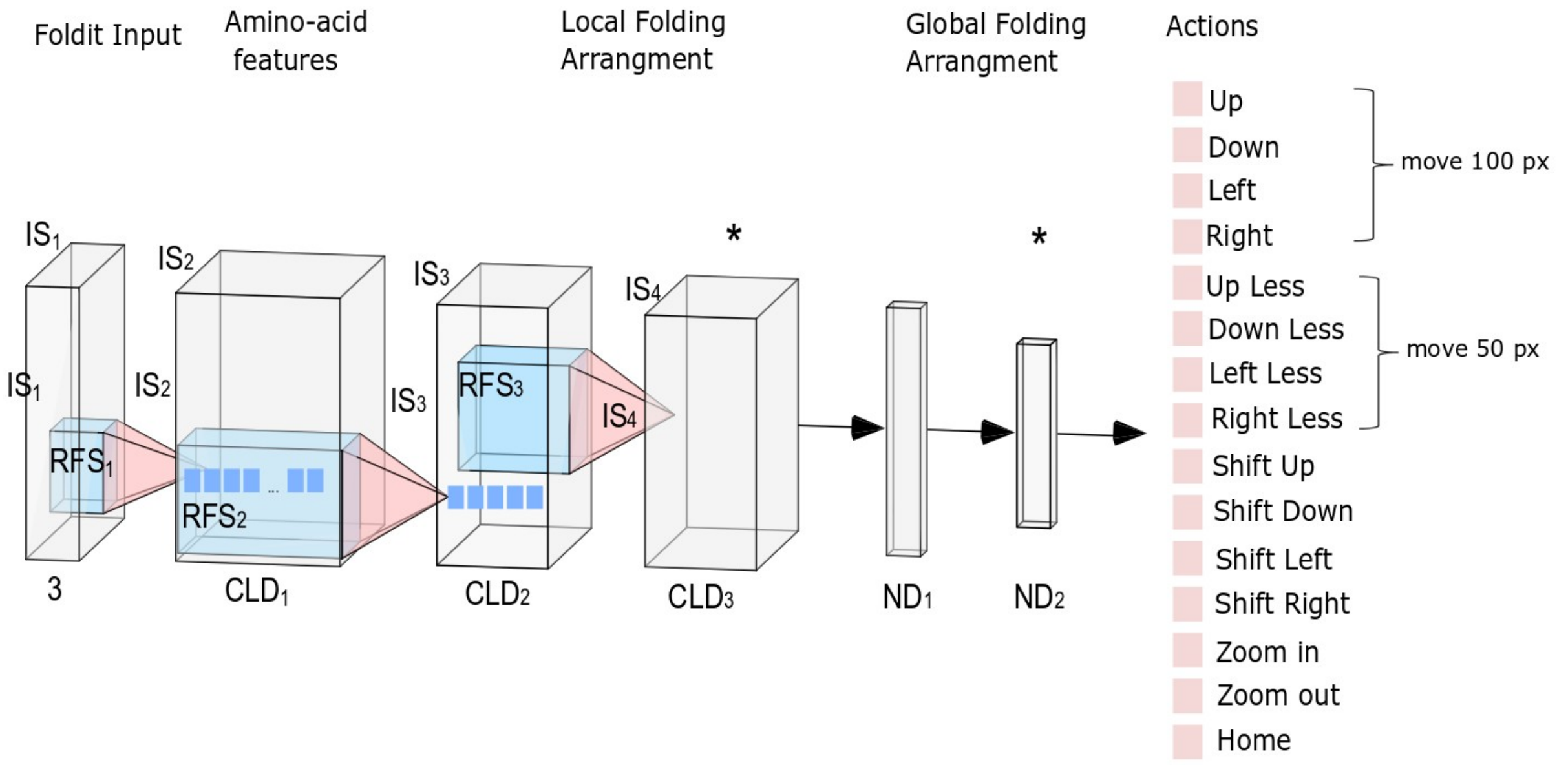

**Figure 22: Model Architecture (designed using NN-SVG tool [148])**

The behavior policy during training was ε-greedy, decreased the ε linearly from 1 to 0.1 over the first million frames, and fixed at 0.1 thereafter. We trained for a total of 30 thousand frames and used a replay memory of the 8 thousand most recent frames. All of these CNNs use a RELU activation function in all layers except for the output layer where a ReLU activation function is applied. Furthermore, a learning rate of 1e−4 was chosen. Moreover, the “Nestrovs Updater” with a momentum of 0.9 was used. As the optimization function “Stochastic Gradient Descent” was used and the loss function in the output layer was “Mean squared error”. The weights were initialized using the “Identity Weight Initializer”.

### 3.4 Tools and Libraries

In order to execute and reproduce all the tests and the implementations that are explained in this thesis, a computer with enough processing power was required. In the case of this project the server used was provided by the Bioinformatics Group at the BSRC “Alexander Fleming”. The server has 2 CPUs with 20 cores per CPU and 256 GB of accessible RAM. We used 20 CPUs to train the network and 10 GB memory is needed to store the temporary files. We also requested the use of the supercomputer "Aris" from National Infrastructures [149] to speed up the parameter optimization. There, each run to finish a game training epoch lasted 12 to 20 hours, which is equal to 30000 game timesteps. GRNET "Aris" computing system does not support graphic environments and the libraries needed by Foldit, so we built a container using Singularity. Singularity is an open source software library that implements virtualization at the Operating System level. A container was built with the game and all the necessary packages within. Using virtual displays in virtual framebuffer X servers (Xvfb) we interacted with the Foldit-standalone session making the NN learning faster. We also used xdotool lib, which allows for programmatically or manually simulated keyboard input and mouse activity, by using X11's XTEST extension and other Xlib functions, to construct the interactive actions with FolditStandalone.

**Implementation Architecture**

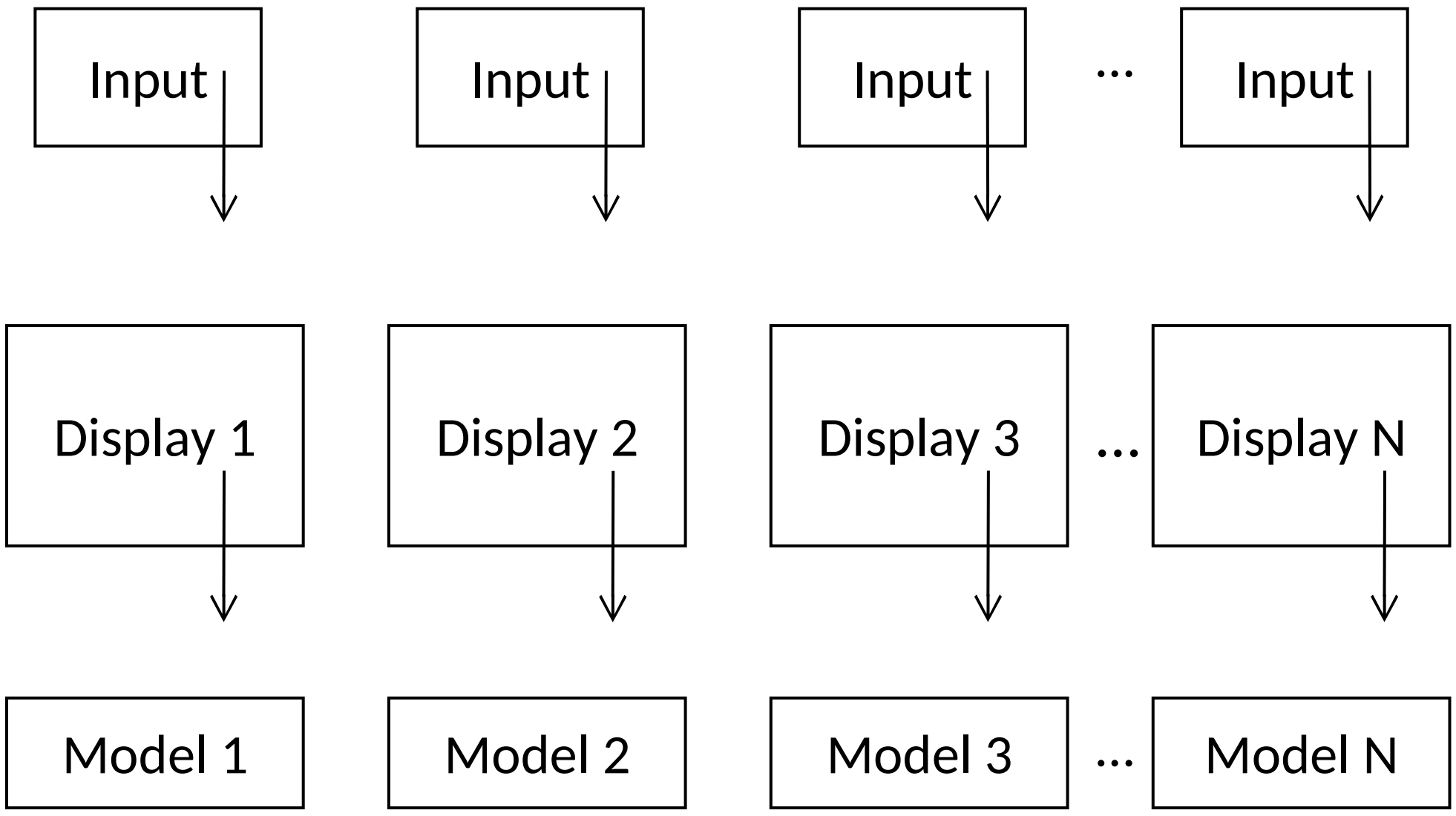

**Figure 23: Parallel system architecture for parameter optimization**

We consider two system architectures. The first architecture (Figure 23) includes multiple implementations using the same training data, which updates different models. This architecture was implemented to parallelize parameter optimization. In our case 10-20 different models were trained at the same time.

The second system architecture (Figure 24), is parallelizing the training process. It is the same implementation, but uses different displays to update the same model. For example, if we use 20 displays, one for each protein, we can pass through all the data 20 times faster. Within this thesis, the first architecture was implemented

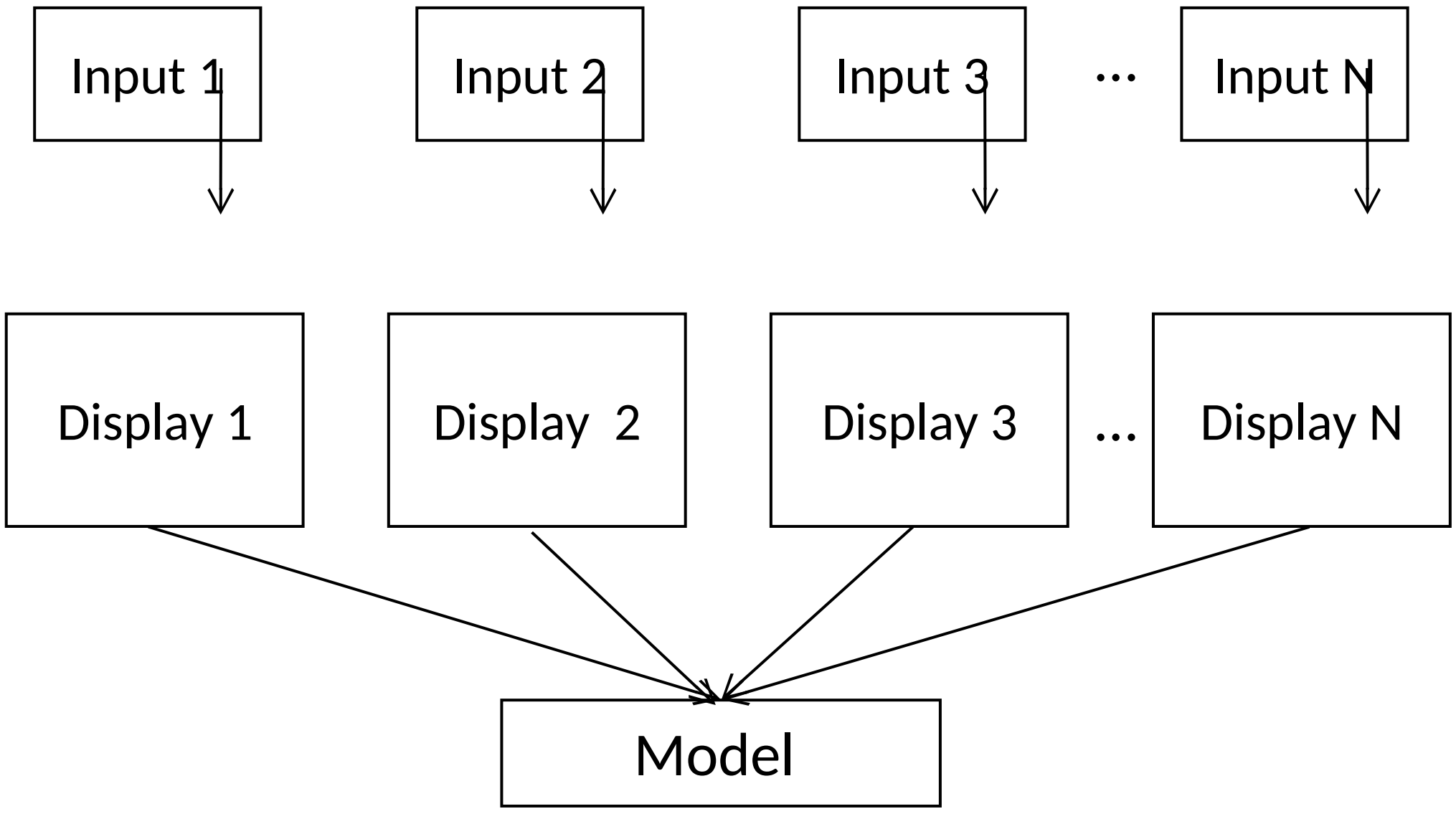

**Figure 24: Parallel system architecture for model training**

All versions of the CNNs were coded and tested in Python using the high-level neural network API Keras [150]. Keras can run on top of different Deep Learning frameworks like Tensorflow, CNTK, and Theano. In this case, the Tensorflow backend was chosen. The main reason we selected TF and therefore Keras was that both systems are optimized for deep learning tasks. Also, all frameworks are implemented in Python, allowing the user to work with them in a lightweight manner without multiple files being used. With Keras, the user first has to define a model which can be chosen between a sequential model or a Graph model. In the first case the layers are stacked and the output from a previous layer feeds the input of a next layer until it reaches the output layer. In the second case, Graph model allows the user to get the output from a desired layer and feed that output to a desired layer, allowing either for multiple output networks to be produced to receive the output in the model's intermediate layer.

The software required is Python 2.7 and Keras with the following libraries:

**Python Libraries**

- Keras — Python Deep Learning library
- Tensorflow — Python Deep Learning library
- subprocess — Subprocess management

- pickle — Python object serialization
- sympy — Hyperbolic functions
- datetime — Basic date and time types
- argparse — Parser for command-line options, arguments and sub-commands
- random — Generate pseudo-random numbers
- collections — High-performance container datatypes
- sys — System-specific parameters and functions
- matplotlib — Python for 2D plotting library
- json — JSON encoder and decode

**Python Packages**

- scikit-image - package for image processing in Python
- numpy — a general-purpose array-processing package
- opencv-python — Image Processing & Feature detection
- freeglut3 — OpenGL Utility Toolkit

**Ubuntu tools**

- xdotool — command-line X11 automation tool
- Xvfb — virtual framebuffer X server for X Version 11
- xinput — utility to configure and test X input devices
- vncviewer — VNC, is a connection system that allows you to use your keyboard and mouse to interact with a graphical desktop environment on a remote server.

# 4. RESULTS

In this study, we focused on the identification of hyperparameters that improve the model in our task. Our target is very specific, which is the construction of a model that can improve the fold of a given protein within 200 iterations.

We used computational resources on the National high-performance computing (HPC) facility – ARIS (https://hpc.grnet.gr/en) to run our implementation. The results are presented as follows, for each model's parameter, different values from a range of suggested options were tested and then compared via boxplots. Every parameter section has a figure with a group of boxplots showing all the value comparisons during the tests. In most cases the first group of boxplots (R1) contains 128 different runs (7 parameters were tested with 2 variants in each parameter, giving $2^7$=128 combinations), while the second column/group of boxplots contains 5 replicates for each value of the parameter we examine. Each boxplot corresponds to a specific parameter value. In section 4.2 we present the results for the best model architecture.

## 4.1 Parameter optimization

The identification of optimal parameter values is of crucial importance and has direct impact on the accuracy of the obtained classification results. Hyperparameters are numerous and difficult to manually tune. The problem gets worse when the hypeparameters are not independent and it becomes necessary to tune them at the same time. Usually, it is impossible to prove the optimality of a solution without testing all solutions. In practice, the accepted solution is the best found in the budget allocated by the user to the search. To demonstrate the importance of some hyperparameters and also to optimize their values we run deepFoldit for different variations using the ARIS-HPC. We run each test for the same time (a day) and then further studied the system with the greatest performance (highest average reward, average positive rewards). Every day, we run 10 variations simultaneously testing one parameter at a time. We completed 80 different runs which are 800 different implementations for variable testing. To describe the procedure, we followed for optimization in detail, we consider a model with standard specifications as a starting point. In most of the variations of those 80 runs, we can meet groups of replicates. Those groups have five repetitions of the same implementation, using different random seeds to assess statistically significant differences between the parameter settings, using a two-sided Wilcox test [151]. Wilcox test is applied to implementations that differ in one parameter and have the same number of samples-replicates (usually 5). Variables that show significant statistical differences are marked in the boxplots and the p-value is indicated.

### Basic model

In the first trials, there is no basic model and we start with hyperparameters derived from related works that use models to play video games. After the first 5 runs, we form a basic model that is updated during the runs. The most common properties along the runs were the following (see Table 6 for the layers' names):

**Table 7: Basic Model**

| | | | |
|---|---|---|---|
| CLD1 | 60 | Replay Memory Size | 8000 |
| RFS1 | 15 | Learning Rate | 1,00E-04 |
| S1 | 1 | Momentum | 0.95 |
| CLD2 | 20 | Decay | 2,00E-06 |
| RFS2 | 20 | Dropout Rate | 0 |
| S2 | 1 | Input Image Size | 150 |
| ND | 100 | Score Scaling | 0.1 |
| Gamma | 0.95 | Random seed | Current time |
| Observation | 100 | Mini Batch Size | 50 |
| Final Exploration Frame | 8000 | Actions per Training | 10 |
| Initial Epsilon | 0.8 | Maximum Running Steps | 29999 |
| Final Epsilon | 0.05 | Log polar | Yes |

### Architecture

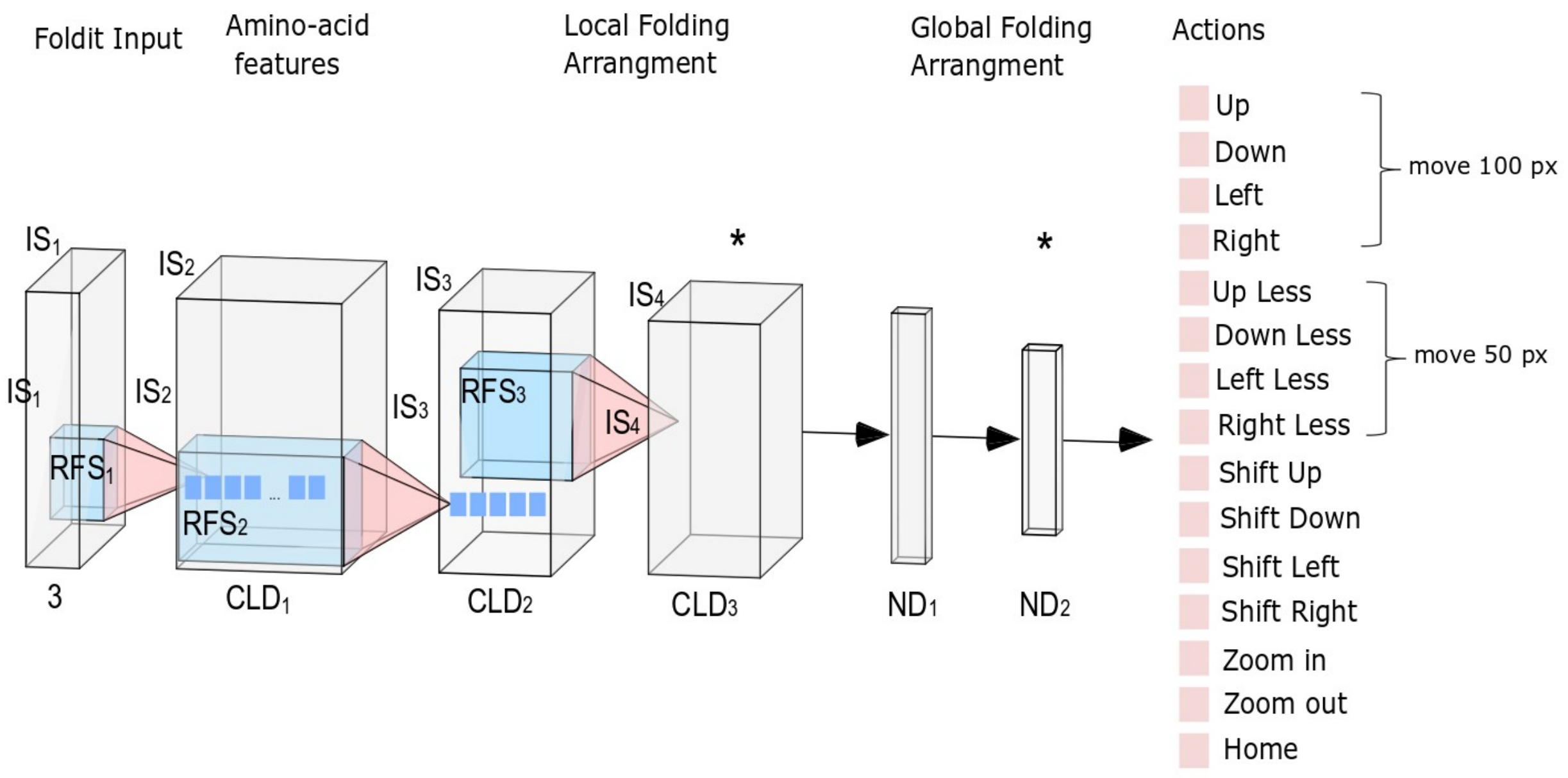

For more about the settings of the tested runs, check ANNEX II.

### a) First convolutional layer depth (CLD1)

Depth represents the dimensionality of the convolutional layer, meaning the number of output filters. The convolutional layer takes as input the raw image as input and then different neurons may activate in the presence of various features such as oriented edges, or blobs of color. We mentioned that depth corresponds to the number of features. Each of them will produce a separate 2-dimensional activation map. In Figure 25 we observe how the depth size of the first layer affects the game score.

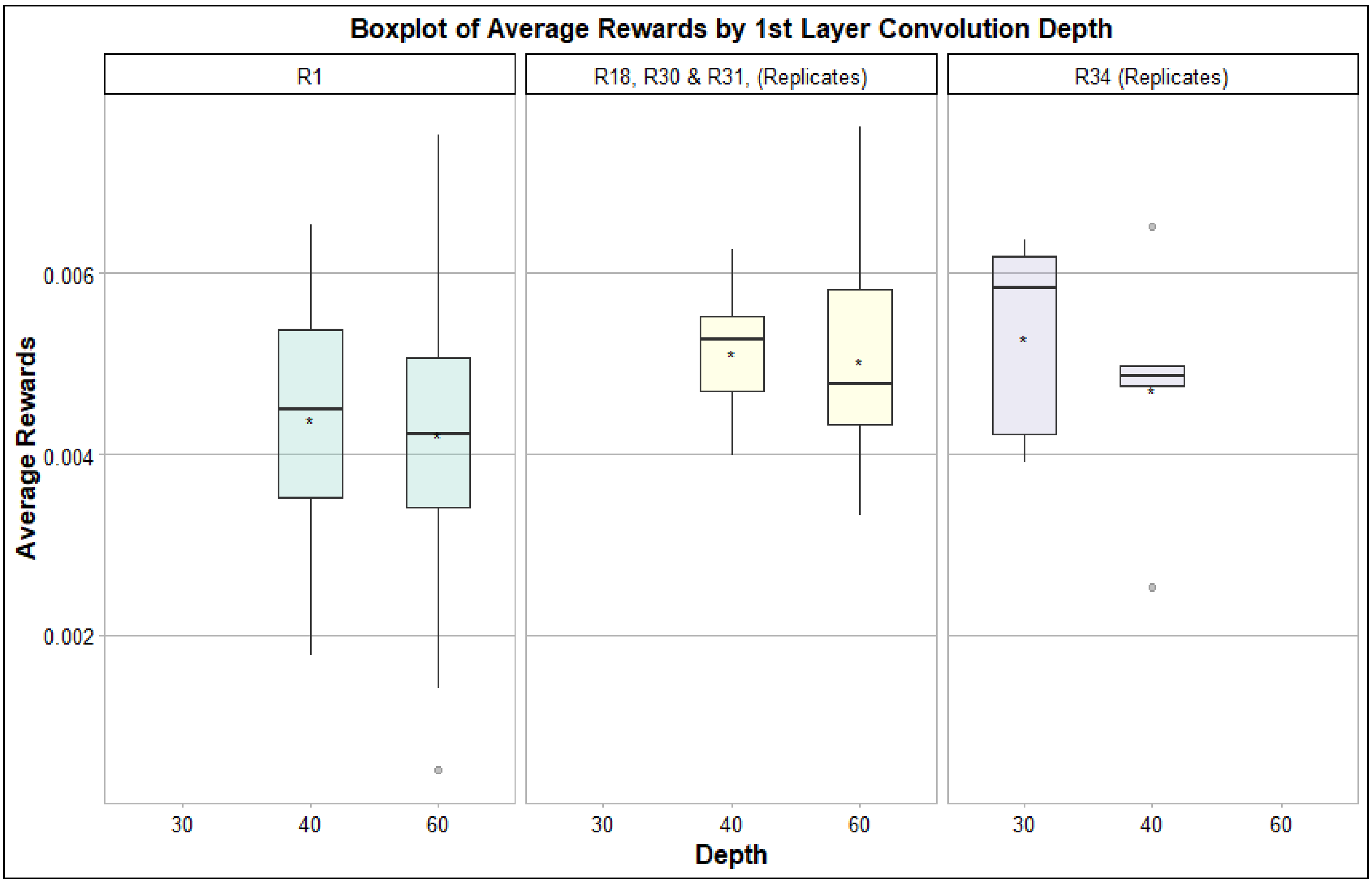

**Figure 25: Convolutional layer's depth comparison for 30, 40 and 60.**

Figure 25 has three different groups, based on the implementations' settings. The tested values for the depth of the first convolutional layer are 30, 40 and 60. In the first group there is only one test, Run 1 (cyan color). It is shown separately because other network architecture parameters are also altered. The cases where Depth is 40 yield better results, while in the second group (Run 18, 30 and 31 all contain 5 replicates with the same hyperparameter settings and different random weight initializations), the layer depth of 60 stands out with a small difference from the depth of 40. The third group (Run 34 in purple) compares 30 with 40, has different hyperparameters (see Annex II) than the other groups and appears to produce better results for Depth of 30.

**b) First layer receptive field (RFS1)**

The receptive field of a convolutional layer is a hyperparameter that shows the partial connectivity of the neurons around a local region of the input volume. In other words, the size of the applied filter and therefore the size of the feature. This parameter along with the stride control the size of the output volume. For example, in Figure 26 we can see the performance of three different receptive fields of values, 15, 20 and 30. The input image has a size of 160x160x3. The filters have sizes of 15x15x3 (3 is for color channels), 20x20x3 and 30x30x3 for each case, so the output of the first layer for stride of 1, would be 145x145, 140x140 and 130x130. Every neuron in the Conv Layer will now have a total of (receptive-field * receptive-field *depth) connections to the input volume, that are local in space (15x15).

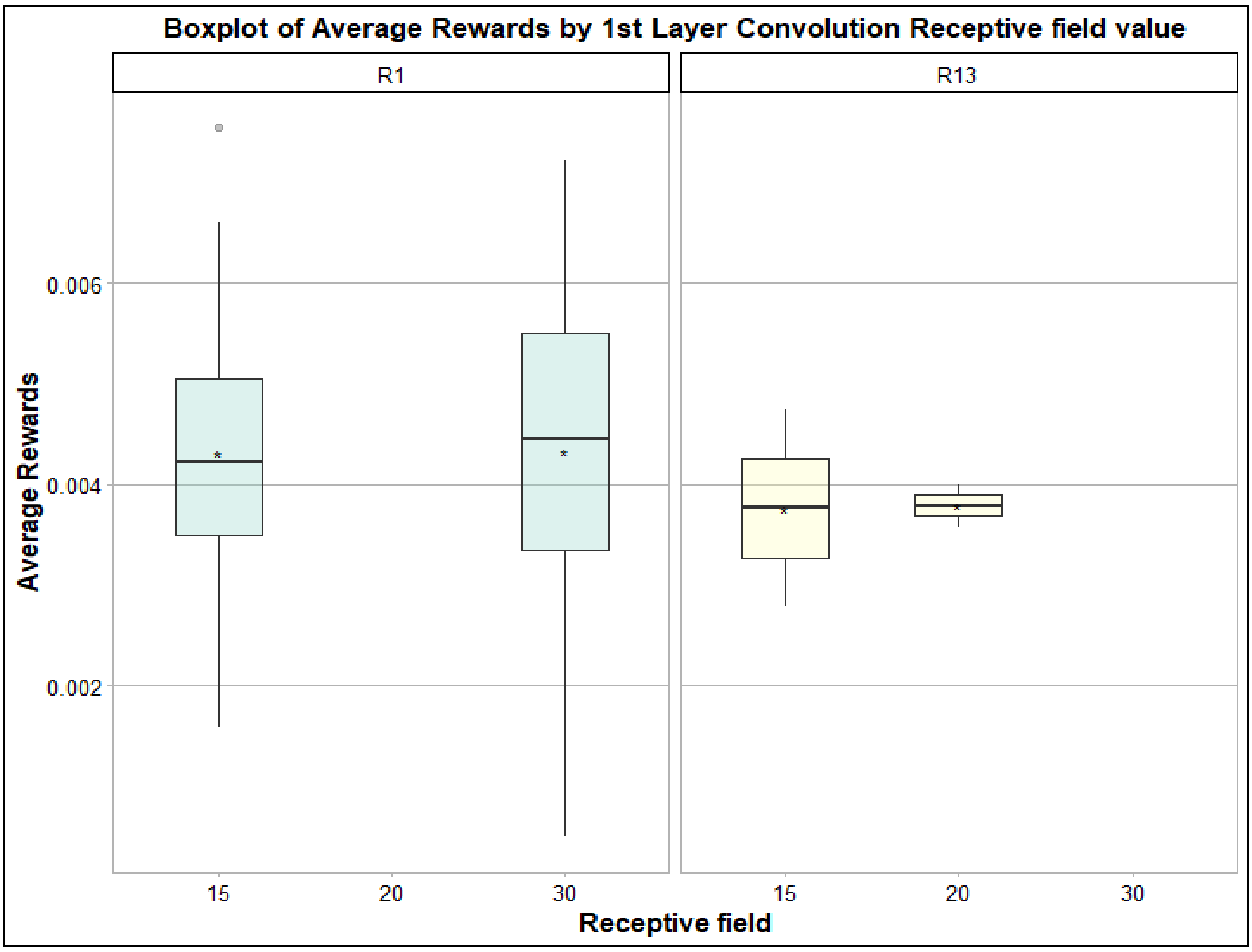

**Figure 26: Presents the value of the first layer's receptive field as a function of Average rewards. The optimal value for receptive field is 15.**

Figure 26 demonstrates how the receptive field of the first layer influences earning positive rewards. There are 2 different run sets (Run1, Run13). In the first group, (Run 1), the receptive field parameter is 15 and 30, with better results for the receptive field equal to 30. In the second group, Run13, we test values 15 and 20, with better results with no significant difference, for receptive field equal to 20. Receptive field 15 is preferred in the following runs.

**c) First layer convolution subsample (Stride S1)**

Stride is another hyperparameter that controls the size of the output volume in a convolutional layer. Stride expresses how many pixels we slide the filter. When the stride is 1, we shift the filter one pixel at a time. When the stride is 2 or more, the filter jumps 2 pixels at a time as we slide it around. This will produce smaller output volumes spatially. There are some limitations to the selection of the stride, as the size of the output volume is calculated as the difference of input size with the receptive field and then all divided by stride.

$$\frac{IS - RFS + 2P}{S} + 1$$

However, the size of the output must be an integer, so if we set the value to 2, the numerator has to be divided by 2. This is why large values are avoided.

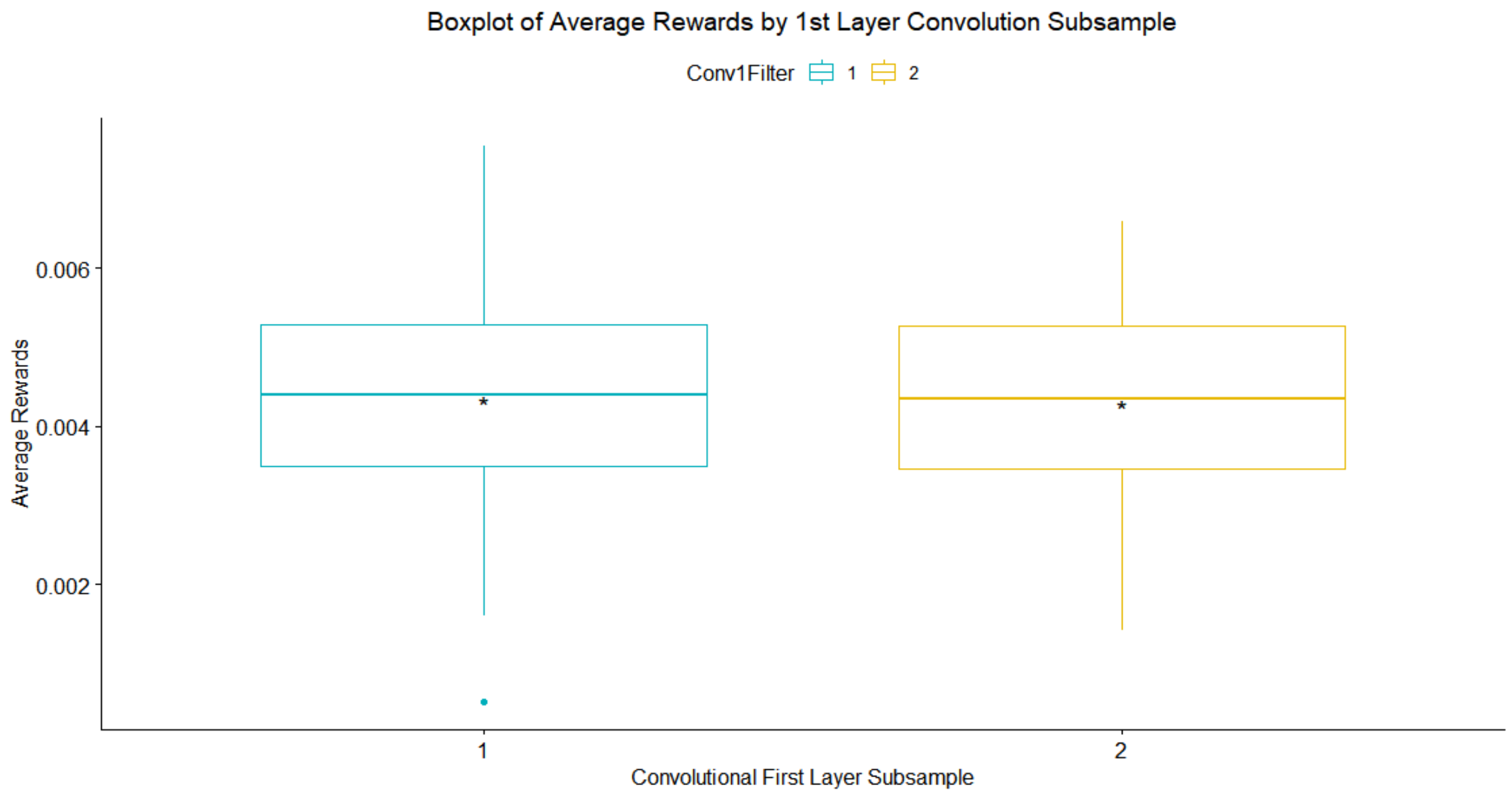

**Figure 27: This boxplot outlines the first convolutional layer's stride size as a function of Average Rewards. The tested values were 1,2 no differences in their performance.**

As we do not observe big difference in the results, we select the stride with value 1 for our tests.

**d) Second layer convolution depth (CLD2)**

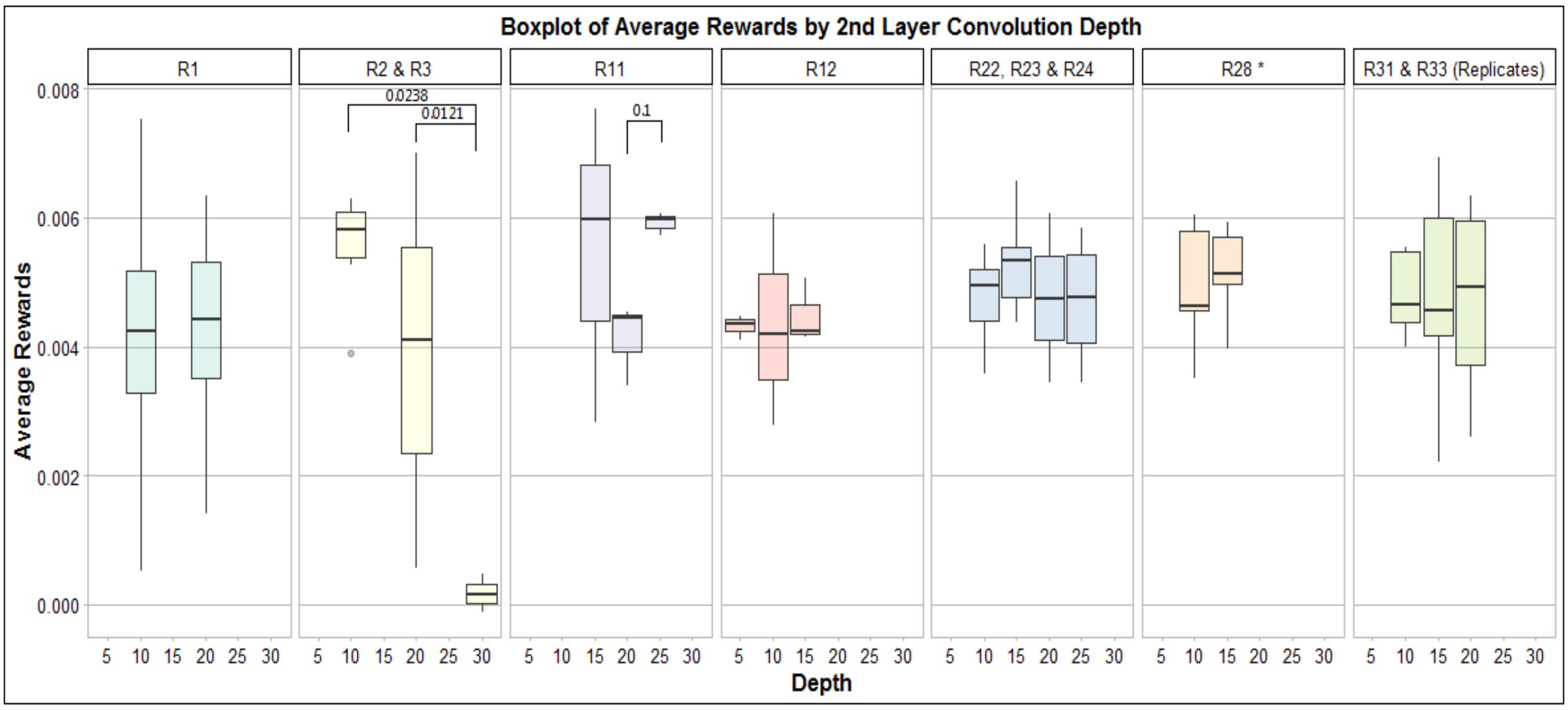

**Figure 28: Depth of the second convolutional layer as a function of Average Rewards. Group 2 (yellow), group 3 (purple), group 5 (light blue), group 6 (yello) and group 7 (light green) boxplots show that depth = 15 had better performance. In the Figure the significance level is annotated.**

The above Figure shows the hyperparameter depth in the second convolutional layer as a function of Average Rewards. There are 7 different groups based on their runs' settings.

- Group 1 (see Figure 25) compares values 10 and 20. Between these two, slightly better results appear for depth of 20.
- Group 2 (in yellow with runs R1, R2, R3) examines 3 different values for depth (10, 20, 30). Best results observed for 10 and worse for 30. Values 10 and 20 have significant difference with 30.
- Group 3 (purple) is Run11 with depths are 15, 20 and 25. Best was depth = 15, while depth = 20 has significant difference with depth = 30. Worst case is also here depth = 30.
- Group 4 (Run 12) the Depth values are 5,10 and 15 respectively with best results value 10.
- In Group 5 (Runs 22, 23, 24) the best Depth value with small difference is again 15. Depth 25 and 30 give similar results and Depth 10 is the worst.
- Group 6 (Run 28) takes two different values 10 and 15. Highest value for average rewards is for depth = 10.
- Runs 31 and 33 have replicates and the best Depth here is also 15. The optimal value among all the examined groups was for depth = 15.

### e) Second layer convolution receptive field (RFS2)

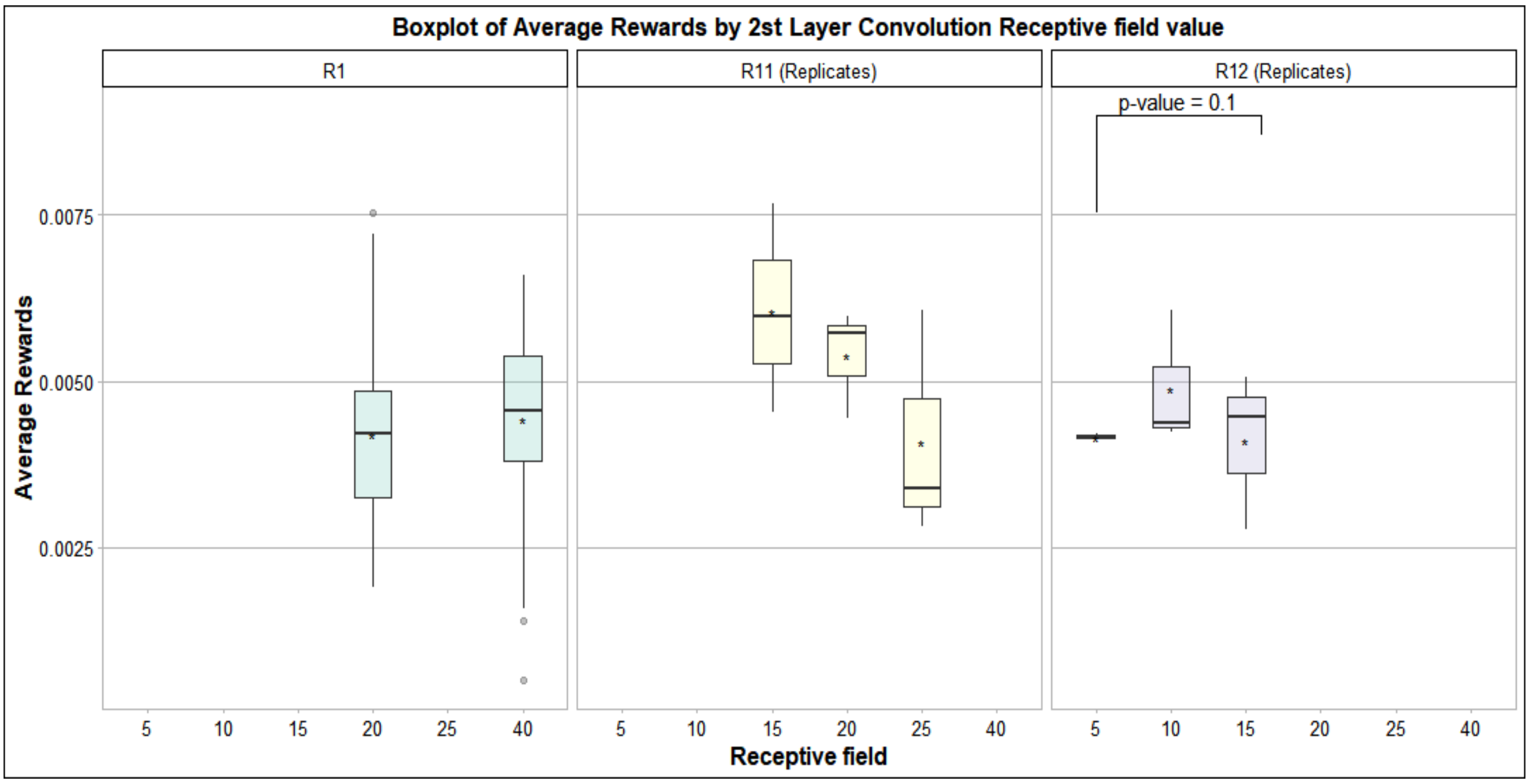

**Figure 29: Second Convolutional layer receptive field (filter size) comparison for values 5, 10, 15, 20, 25 and 40. The significance level is annotated.**

The above Figure shows the receptive field of the second convolutional layer as a function of average wined rewards. There are 3 different run sets Run1, Run 11 (multiple runs with the same implementation) and Run 12 (also with replicates). In Group 1 we examined receptive fields 20 and 40 and both had similar results. In Run 11 values 15, 20, 25 were tested for the receptive field and average rewards are decreasing respectively as is increasing the receptive field. In Run 12 we tested values 5,10,15 for receptive field with different settings than the previous groups (color purple). The optimal receptive field is 10. Between values 5 and 15 there is a significant difference with p-value 0.1.

### f) Second layer convolution subsample (Stride S2)

Contrary to the first layer subsample size, the second layer subsample, has a reduction in average rewards using filter of size 2. We didn't experiment more with other stride sizes and we kept the smallest for the next runs.

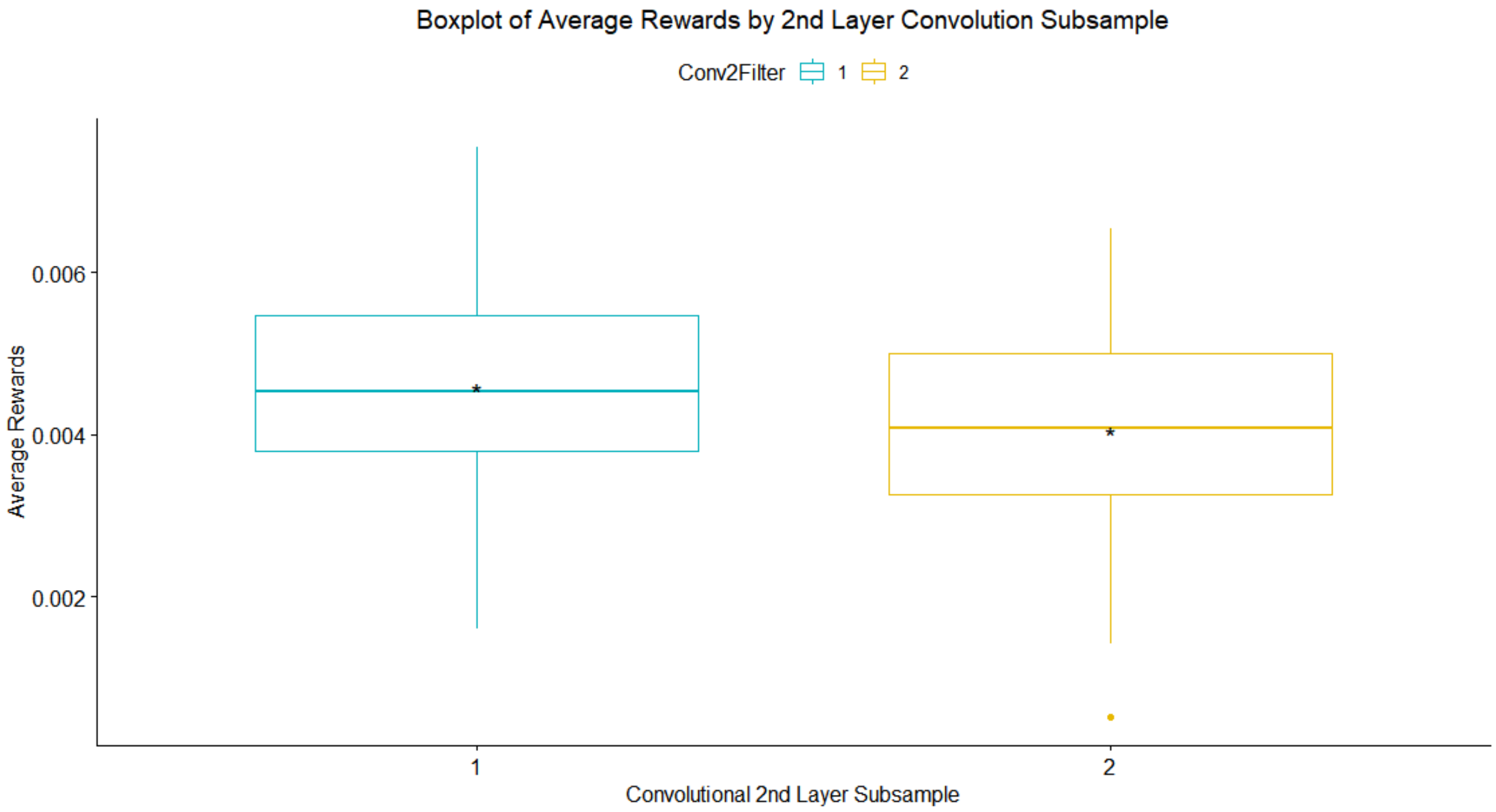

**Figure 30: This boxplot outlines the second convolutional layer's stride size as a function of Average Rewards. The tested values were 1,2 with slightly difference in their performance.**

### g) Dense layer size

Dense is the fully connected layer of a CNN, a linear operation in which every input is connected to every output by weight. For dense dimensions, we evaluated the values 32, 64, 100, 128, 200 and 256. Figure 1 shows a list of boxplots of the average rewards by the dimension of the fully connected layer. Three different groups have been formed and declare which values/metrics can be compared, as long as those in the same group share common settings. In the first group, there is only one test Run 1 (cyan color). It is computed separately because the variations lie in all the network's architecture parameters. Second, third and fourth Run form the second group (yellow). Here, only 2 parameters differentiate and dense = 100 had the highest average rewards.

The rest runs (35, 36, 37) (in purple) are replicates of the same implementation and only examine the dense parameter. Results showed that dense equals to 200 had better performance.

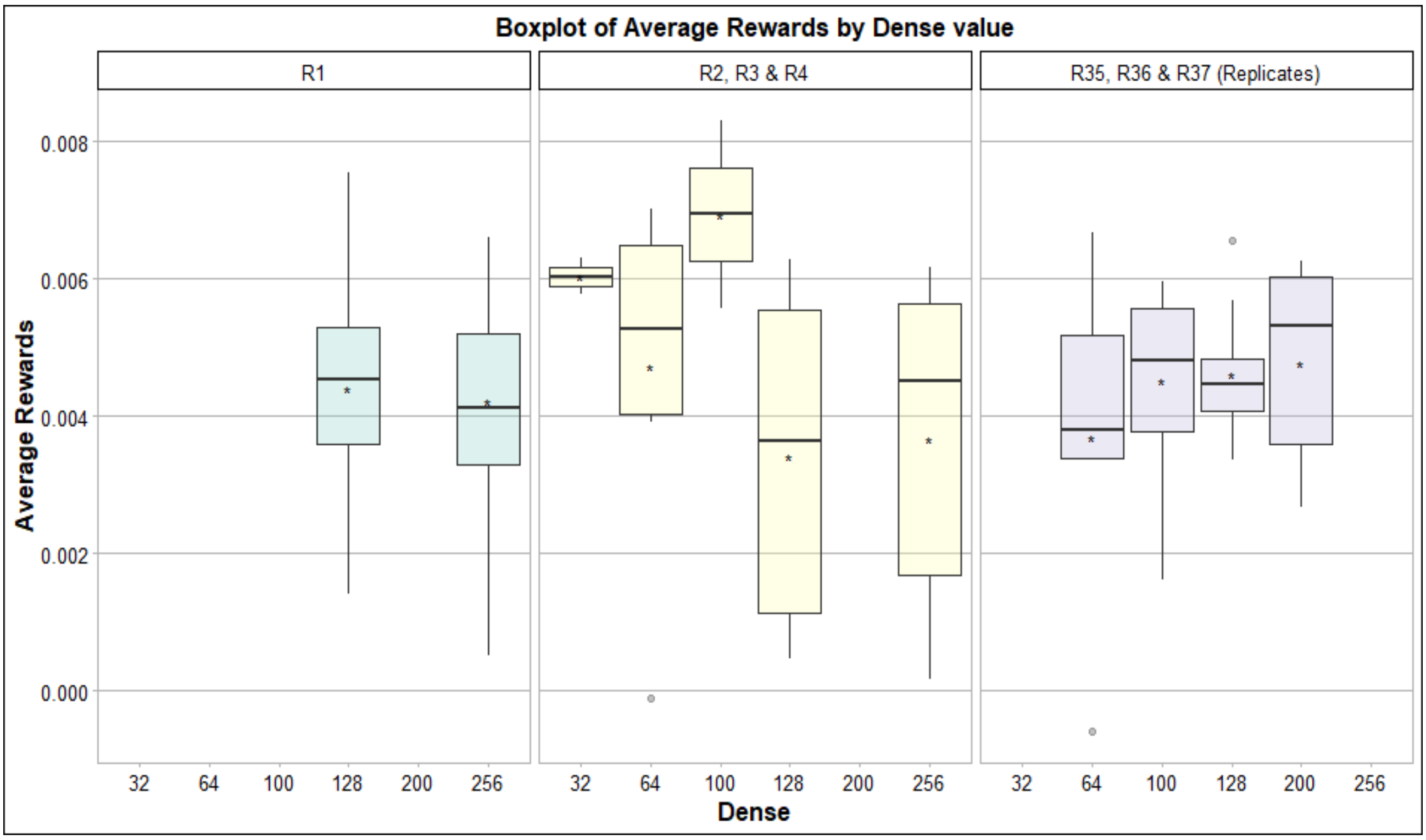

**Figure 31: Fully connected layer's dimensions comparison. Boxplots with the same color can be compared with each other. For dense = 100 (yellow) and for dense = 200 (purple) we had the best results**

### h) Input image size (IS)

During the preprocessing, the game screenshots are reshaped to smaller size images in order to reduce the computational cost of training.

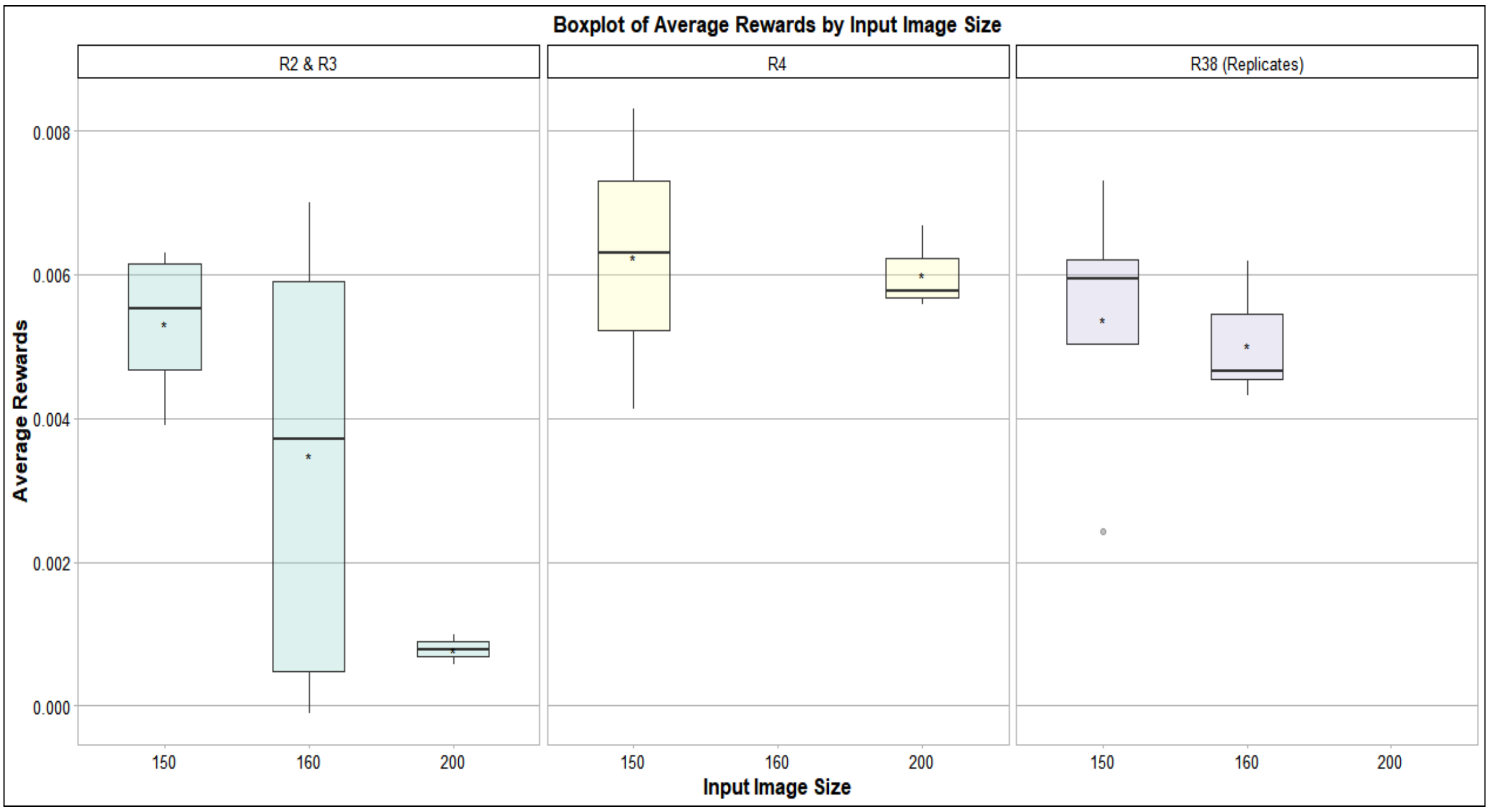

**Figure 32: Input size comparison for 150, 160 and 200 in function of the average rewards. There are 3 groups with different properties in each case. In all cases the input size 150 had optimal performance.**

### i) Max running steps

Running steps that required to finish the training procedure. In Figure 33_we can see the tested values (20000, 30000, 40000). For more running steps it seems that we have better results in average rewards. However, during the testing of 40000 parameter some tests failed to finish within the runtime limit of our HPC environment, thus we used 30000 steps in the runs of the following sections.

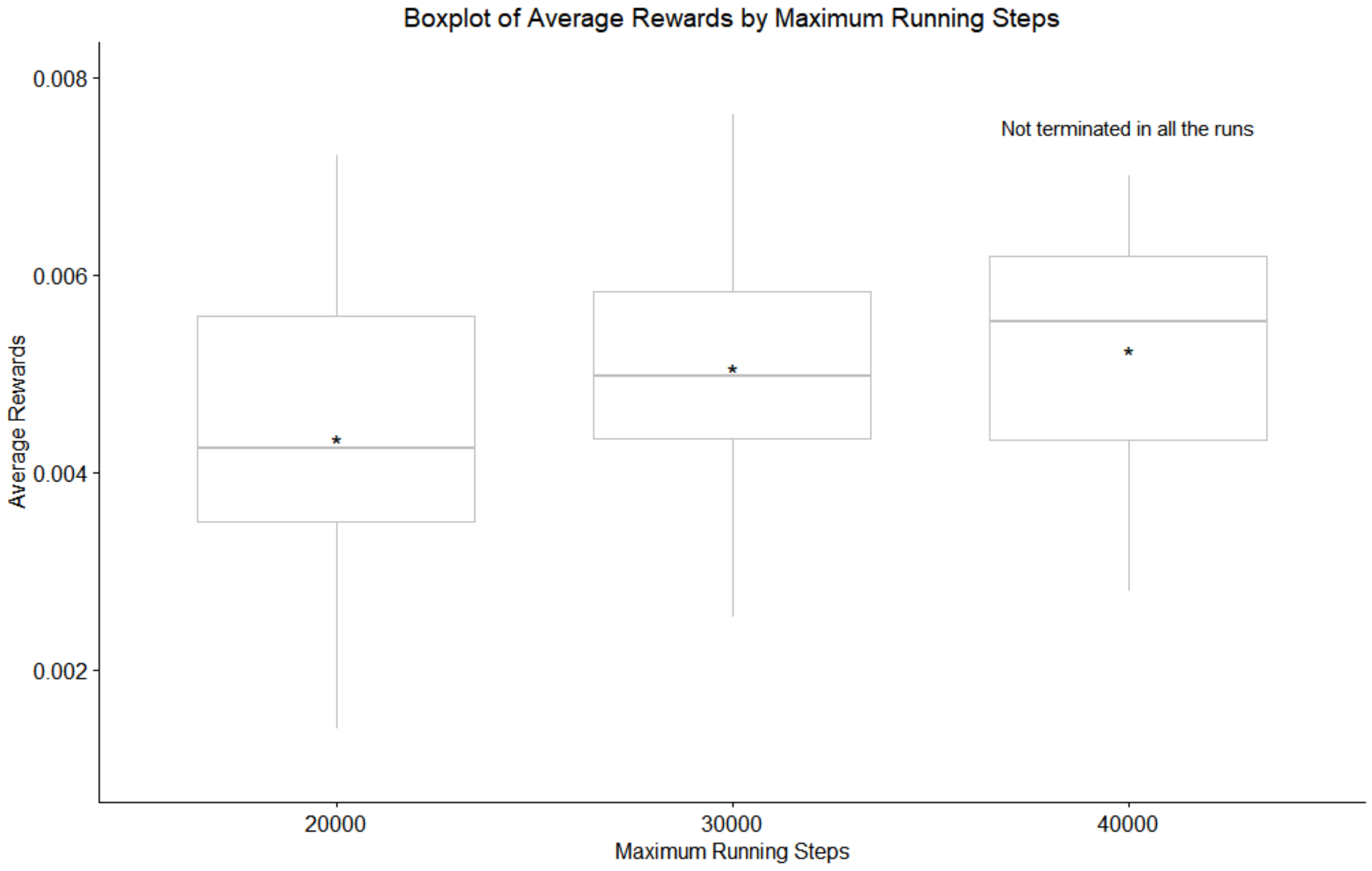

**Figure 33: Required running steps as a function of Average Rewards.**

### j) Number of hidden layers

This is one of the most important hyperparameters to experiment with, as more layers are able to create more complex functions but also make the learning process slower. Adding more hidden layers of neurons generally improves accuracy, to a certain limit which can differ depending on the problem.

In Figure 34, a 2-hidden **layer** architecture is compared with a 3-layer architecture. Both of them have share the same settings for the first two convolutional layers and the extra layer is the last. We observe that better results appear in the smaller network.

In Figure 35, values equal to (10, 20 and 30) for 3-layer convolutional **depth** are compared using as index the average rewards that achieved until the end of the training. We used the 3-layer architecture of Figure 10 and depth=10 had the highest average reward. Figure 10-12 and Figure 15 are referred to the same models hyperparameters, while Figure 13-14 have to do with the position of the extra layer.

Figure 13 shows that deeper and wider neural networks may not always work better. Although larger architectures tend to learn more complex data, in our case it does not seem to work because it is an DRL model [152].

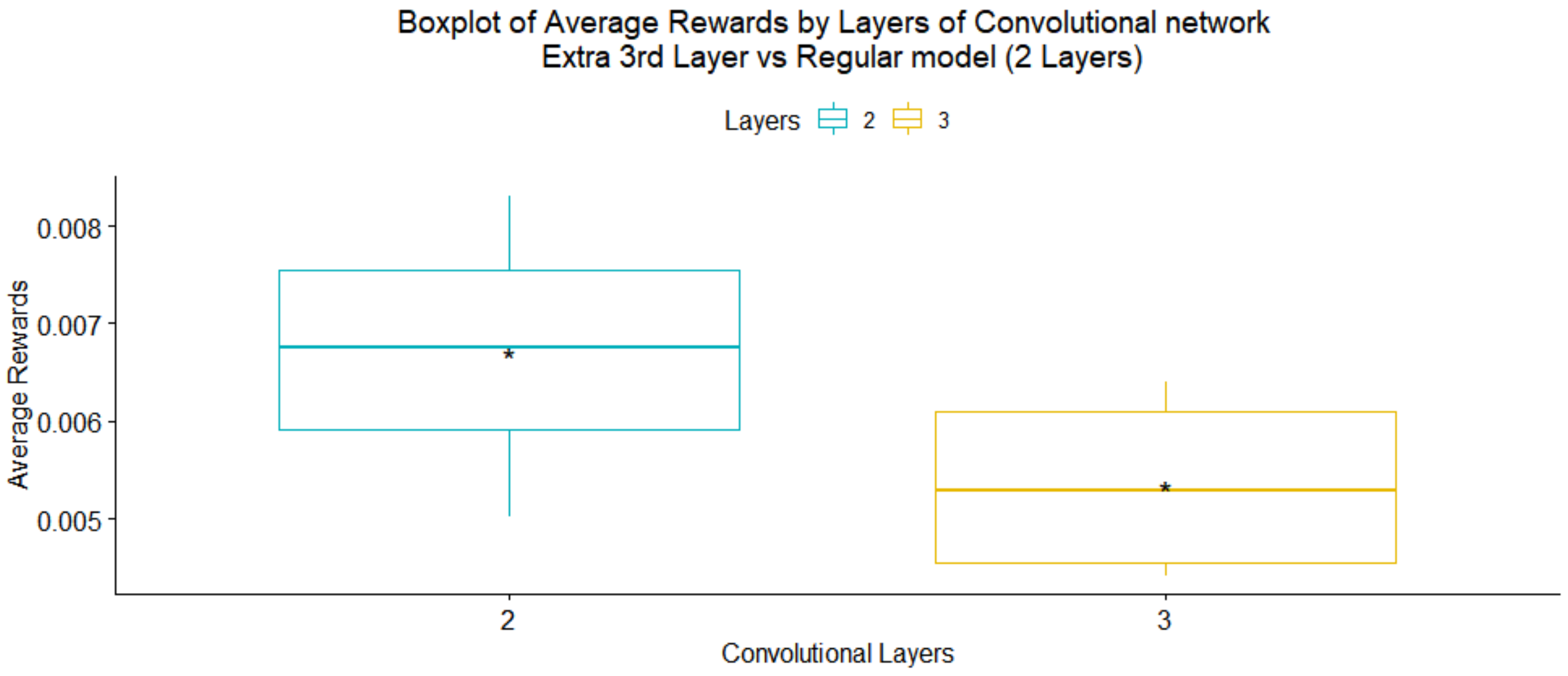

**Figure 34: Comparison of different architectures based on the number of hidden layers. Two-layer architecture versus a three hidden layer network. The two architectures have same settings in the two layers. Better results with no significant difference had the 2-hidden layer architecture.**

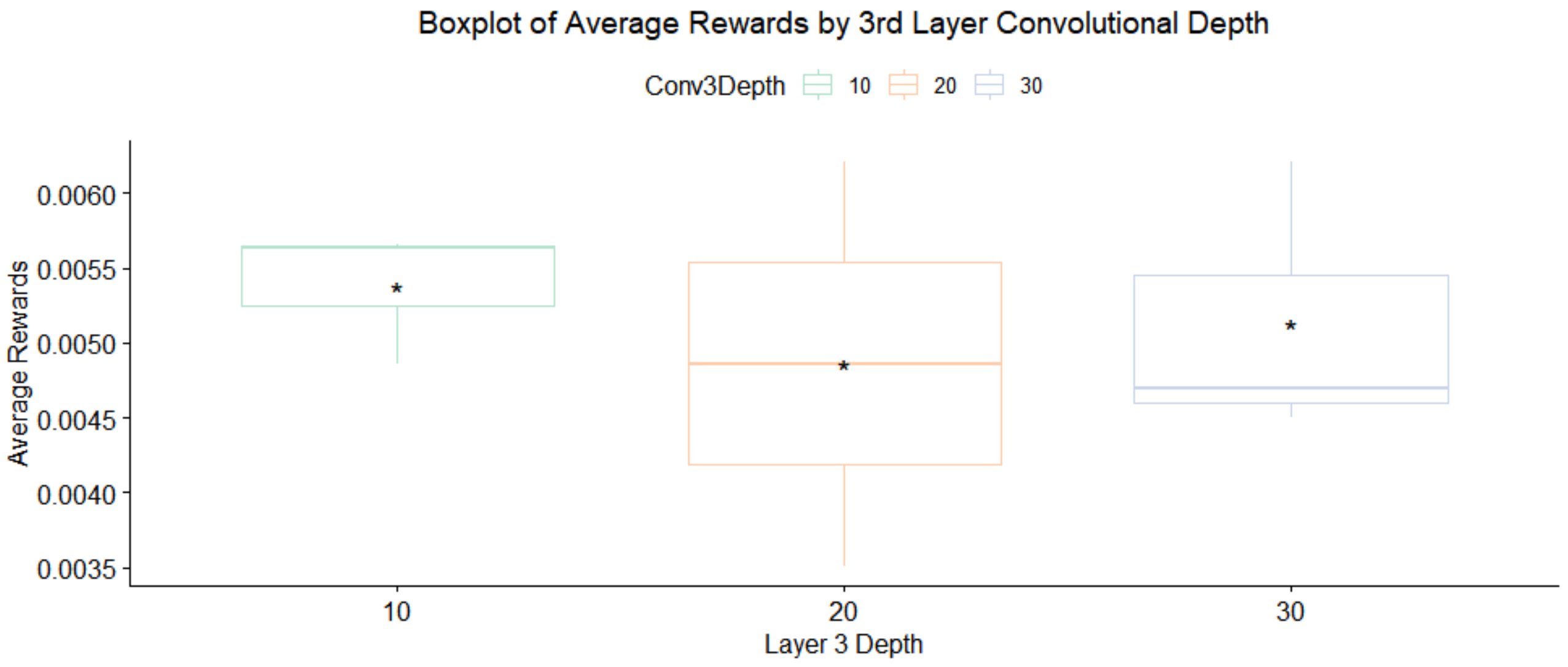

**Figure 35: 3-Layer architecture depth comparison for values 10, 20 and 30.**

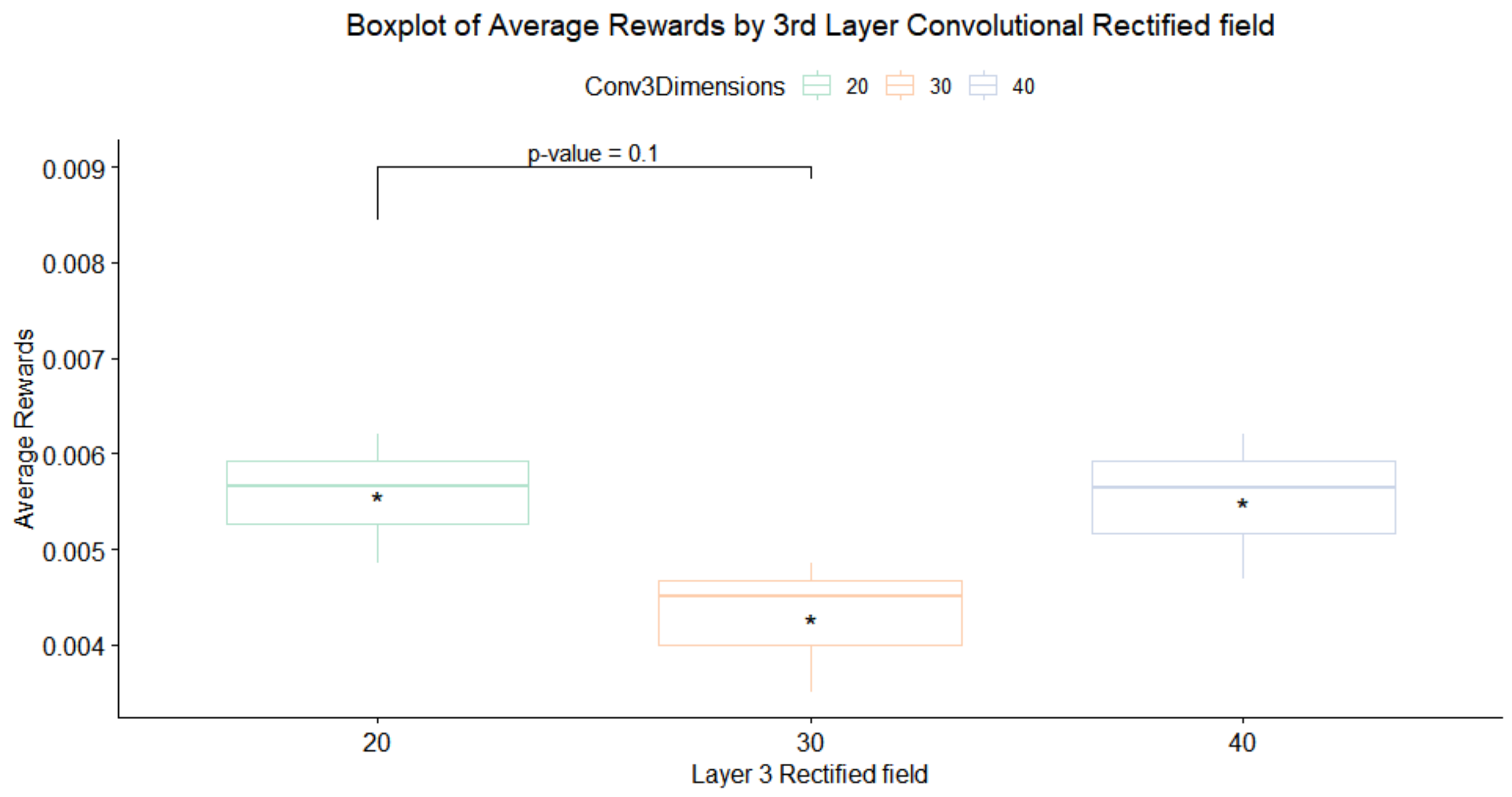

**Figure 36: 3-Layer architecture rectified field comparison for values 20, 30 and 40.**

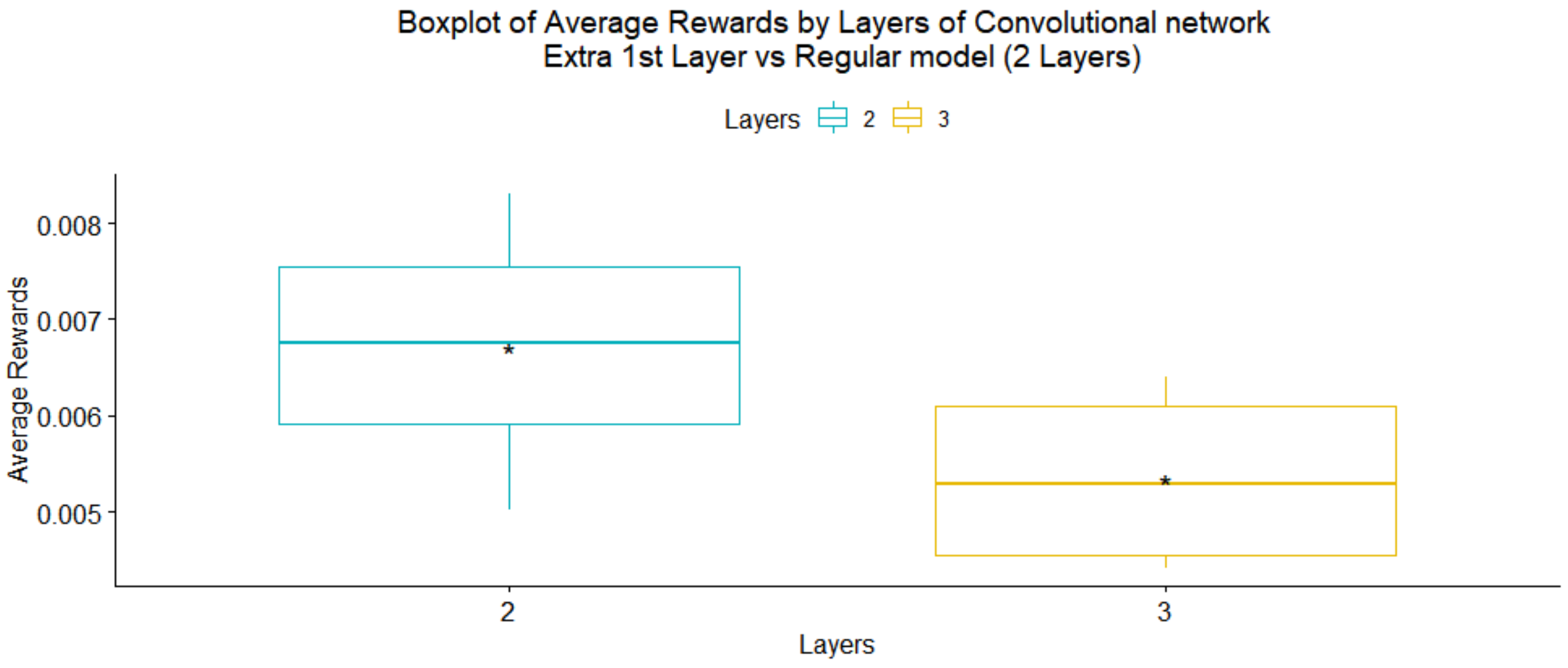

**Figure 37: Architecture of 2-hidden layers compared with 3-hidden layers. Both architectures have same settings for the last two convolutional layers and the extra layer is the first. Better results appear in the smaller network.**

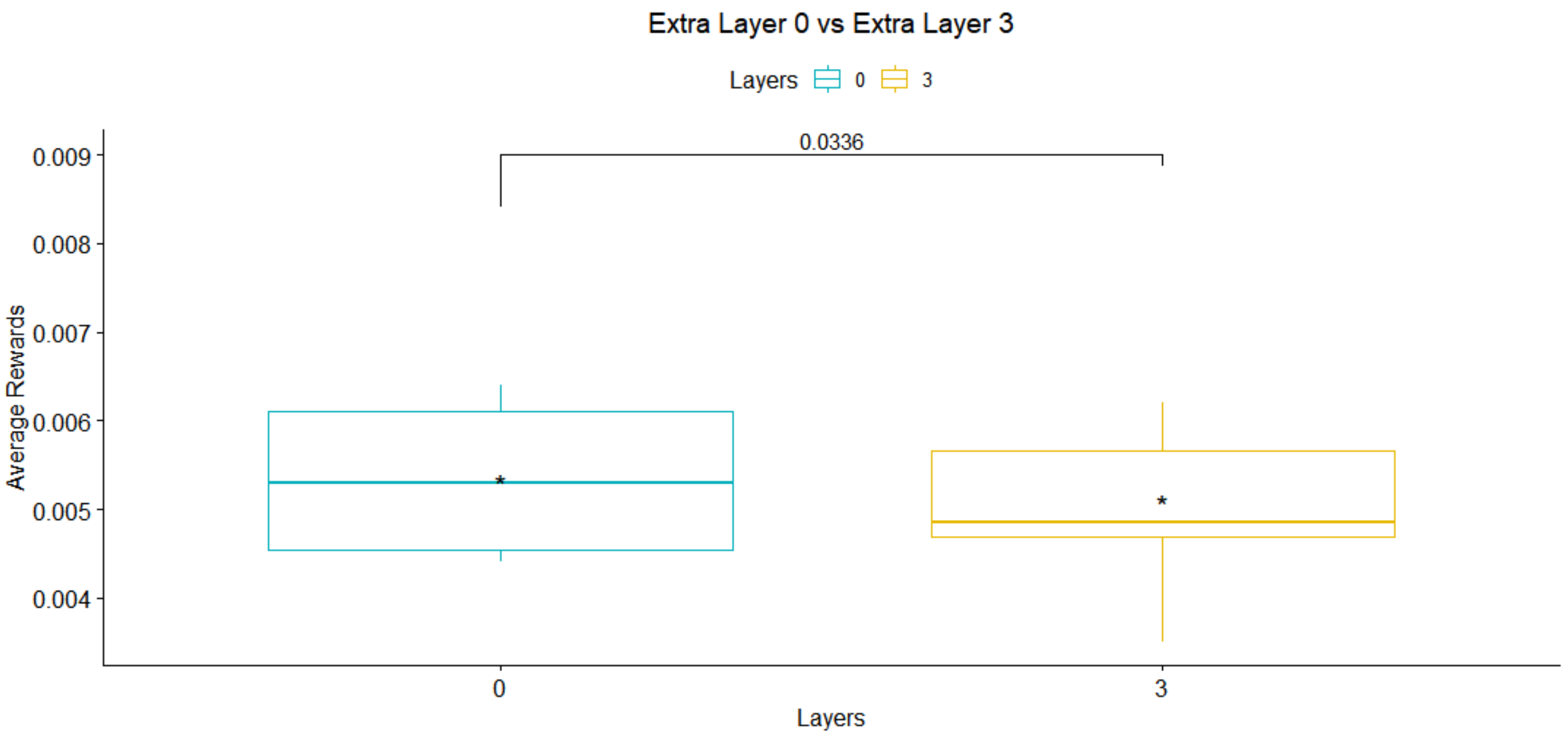

**Figure 38: Optimal position for the extra layer. First layer (with 0) versus last layer (with 3). As first the same extra layer appeared to have better performance with statistically significant difference.**

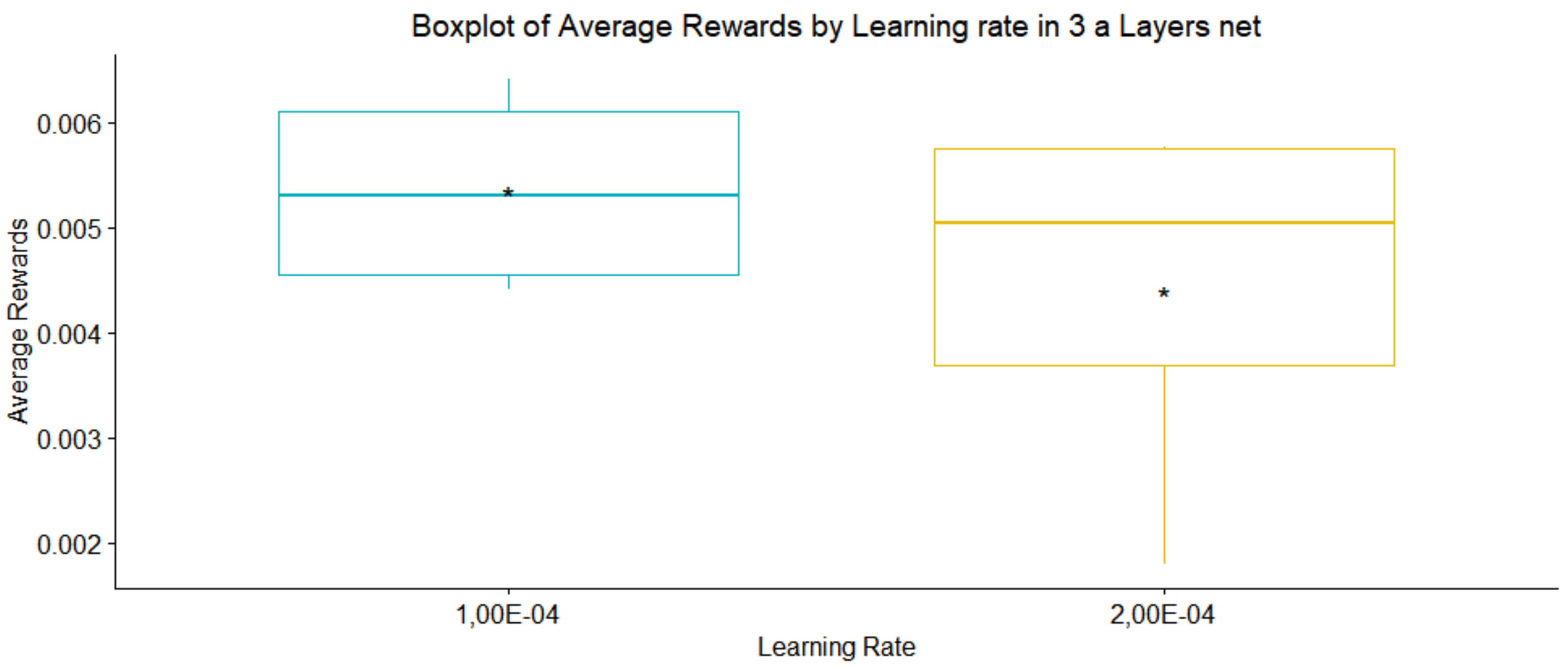

**Figure 39: Learning rate hyperparameter performance in a 3-layer convolutional neural network. The learning rate can take values $10^{-4}$ and $2*10^{-4}$. Better results appear for $10^{-4}$.**

## k) Protein Iterations

Is the number of iterations for each training sample- protein. This number is fixed and equals 200. Firstly, we tested this fixed number, and then we tried to increase the timestep randomly in range [1,200]. This will have as result to load more proteins in less training steps and Q will have random moves in all protein samples. The best results are shown for the step of 200.

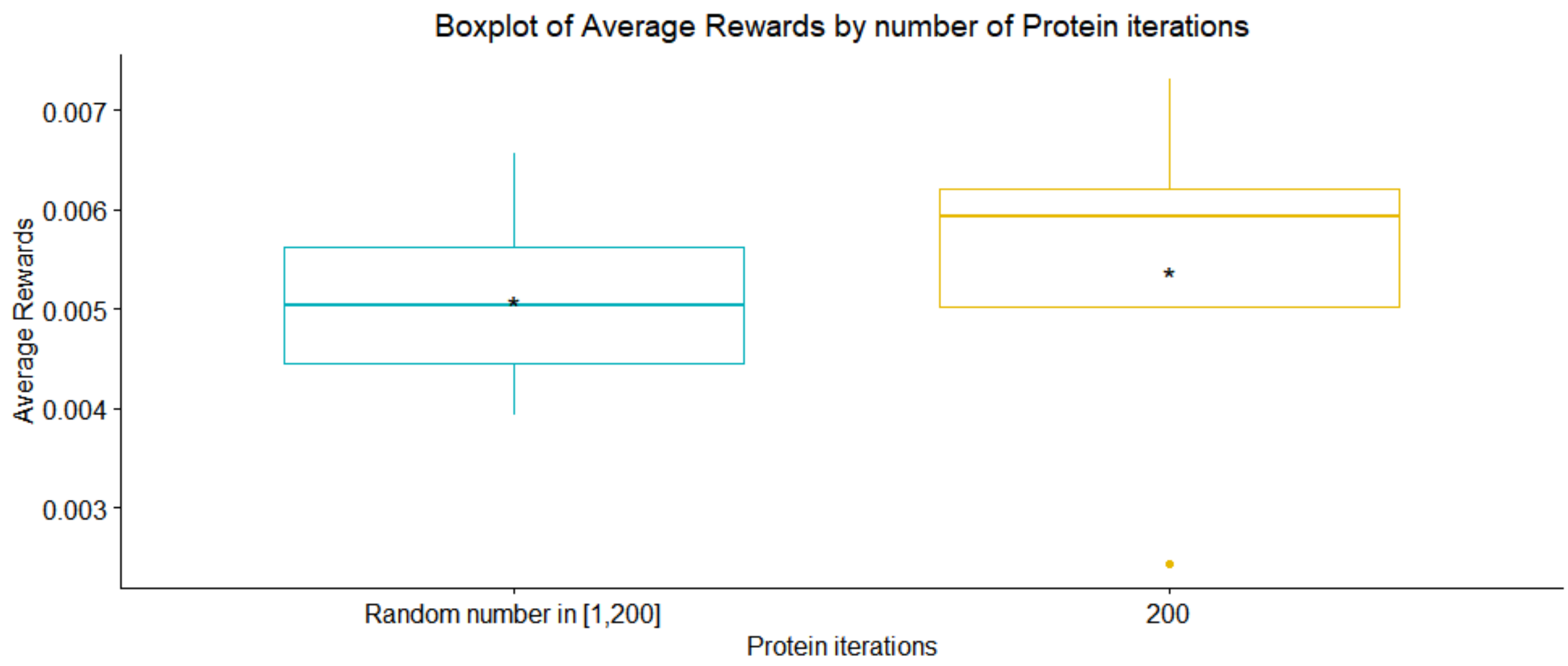

**Figure 40: Protein iterations for each input as a function of Average Rewards.**

### l) Mini batch size

An epoch is a group of samples which are passed through the model together and then run through backpropagation (backward pass) to determine their optimal weights. If the epoch cannot be run all together due to the size of the sample or complexity of the network, it is split into batches, and the epoch is run in two or more iterations. The number of epochs and batches per epoch can significantly affect model fit and determine the rate at which samples are fed to the model for training.

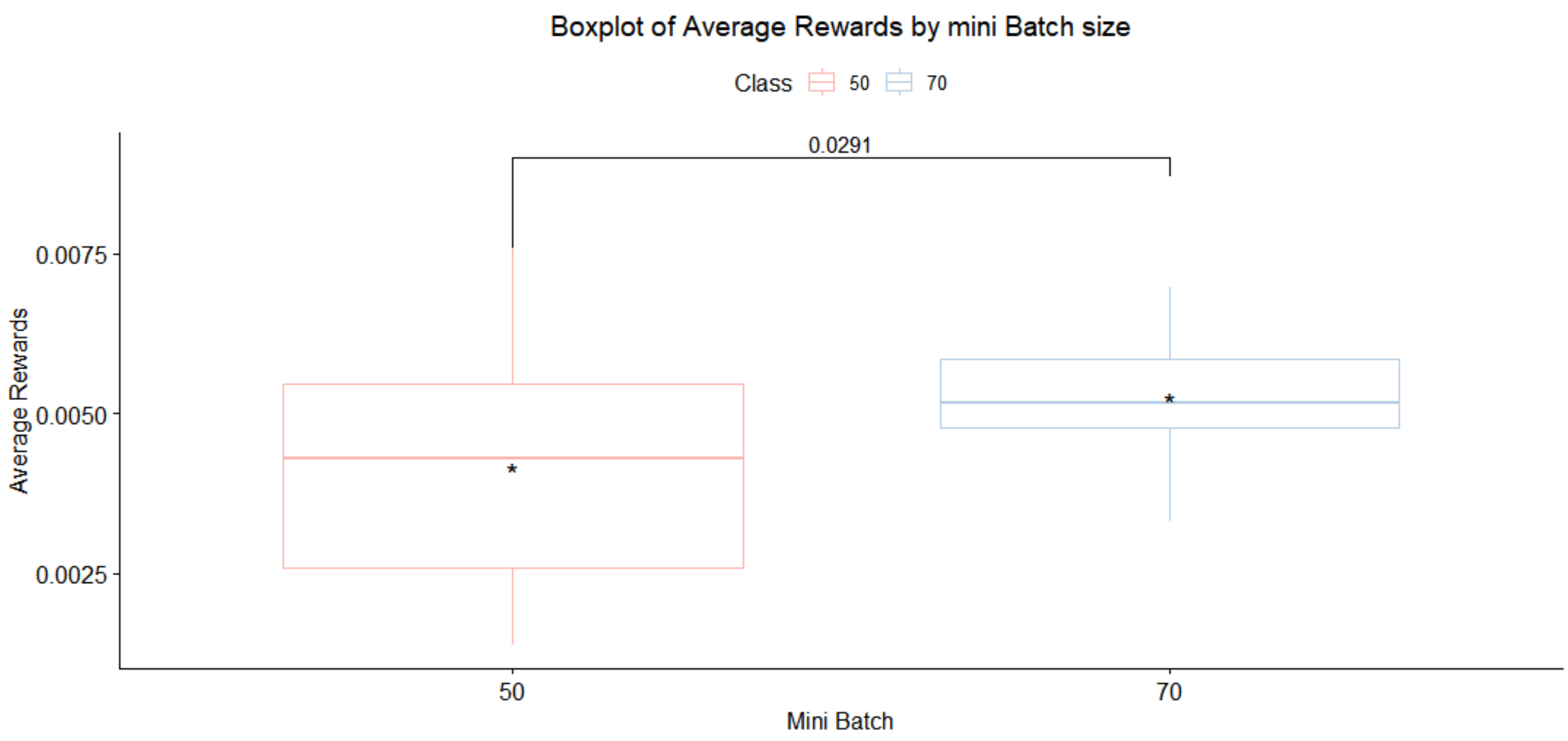

**Figure 41: Mini batch size comparison for values 50 and 70. The significance level is annotated.**

In Figure 41, we compare the running tests for Mini Batch value. We use Run 38, Run 6 and Run 25 for the first value (batch size = 50), while Run 17 and Run 18 are used to compute batch size 70. Better results appeared for size 70 where, the p-value is less than 5% so we can consider statistical significance.

## m) Actions per Training and initial dropout rate tests

Figure 18 shows actions per training as a function of average rewards. Four groups were examined. The first group (light blue) contains only Run 18 with 10 actions per training and no use of dropout). Second group (in yellow) we find two runs (Run 20 and Run 21) with similar properties. We used 0.1 dropout in that case for 3 and 15 actions per training. Between those two, 15 actions per training had better results in the gained rewards, with p-value = 0.0545 < 0.1 so, we have statistically significant difference. The next group contains Run 20 and Run 21 with a dropout rate of 0.2 and we performed 6 and 20 actions per training in each group of runs respectively. The optimal value for actions in this case was 6. In R45, no dropout was used and 5 actions per training was compared to 15 actions. Better results were obtained with 5 actions, but overall 6 actions per training is suggested as the optimal value.

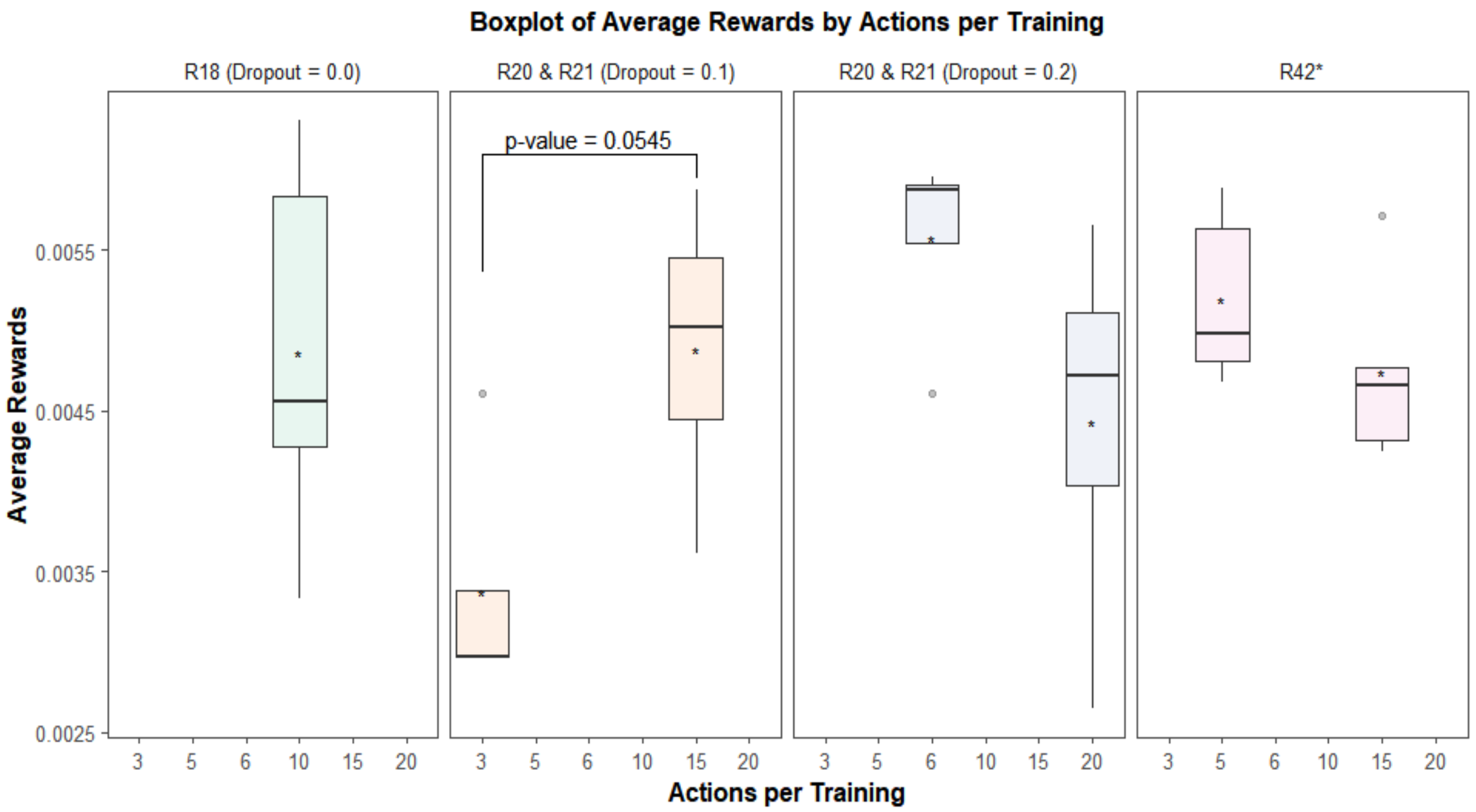

**Figure 42: Number of actions per training in function of Average rewards. Compared values 3, 5, 6, 10, 15 and 20. The level of significance is annotated.**

### n) Dropout rate and actions per training

The dropout rate determines the percentage of neurons in hidden layers that should be randomly ignored during training to prevent overfitting.

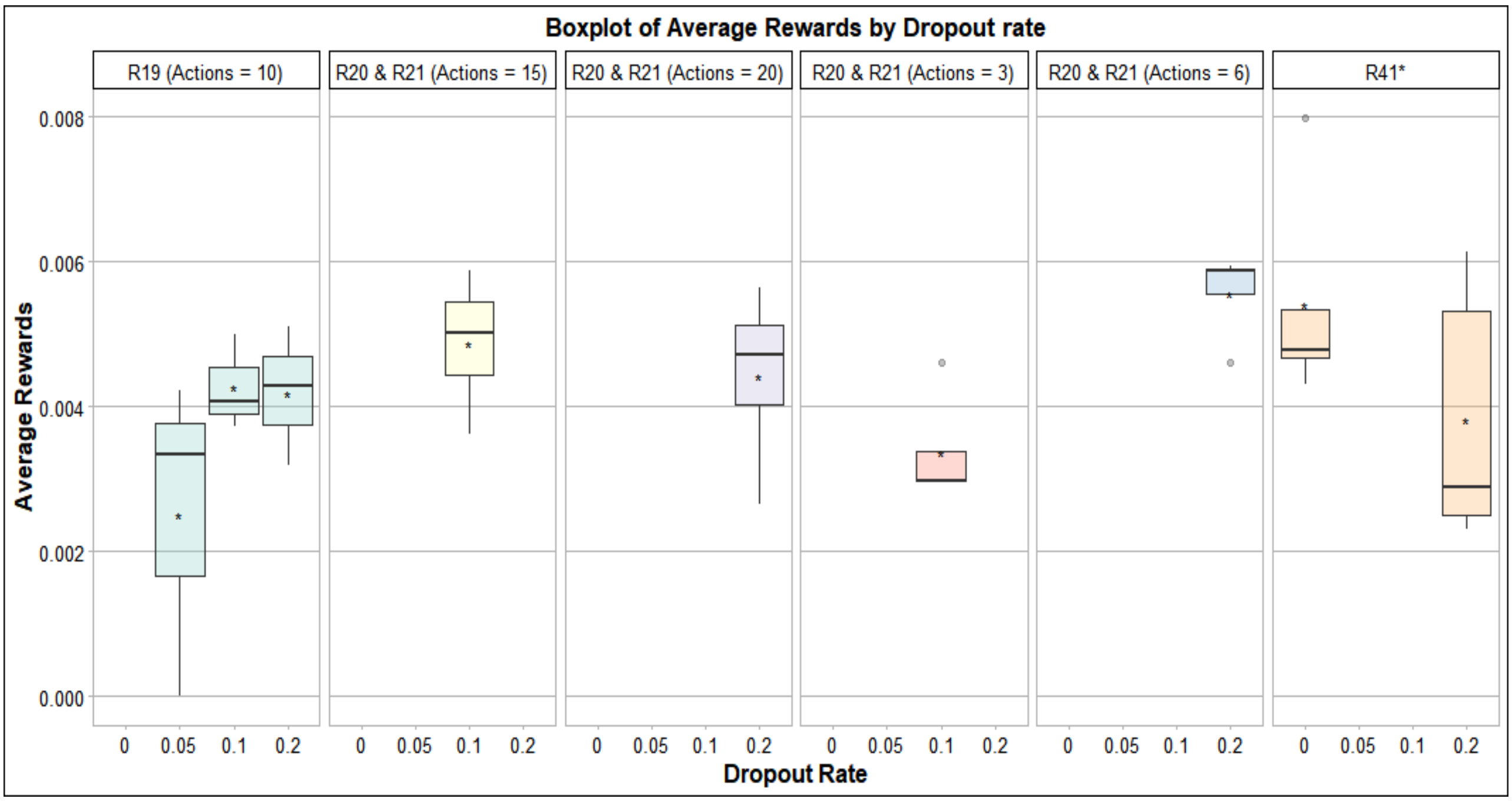

**Figure 43: Dropout rate performance**

In Figure 43, we show how game rewards are affected by the dropout rate. There are six different groups having different number of actions per training. The first group (light blue) performs 10 actions and uses for dropout rate the values 0.05, 0.1 and 0.2. In the next four groups we use the same hyperparameter settings (Run 20, Run 21) but different actions per training. The best results found for dropout rate 0.2. To verify this observation, we compare implementations with dropout rate = 0.2 with implementations with no use of dropout in Run 41. We observe slightly better average rewards without the use of dropout. This result can be explained as dropout rate essentially introduces a bit more variance and in reinforcement learning, additional variance is an issue for learning stability and for learning speed. Randomness in actions along with randomness inherent to the environment itself, leading to variance in observations. A similar observation is reported in [152], lecture 6.

## o) Score scaling

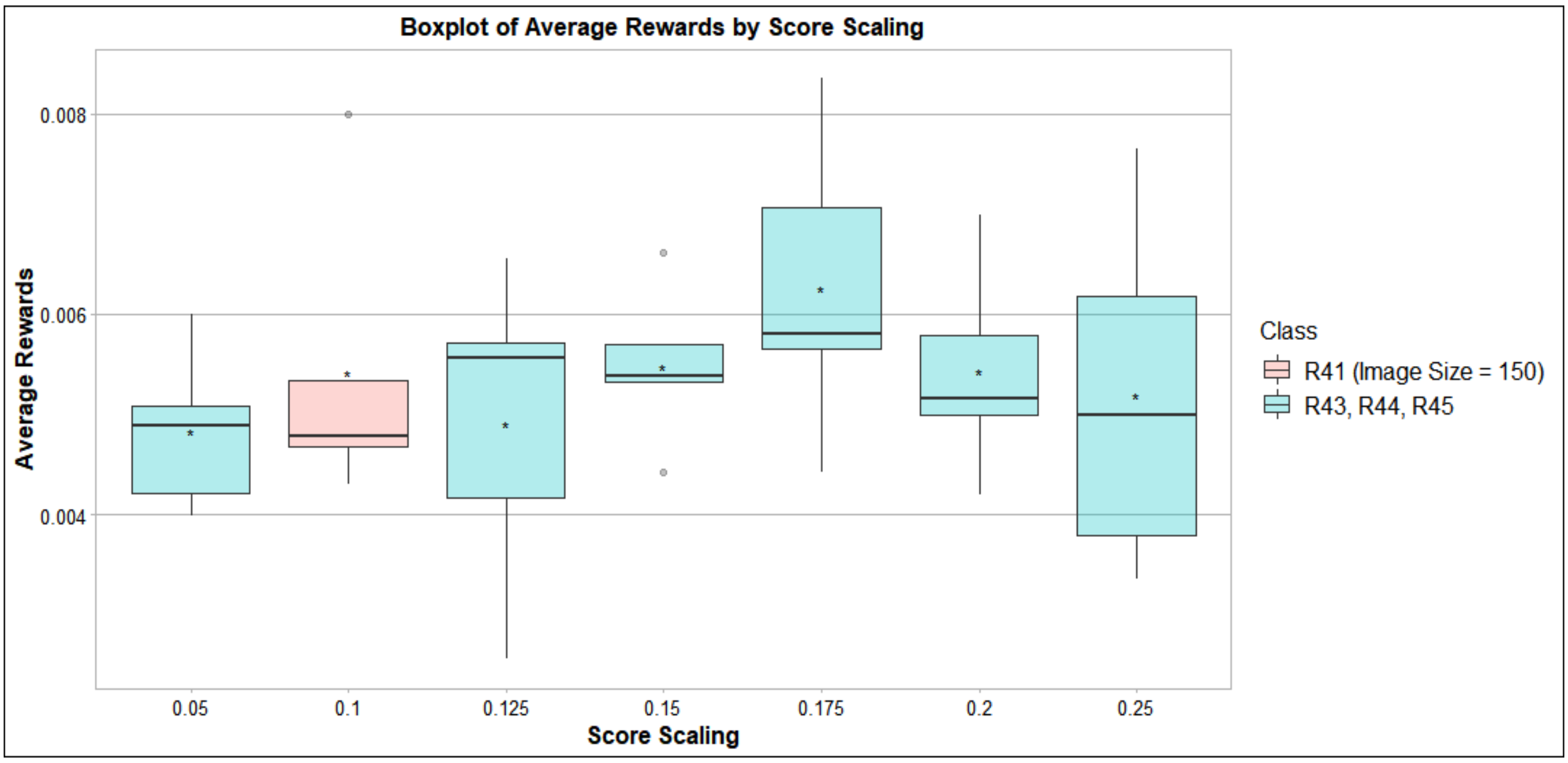

**Figure 44:Score scaling comparison. With the same color are the implementations with similar settings in the other hyperparameters.**

In Figure 44, we focus in finding the optimal value for score scaling, the value set to 0.1 in equation (1) in section 3.2.3 . With the same color are the implementations with similar settings in the other hyperparameters. Run 41 has different the image size (150) while the rest implementations have input size 160x160. Run 43, 44 and 45 tested 6 different values 0.05, 0.125, 0.15, 0.175, 0.2 and 0.25 for score scaling. The best performance was observed for score scaling = 0.175

### p) Learning rate

The Learning rate expresses how steep the backpropagation algorithm performs gradient descent. The learning rate is an important hyperparameter when configuring a neural network. During the parameter optimizing we tested three different values for learning rate ($0.7 \cdot 10^{-4}, 1 \cdot 10^{-4}, 3 \cdot 10^{-4}$). Figure 45, shows how fast the model is learning. Best results appear for learning rate $10^{-4}$. The p-value is 0.0556, less than 0.1 that means that we can also consider statistical significance between the second-best values.

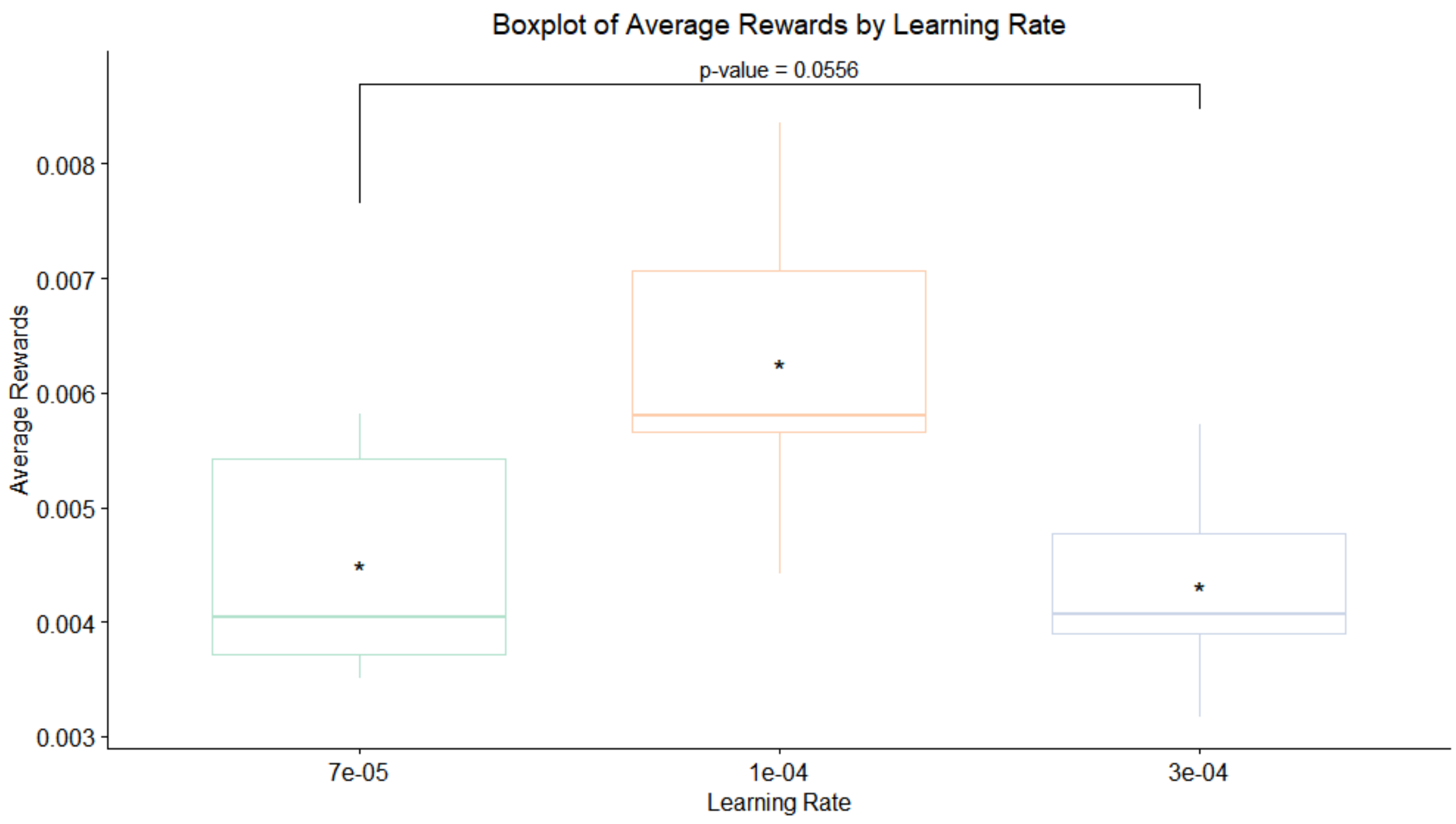

**Figure 45: Learning rate performance. Tested runs: 46, 56-58.**

## q) Momentum

Momentum exists to increase the speed of training gradually, with a reduced risk of oscillation. We used Nesterov momentum update in most runs.

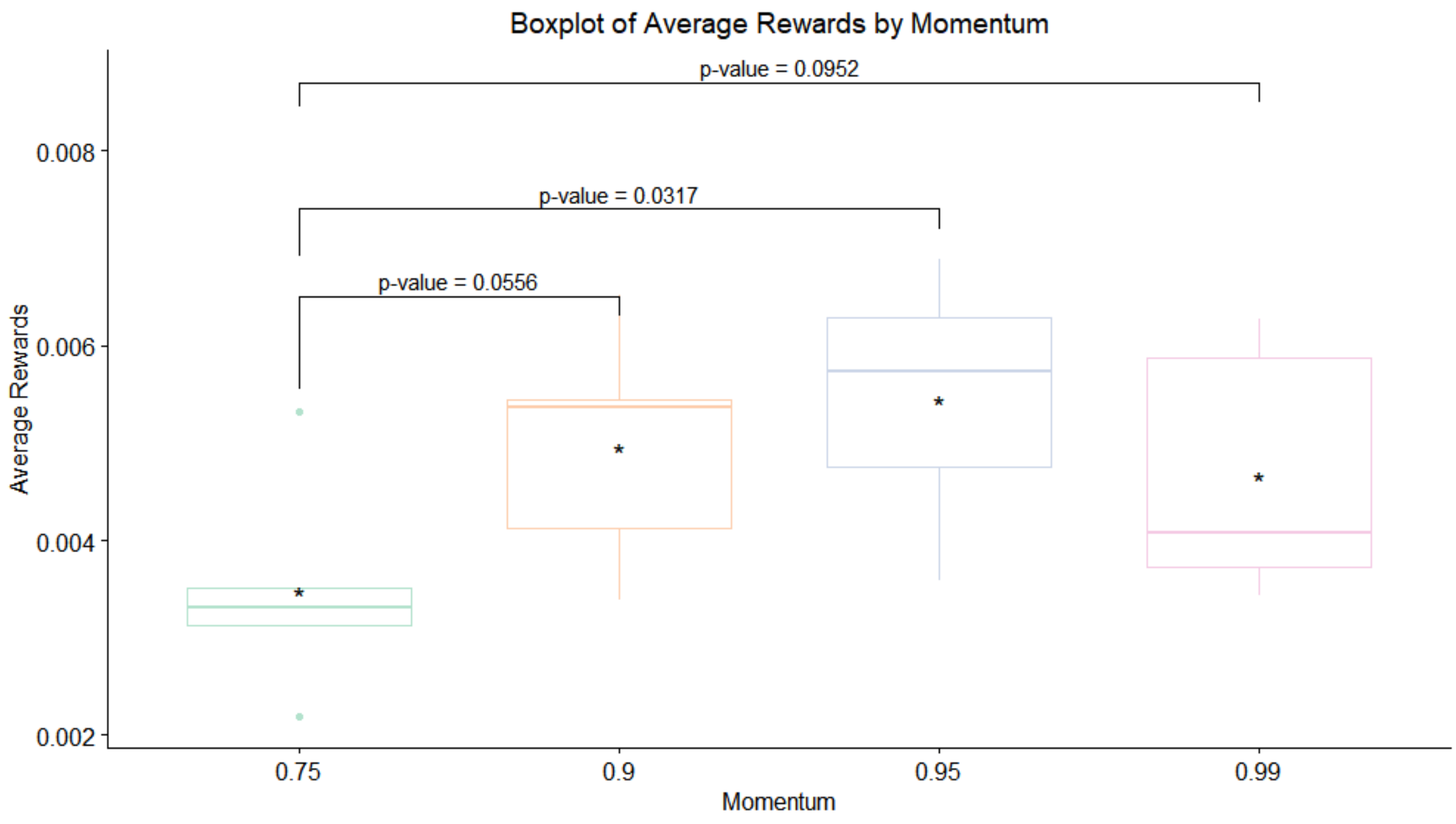

**Figure 46: Momentum as a function of Average Rewards. Runs for testing: 47, 48.**

All indicated p-values show statistically significant differences. The best results were observed for momentum = 0.95.

### r) Gamma

Gamma is a hyperparameter that controls the percentage of future awards that count for the computation of Q value. In this Figure we compared 4 different values (0.5,0.75,0.95,0.99) for γ. Optimal performance was observed for γ = 0.95 .

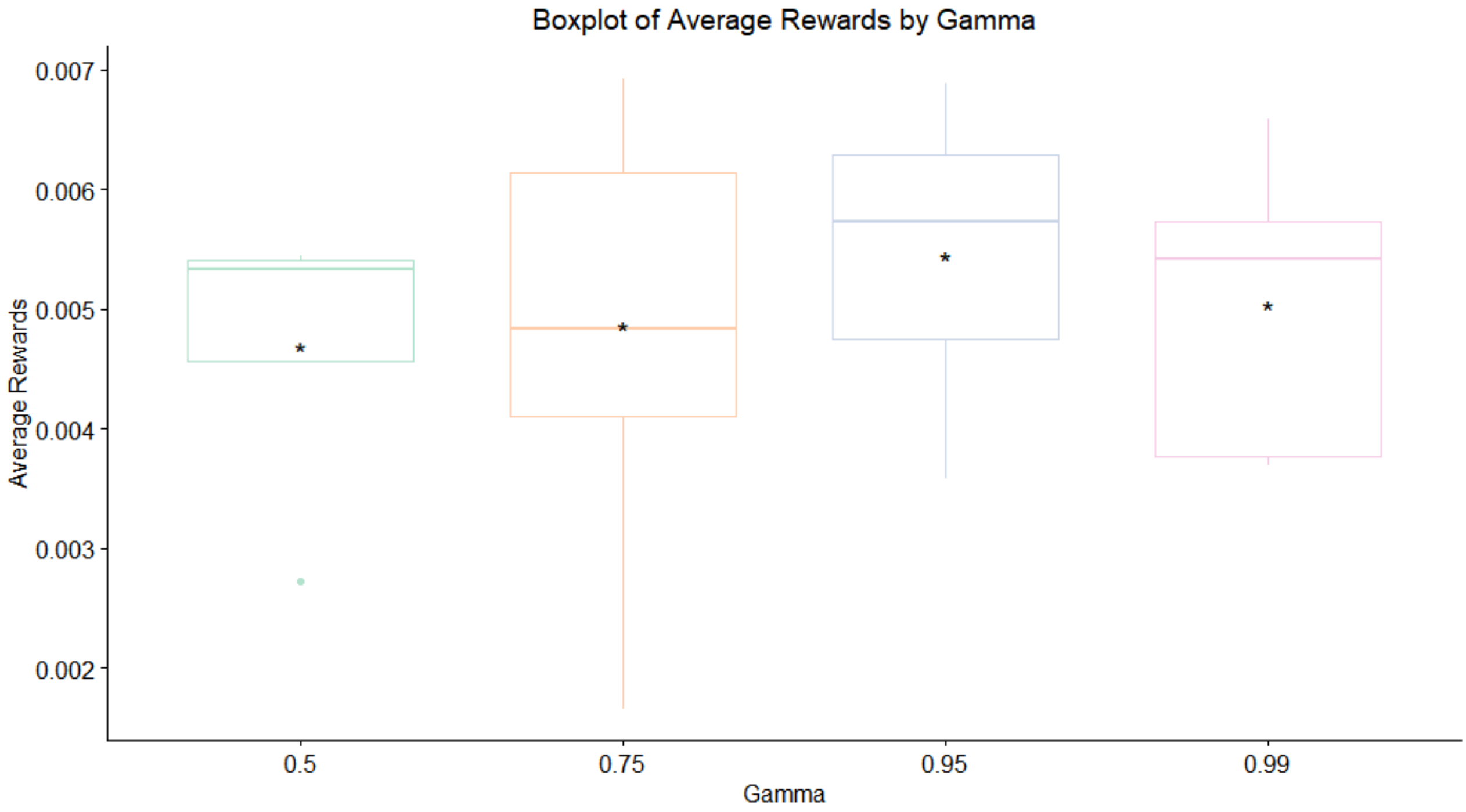

**Figure 47: Gamma value as a function of Average Rewards. Runs for testing: 49-52.**

## s) Initial epsilon

Initial value for exploration hyperparameter which controls the probability to perform a random action. This value is reduced during training (until it reaches final epsilon).

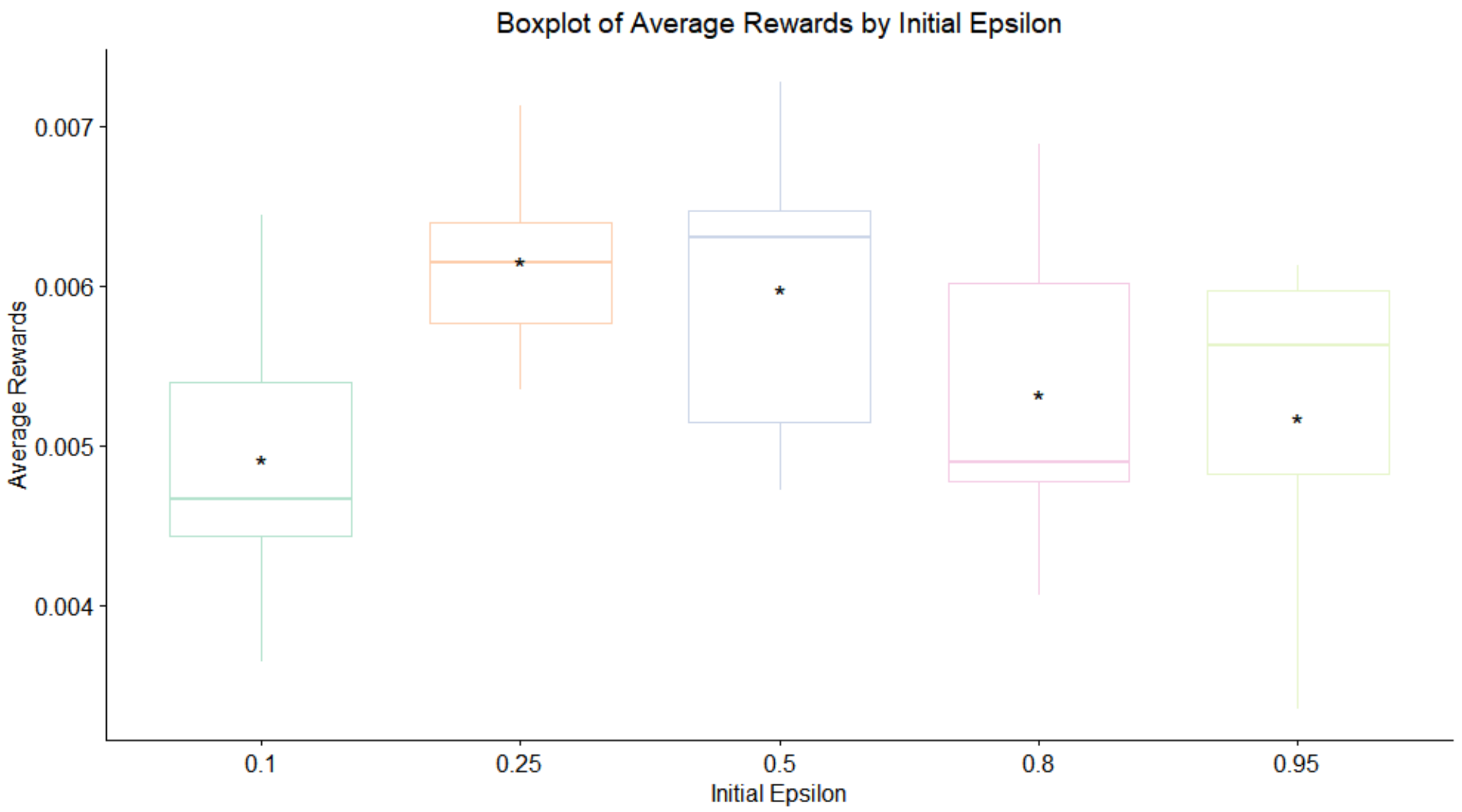

**Figure 48: Average rewards by initial ε. Five values were tested with best results for initial ε is 0.5. Details in Annex II: R54, R55.**

### t) Final epsilon

Final epsilon Is the minimum value that reaches epsilon during training. Final epsilon also has a connection with the random moves. In our implementation the tested values for ε are (0.005, 0.5, 0.2). The optimal final ε is 0.2 .

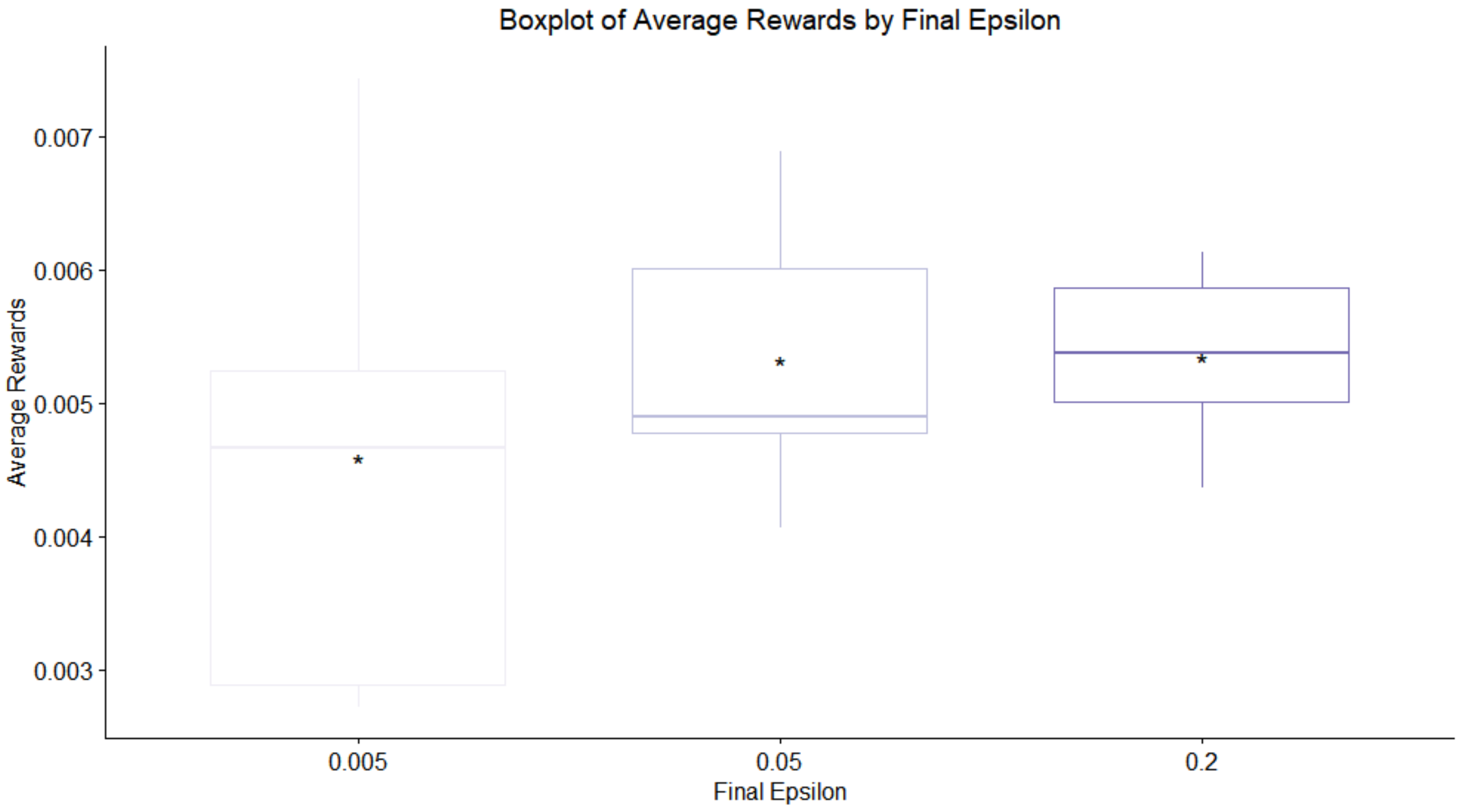

**Figure 49: In this boxplot, it appears how the average rewards are affected by the value of final ε.**

## u) Final exploration frame

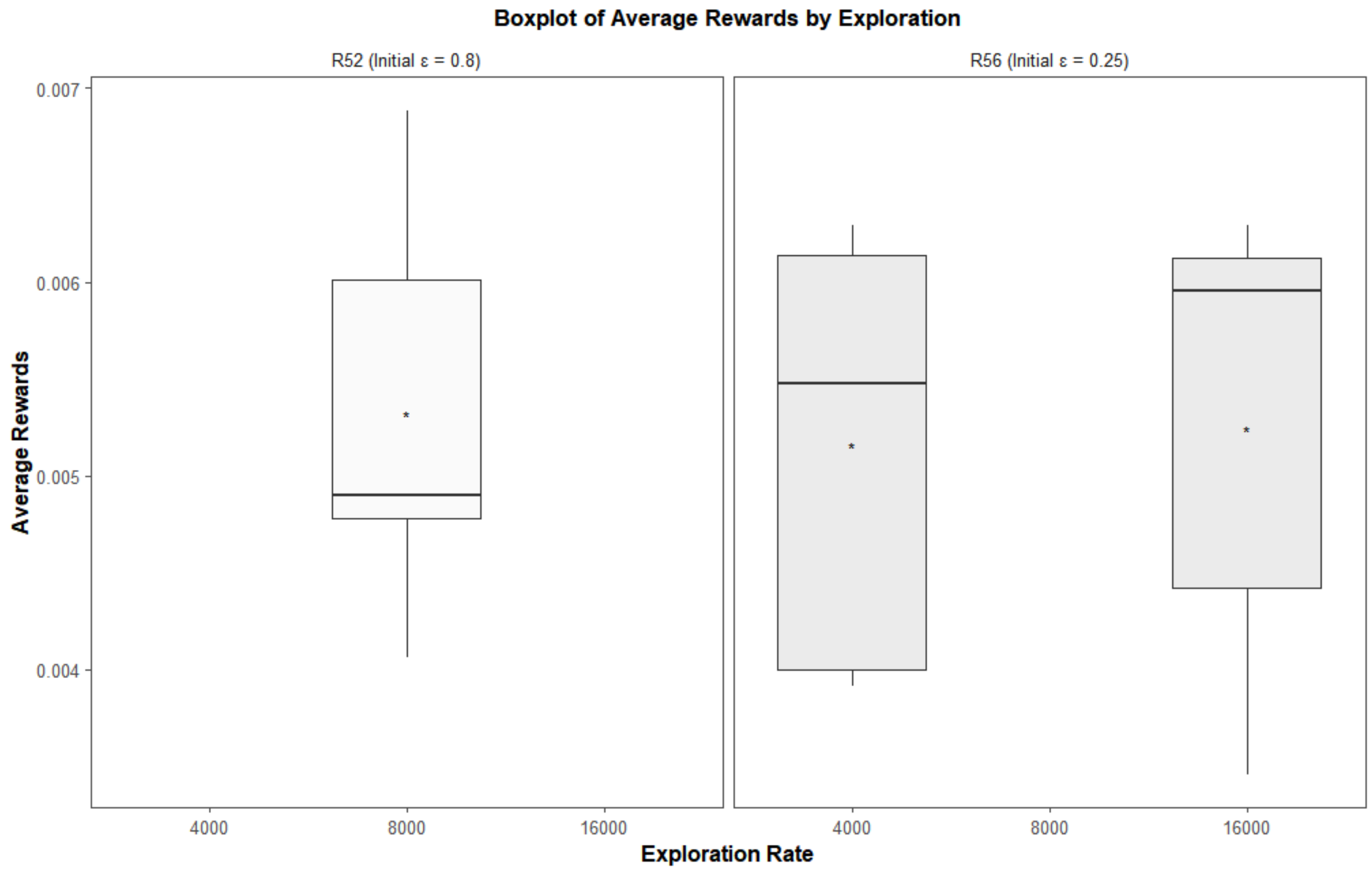

**Figure 50: Average rewards by final exploration frame. Five values were tested with better results for ε is 0.25. Runs for testing: 54, 55.**

Figure 50 presents the final exploration frame variations. There are 2 groups with different settings. In run 52, the initial ε is 0.8 and the final exploration frame is 8000. In Run 56, the initial ε= 0.25 and the final exploration frame takes values 4000 and 16000. Results show that variations in the final exploration frame make no significant difference in the gained rewards with a preference for a later final exploration frame.

### v) Replay memory

Replay memory is a parameter works as a buffer and stores the most recent transitions that the agent observes, allowing us to reuse this data later. The size of replay memory controls the history that the network can use.

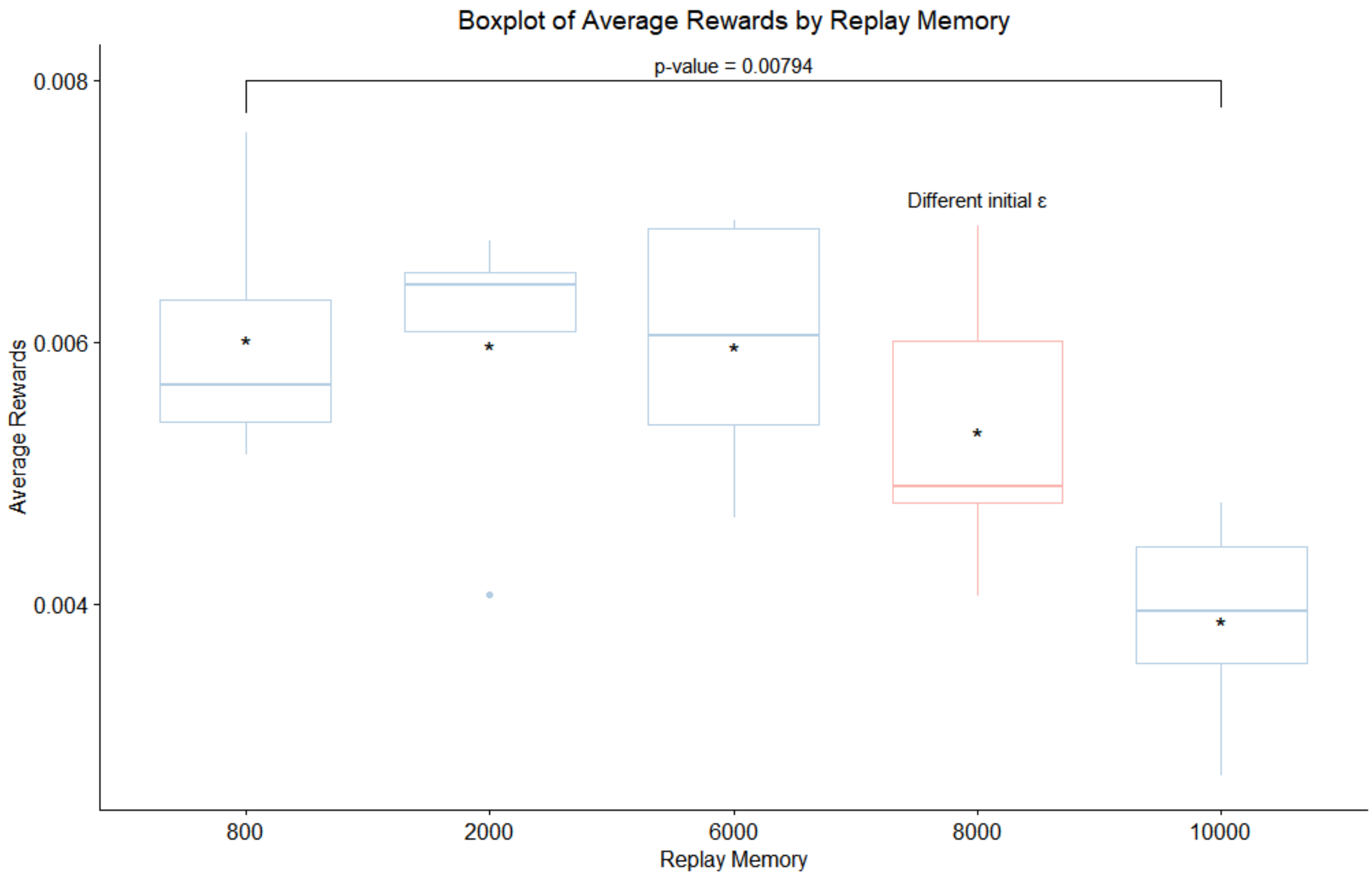

**Figure 51: Replay memory as a function of average rewards**

We have 5 different tested values for replay memory that range from 800 to 10000. The best results were observed for replay memory = 6000

### w) Batch normalization

Batch normalization [153] normalizes the output of a previous activation layer by subtracting the batch mean and dividing by the batch standard deviation. This increases the stability of a neural network. For better testing in batch normalization, we created four architectures. Architecture 1 applies batch normalization before the activation function, while in architecture 2 the batch normalization step, comes after the activation function. In Architectures 3, 4 we apply once the batch normalization, one in the final layer and the other in the first layer. Figures 51-55 presents the comparison of the following architectures:

### Architecture 1: Runs 390-394

| |
|---|
| Conv2D |
| Batch Normalization |
| Activation ReLU |
| Conv2D |
| Batch Normalization |
| Activation ReLU |
| Dense |
| Batch Normalization |
| Activation ReLU |

### Architecture 2: Runs 395-399

| |
|---|
| Conv2D |
| Activation ReLU |
| Batch Normalization |
| Conv2D |
| Activation ReLU |
| Batch Normalization |
| Dense |
| Activation ReLU |
| Batch Normalization |

### Architecture 3 : Runs 400-404

| |
|---|
| Conv2D |
| Activation ReLU |
| Conv2D |
| Activation ReLU |
| Dense |
| Activation ReLU |
| Batch Normalization |

**Architecture 4: Runs 405-409**

| |
|---|
| Conv2D |
| Activation ReLU |
| Batch Normalization |
| Conv2D |
| Activation ReLU |
| Dense |
| Activation ReLU |

The results in Figure 54 indicate that batch normalization does not improve the average rewards. A similar observation is mentioned in [152], lecture 6.

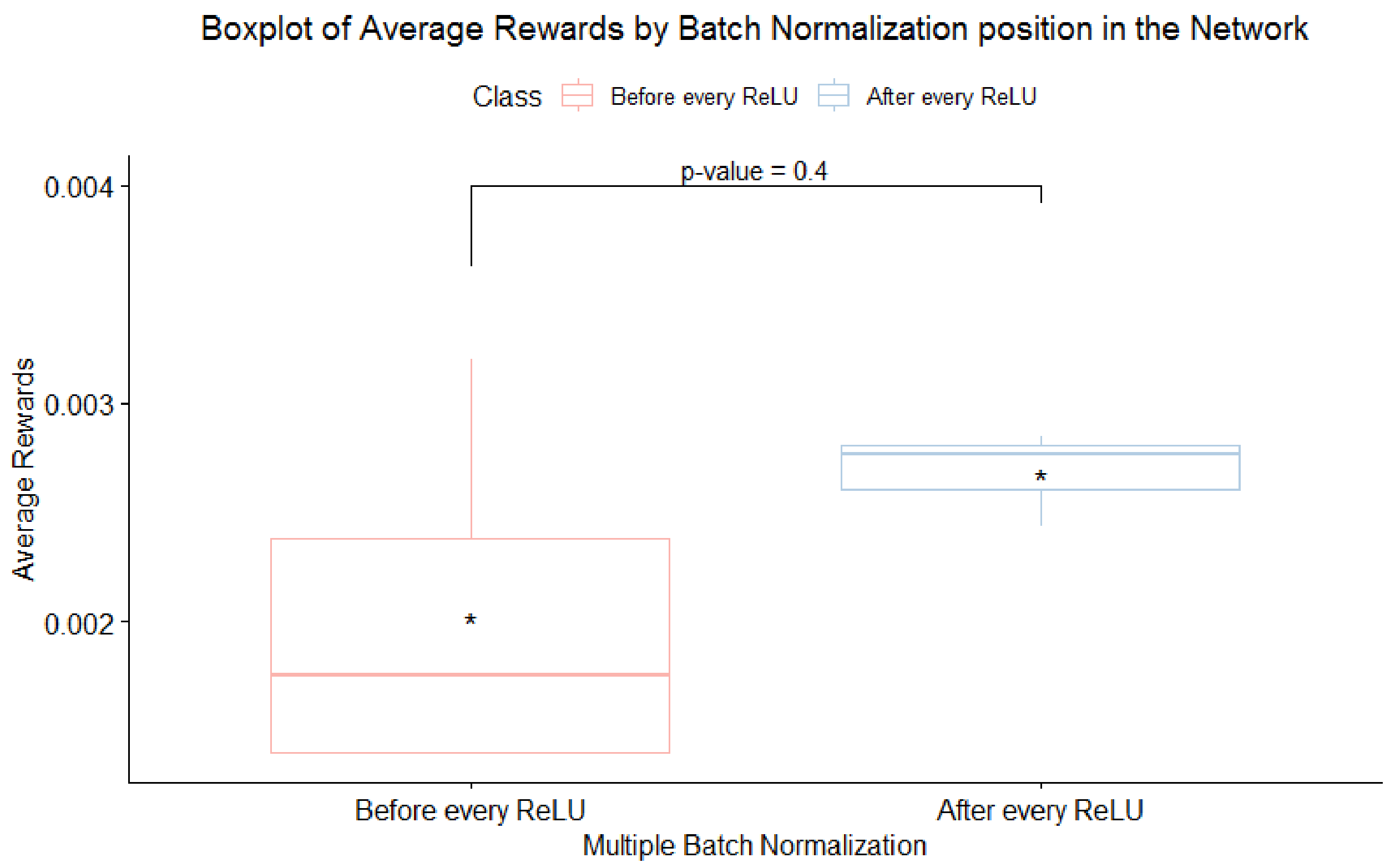

**Figure 52: Architecture 1 vs Architecture 2**

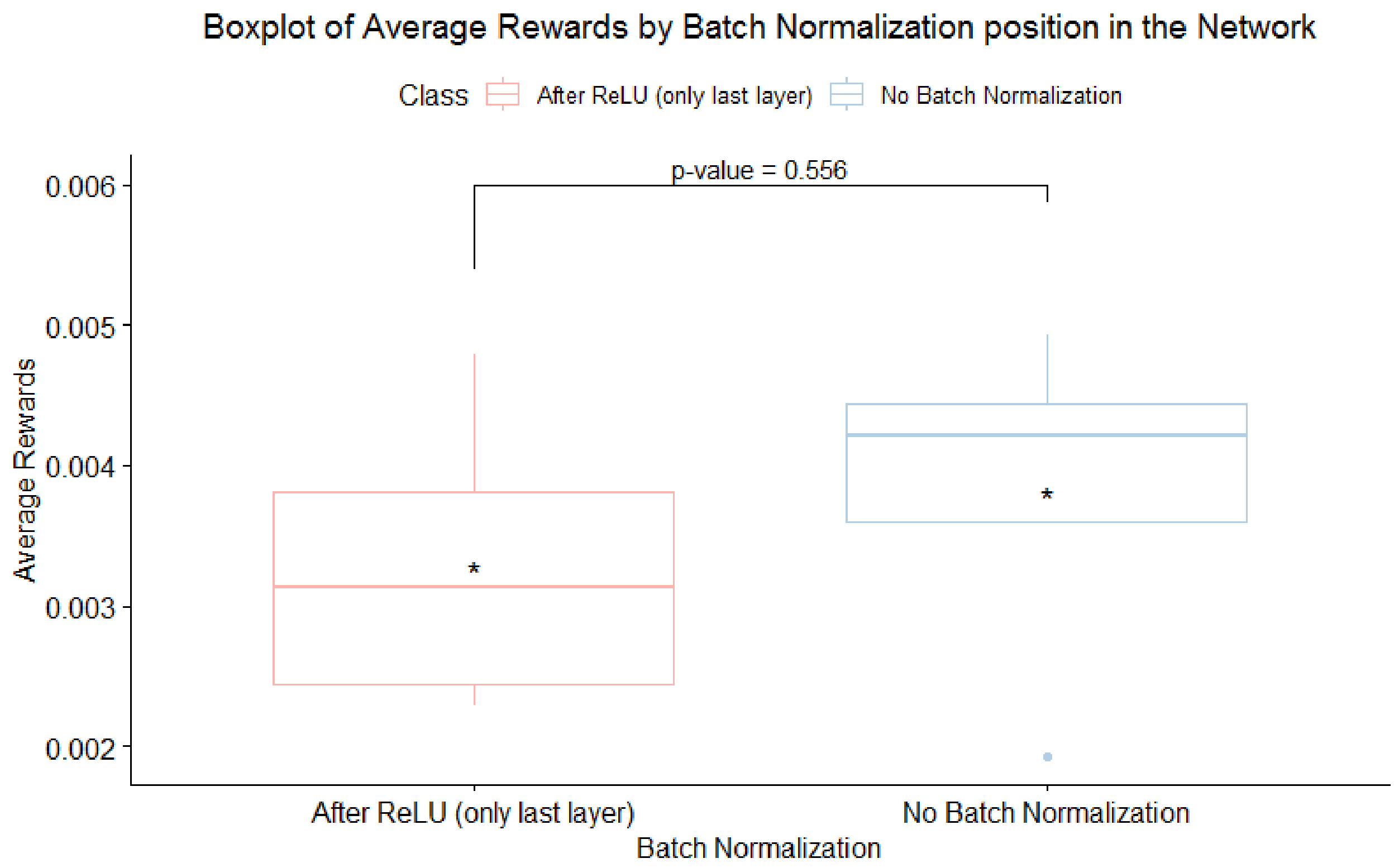

**Figure 53: Architecture 4 vs basic architecture**

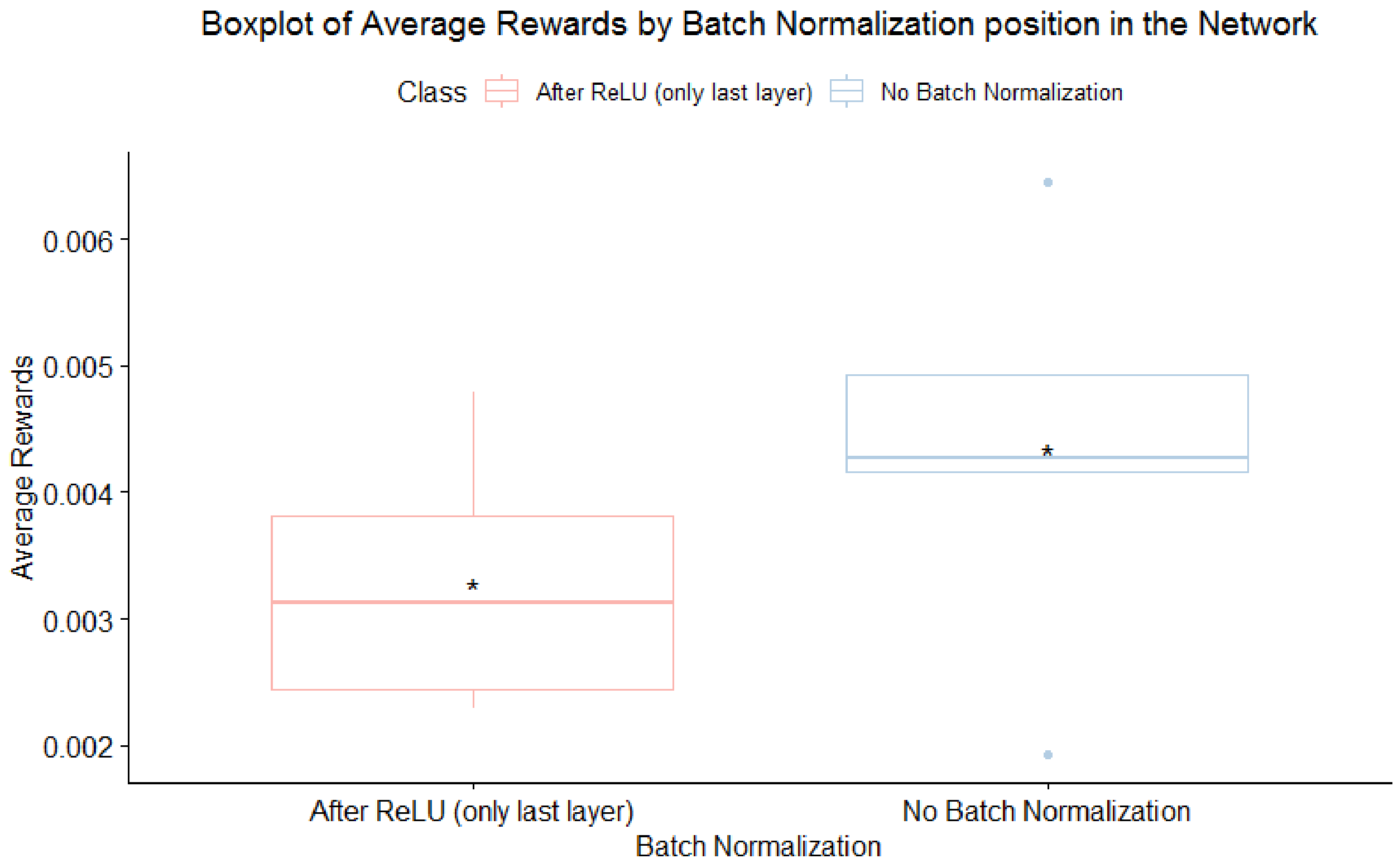

**Figure 54: Architecture 3 vs basic architecture**

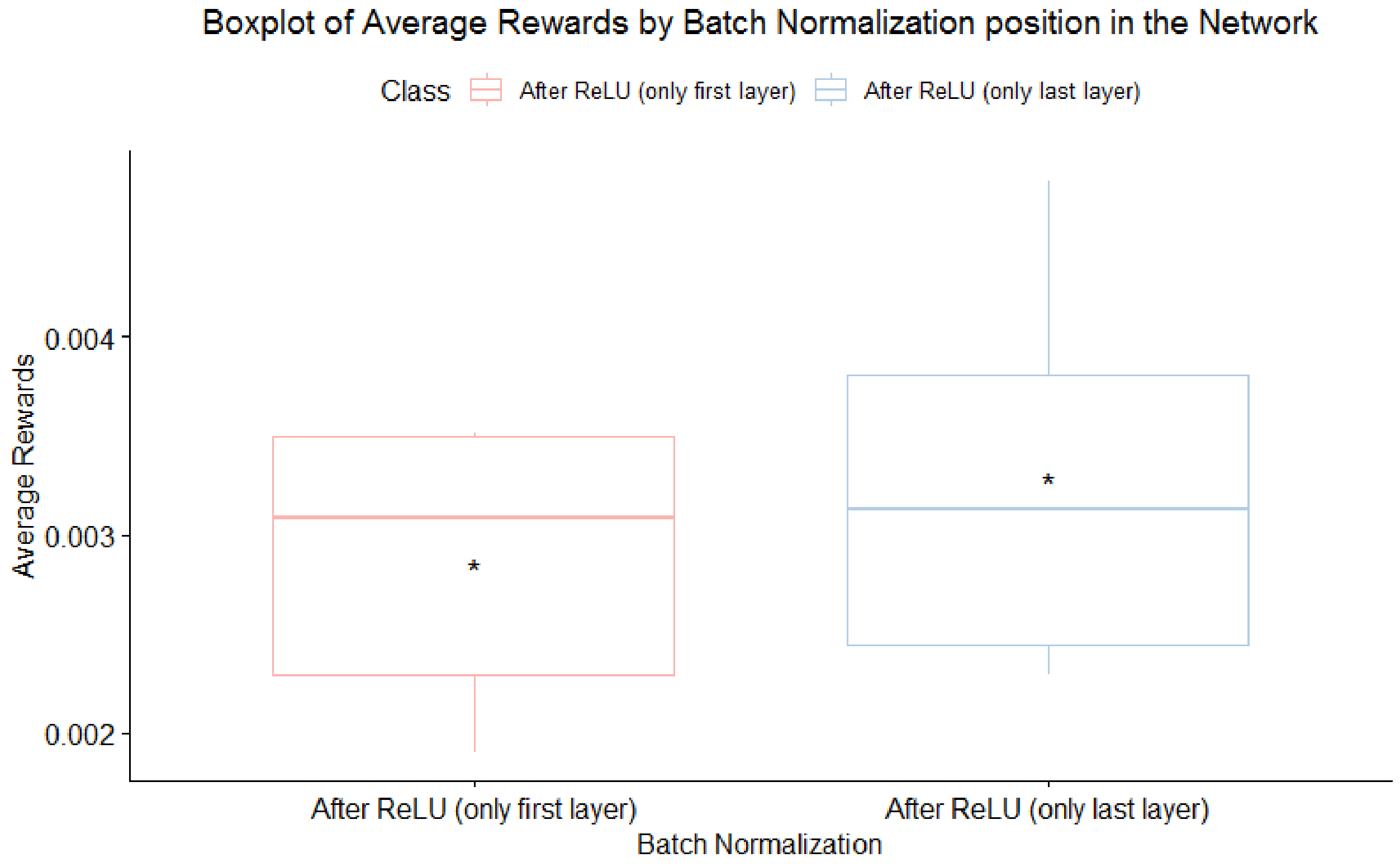

**Figure 55: Architecture 3 vs Architecture 4**

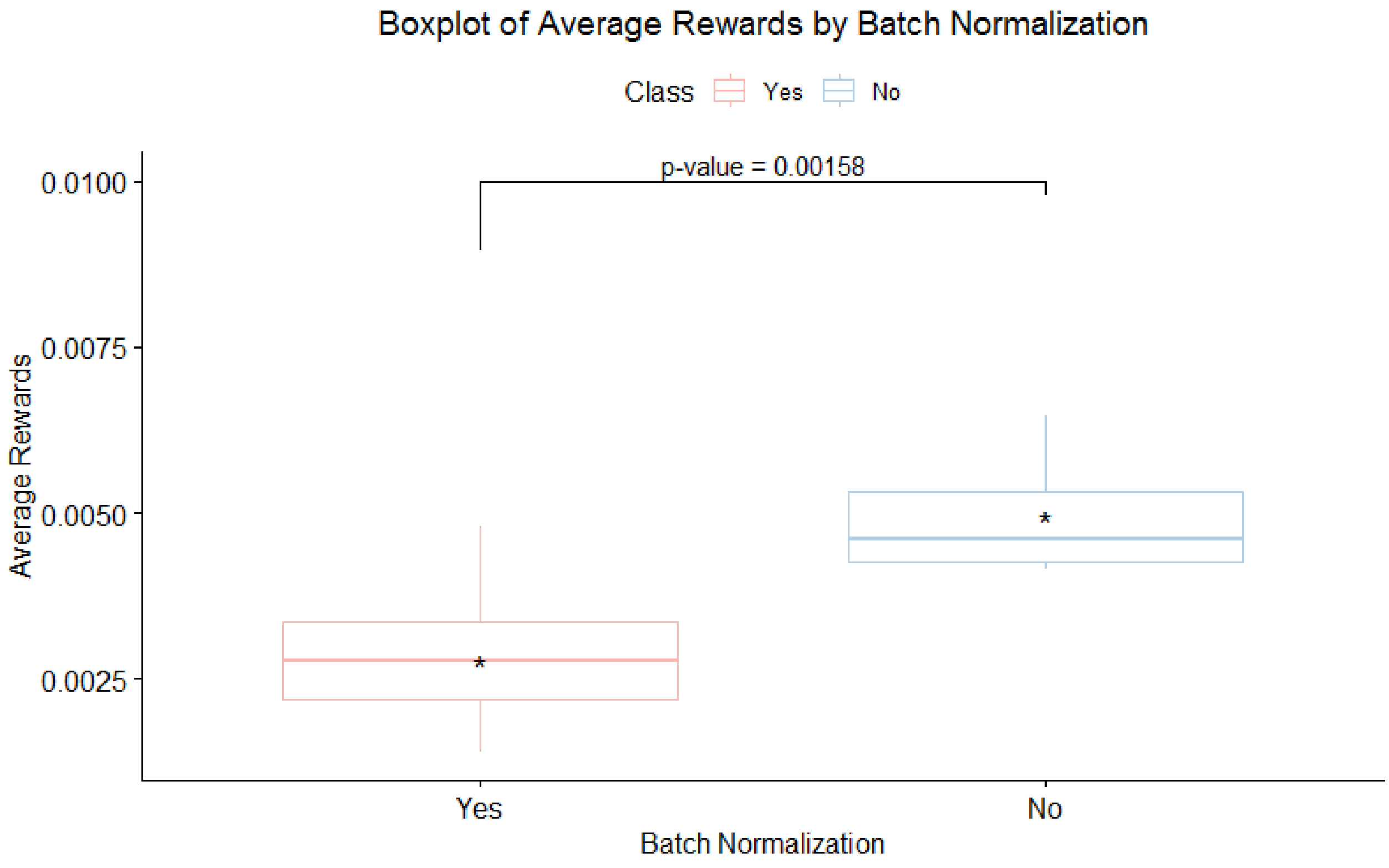

**Figure 56: Batch normalization vs Basic model**

### x) MaxPooling2D

Max pooling is a down sampling strategy in CNN. For each of the regions represented by the filter, we will take the max of that region and create a new, output matrix where each element is the max of a region in the original input. No significant performance differences were observed when including max pooling layers.

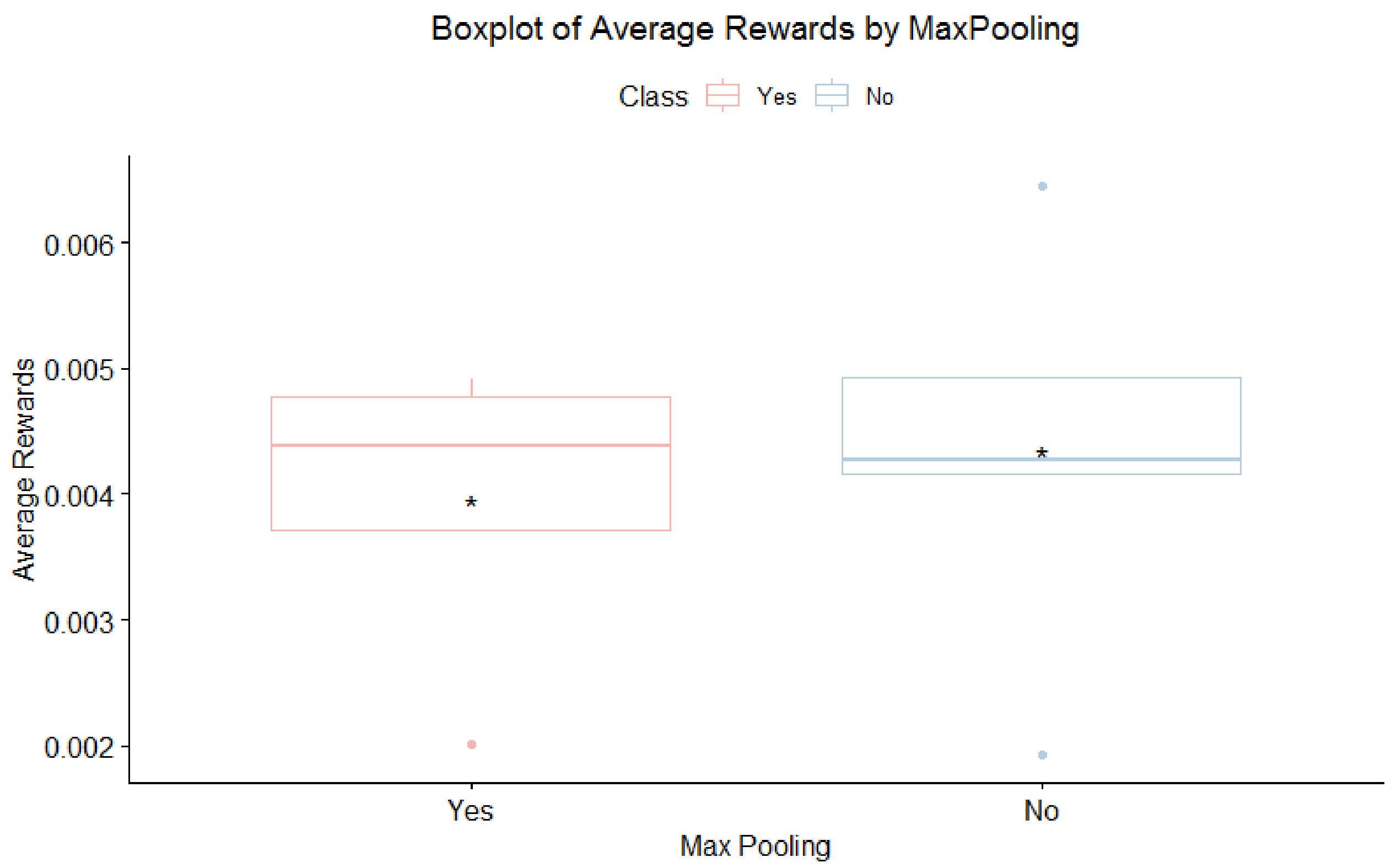

**Figure 57: MaxPooling vs Basic model**

### y) Log-polar transformation

The log-polar transformation is described in preprocessing (section 3.2.1). We trained 5 replicates with the log-polar preprocessing step and other 5 without this. During the first 30000 iterations, the log-polar replicates had worse results than those without the use of log-polar transformation. However, we retrained our networks for other 30000 iterations and log-polar had better results (average 0.00622) than non-log polar (0.00524) results with more training steps. Remarkably, when we tested those 2 sets of implementations using the test set as input and logpolar implementation had average positive rewards 0.0107 in 5 replicates and non-logpolar had average positive rewards 0.005568. This indicates that log-polar transformed input has better generalization performance than the unprocessed image, verifying our assumption that using a representation invariant to rotations and scalings is beneficial.

## 4.2 Optimized Model

After the previous tests and studying the hyperparameters to optimize them, we decided to use the following model parameters and architecture for an extended training:

| | | | |
|---|---|---|---|
| CLD1 | 30 | Replay Memory Size | 8000 |
| RFS1 | 15 | Learning Rate | 1,00E-04 |
| S1 | 1 | Momentum | 0.95 |
| CLD2 | 15 | Decay | 2,00E-06 |
| RFS2 | 20 | Dropout Rate | 0 |
| S2 | 1 | Input Image Size | 160 |
| ND | 200 | Score Scaling | 0.175 |
| Gamma | 0.95 | Random seed | Current time |
| Observation | 100 | Mini Batch Size | 50 |
| Final Exploration Frame | 8000 | Actions per Training | 10 |
| Initial Epsilon | 0.25 | Maximum Running Steps | 29999 |
| Final Epsilon | 0.05 | Log polar | Yes |

### 4.2.1 Model Architecture

The architecture of the model and the activations of the first two layers are shown in Figure 58.

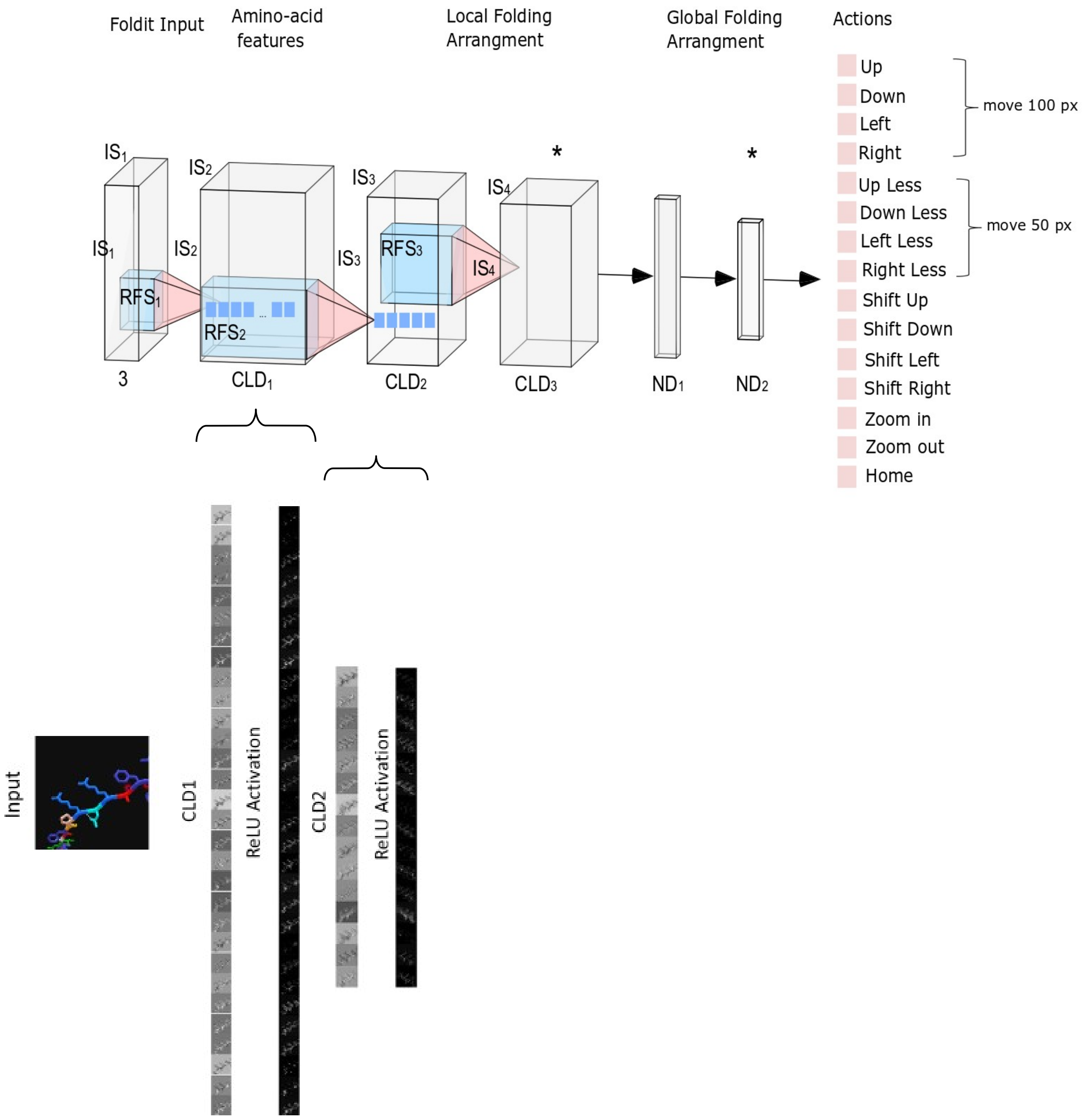

**Figure 58: Model architecture and layers' structure**

### 4.2.2 Model Convolved features

Each convolved feature in Figure 60, Figure 61 results by applying a filter to the original image. Some filters seem to identify colors, others focus on border detection. In Figure 62 the detailed activation maps are shown for the first two layers.

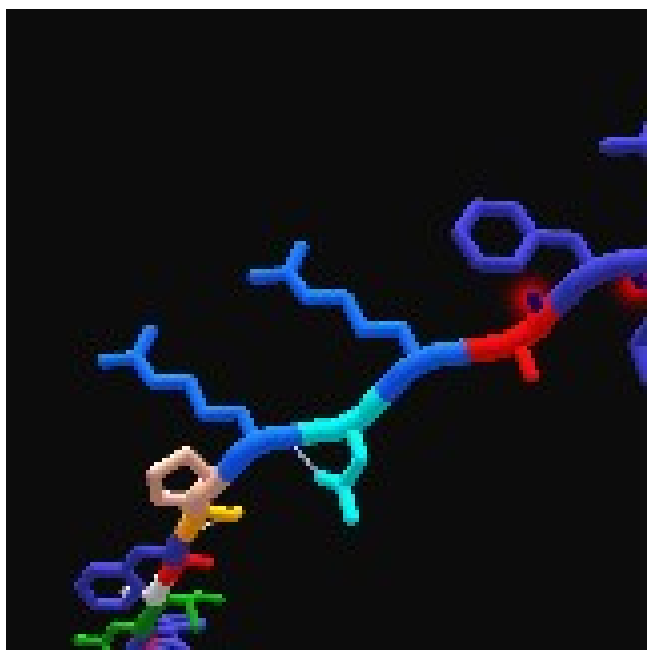

**Figure 59: Input Image**

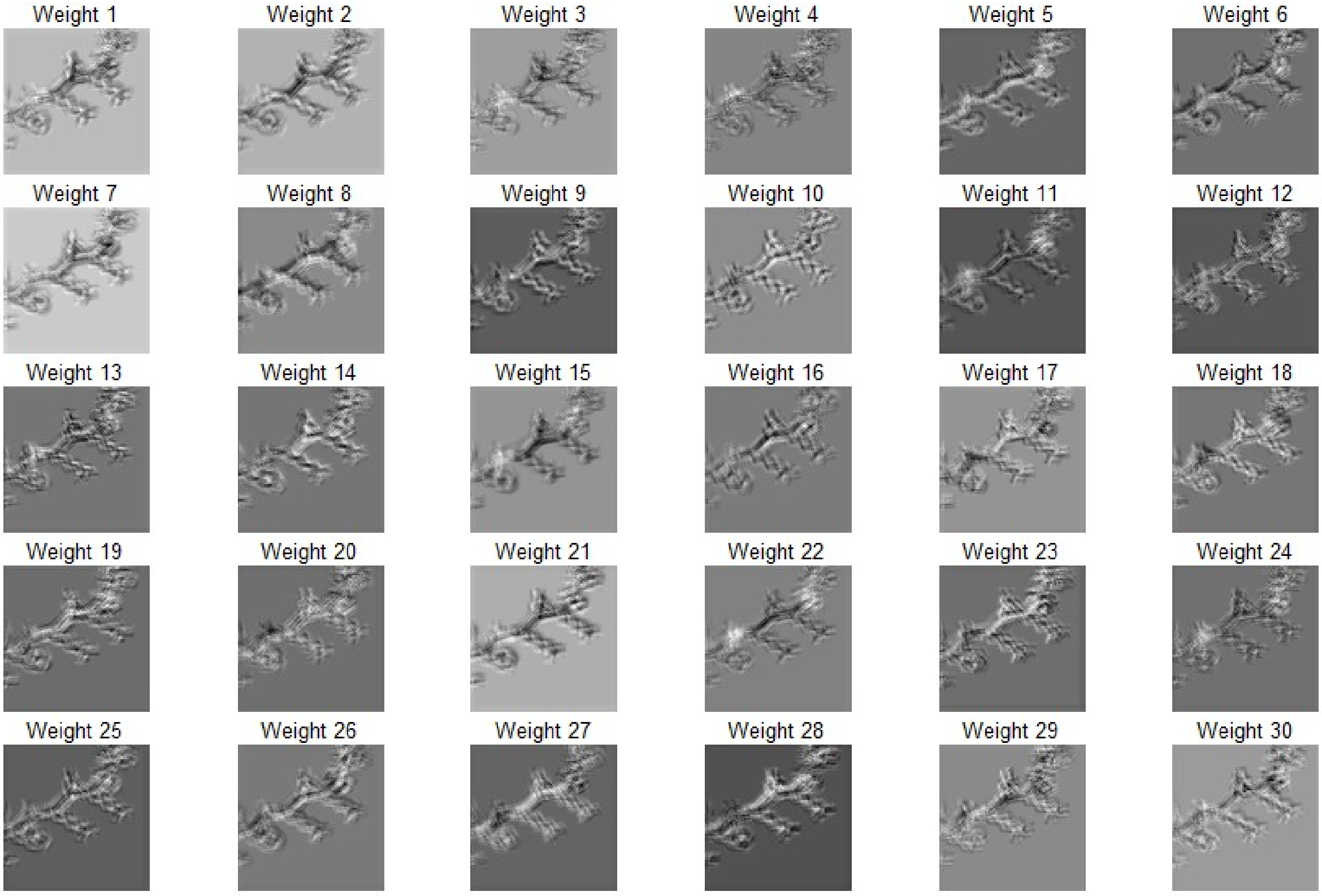

**Figure 60: CLD1 convolved features of optimized model**

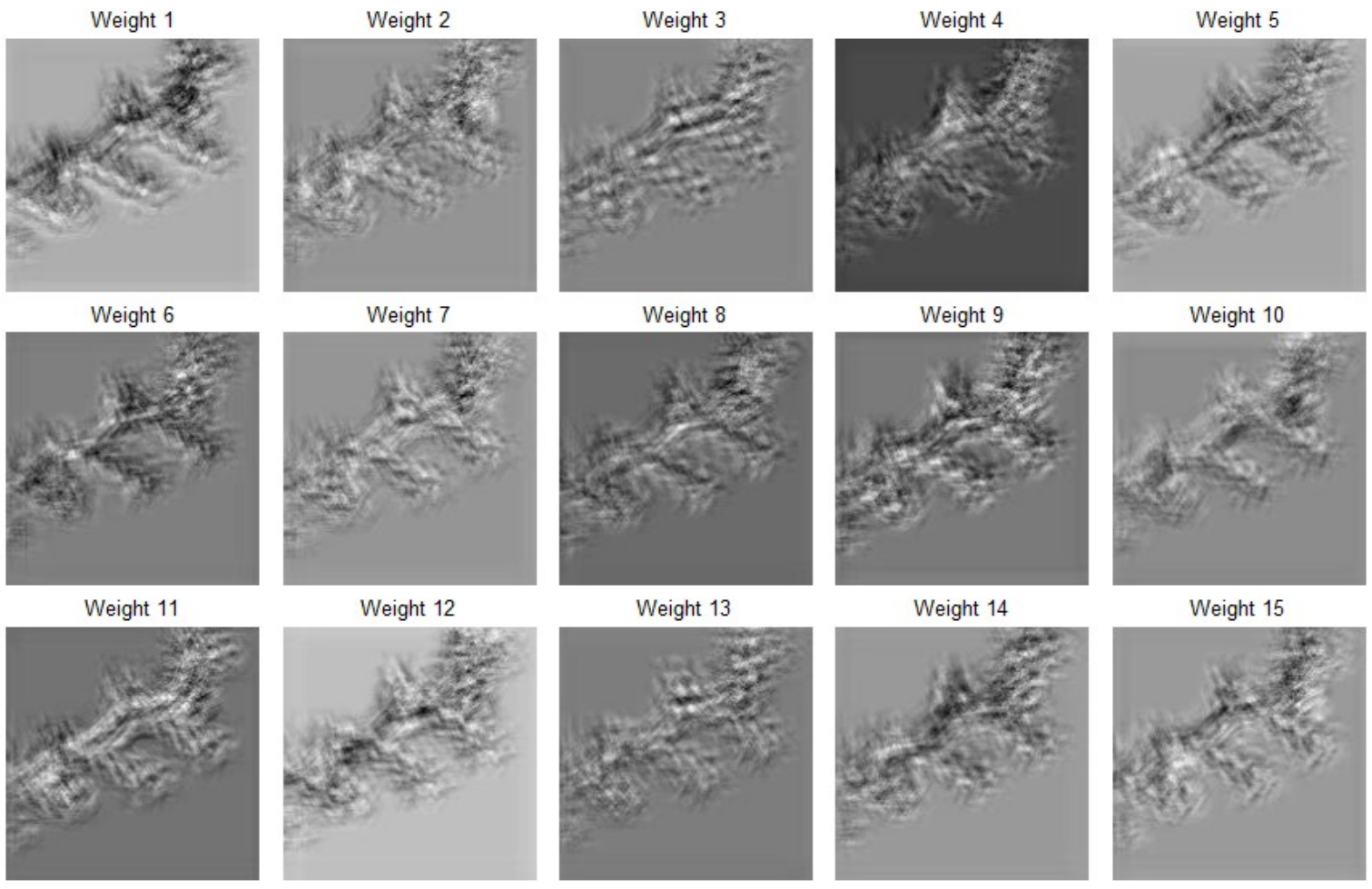

**Figure 61: CLD2 convolved features of optimized model**

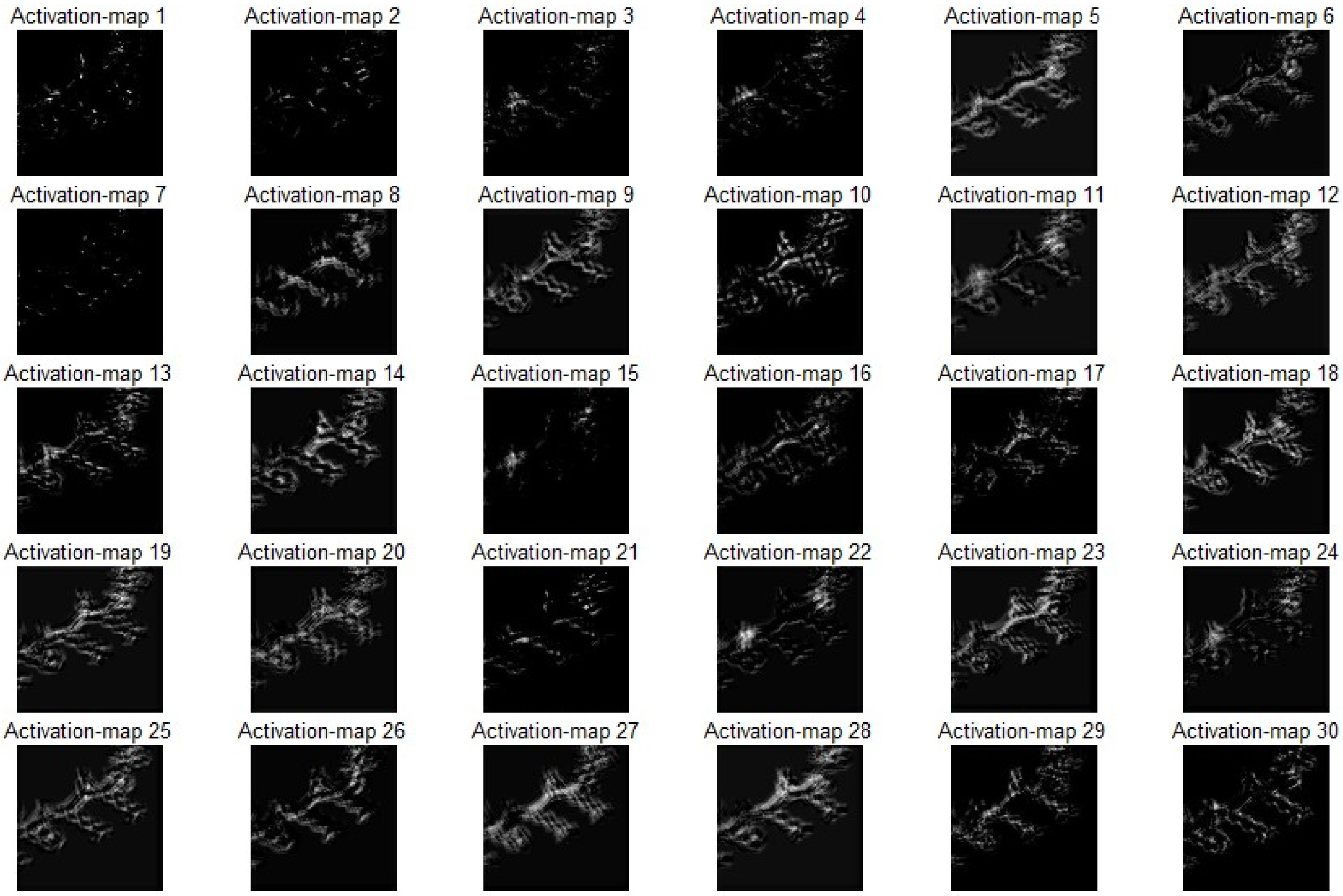

**Figure 62: CLD1 Activation maps**

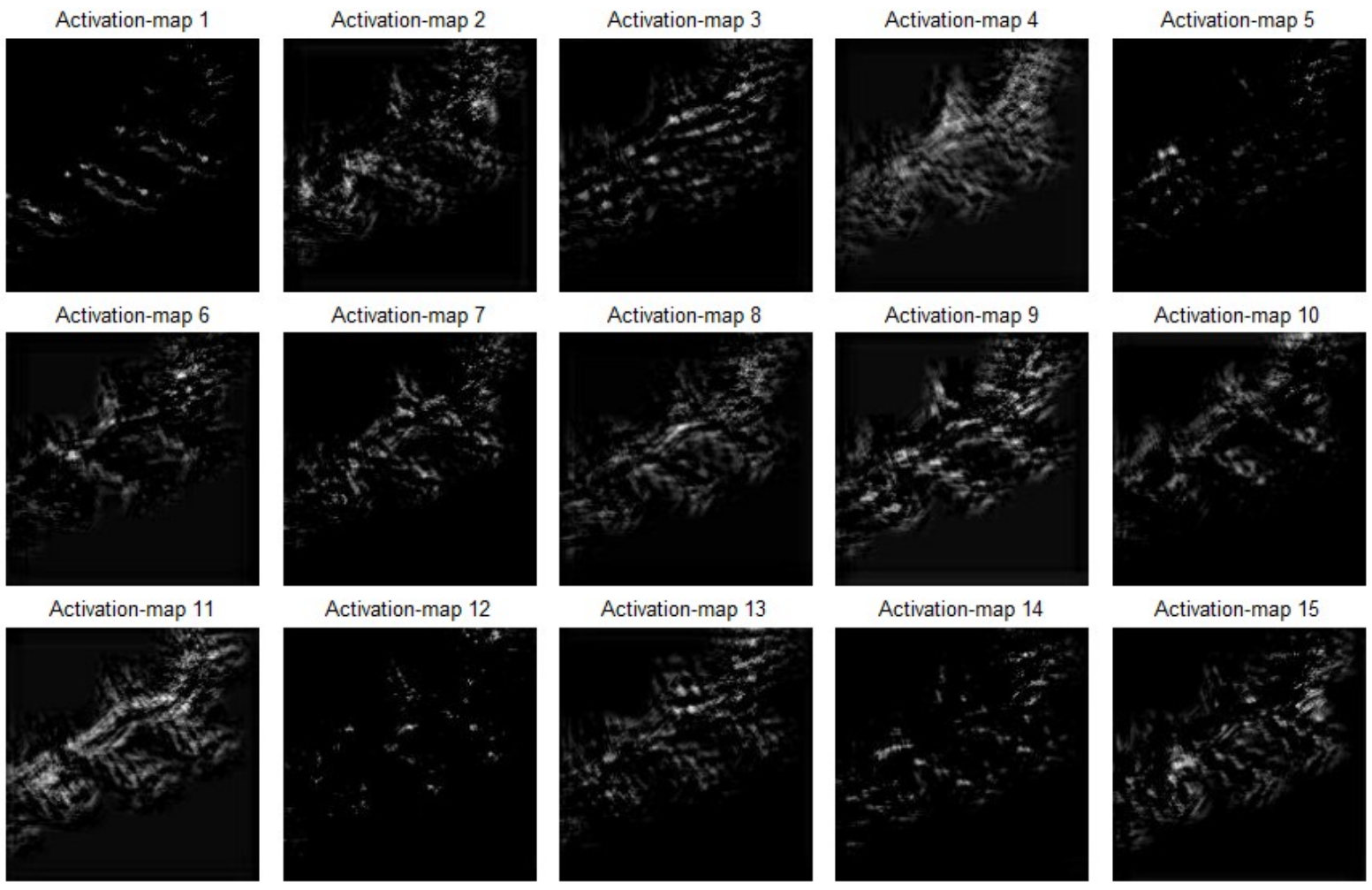

**Figure 63: CLD2 Activation maps**

### 4.2.3 Test set results

We trained a model for 220000 timesteps for the Soloist data (see section 3.1). The input to the neural network is a 160x160x3 RGB image sampled by 3 rectifier hidden layers convolving 1 1x1 filter with stride 1 (for adaptive RGB to gray conversion), 40 20x20 filters with stride 2 and 5 40x40 filters with stride 8. The final hidden layer with 256 fully-connected rectifier units is connected to the output layer of 15 actions. The behavior policy during training was ε-greedy with annealed linearly from 0.8 to 0.2 over the first 8000 frames, and fixed at 0.2 thereafter. We trained for a total of 30000 frames with a learning rate of 1e−4, Nestrov's update rule, a momentum of 0.9 and a replay memory containing the 10000 most recent frames. All of these CNNs use a RELU activation function in all layers except for the output layer where a linear activation function is used.

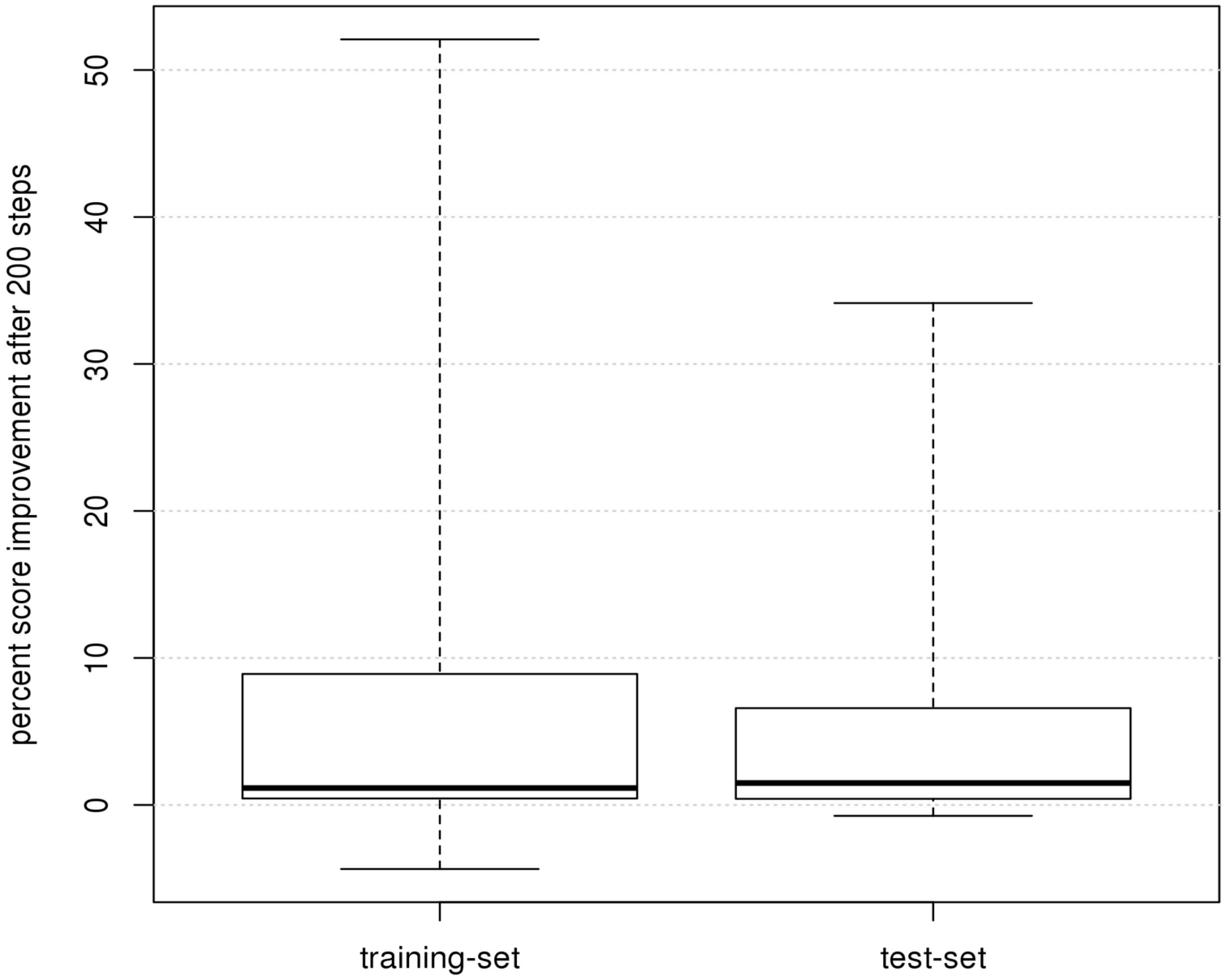

**Figure 64: Relative score improvements**

In Figure 64, the relative score improvements are shown both for the training and test sets (as defined in Table 3) after 220000 training steps. The score improvement is 5.8% on the training set and 4.5% on the test set, indicating that the trained action sequences can generalize to novel proteins with unrelated, novel sequences.

### 4.2.5. Features learned by the model

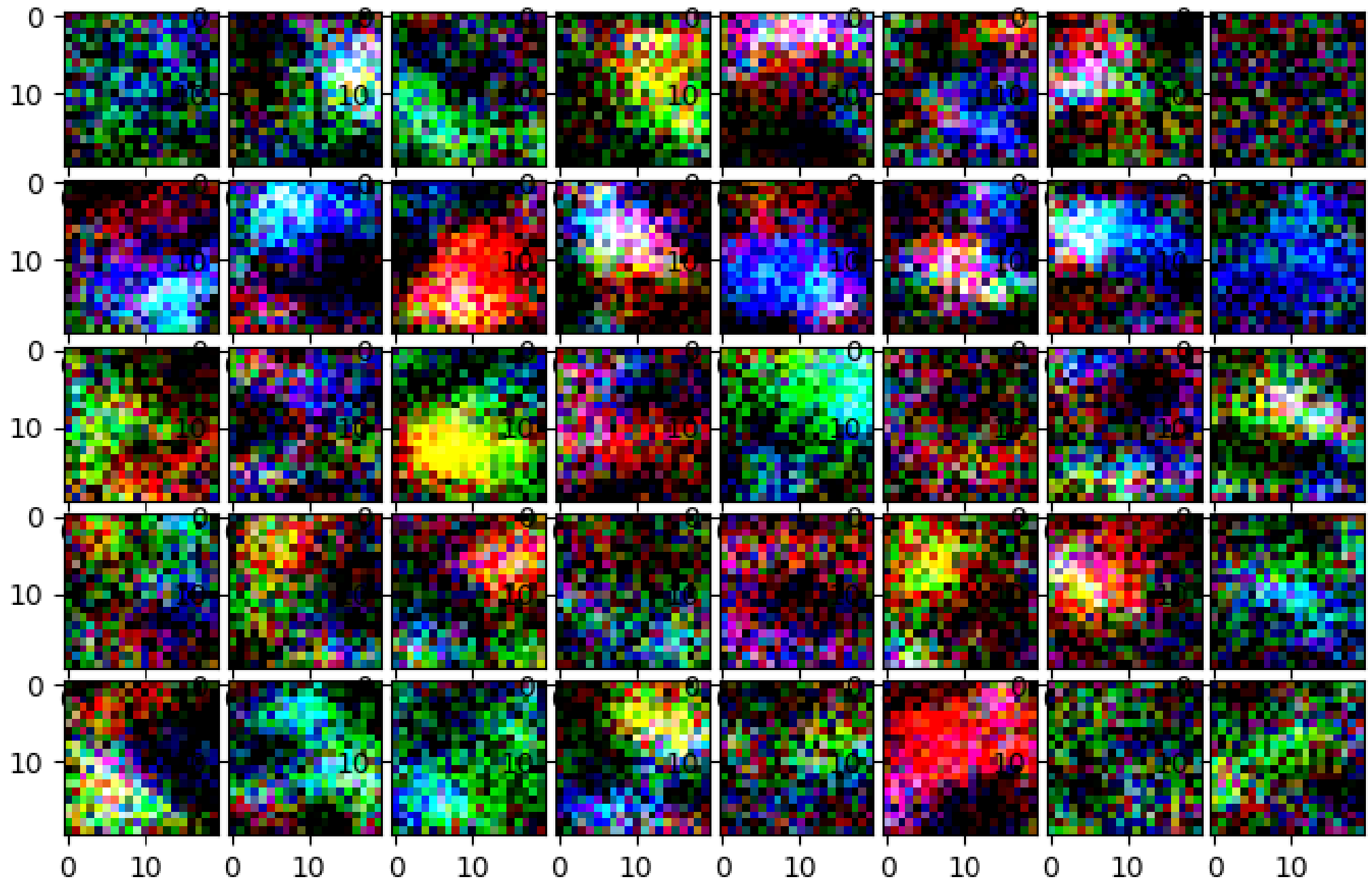

Figure 65: Amino acid feature layer of Evolver dataset

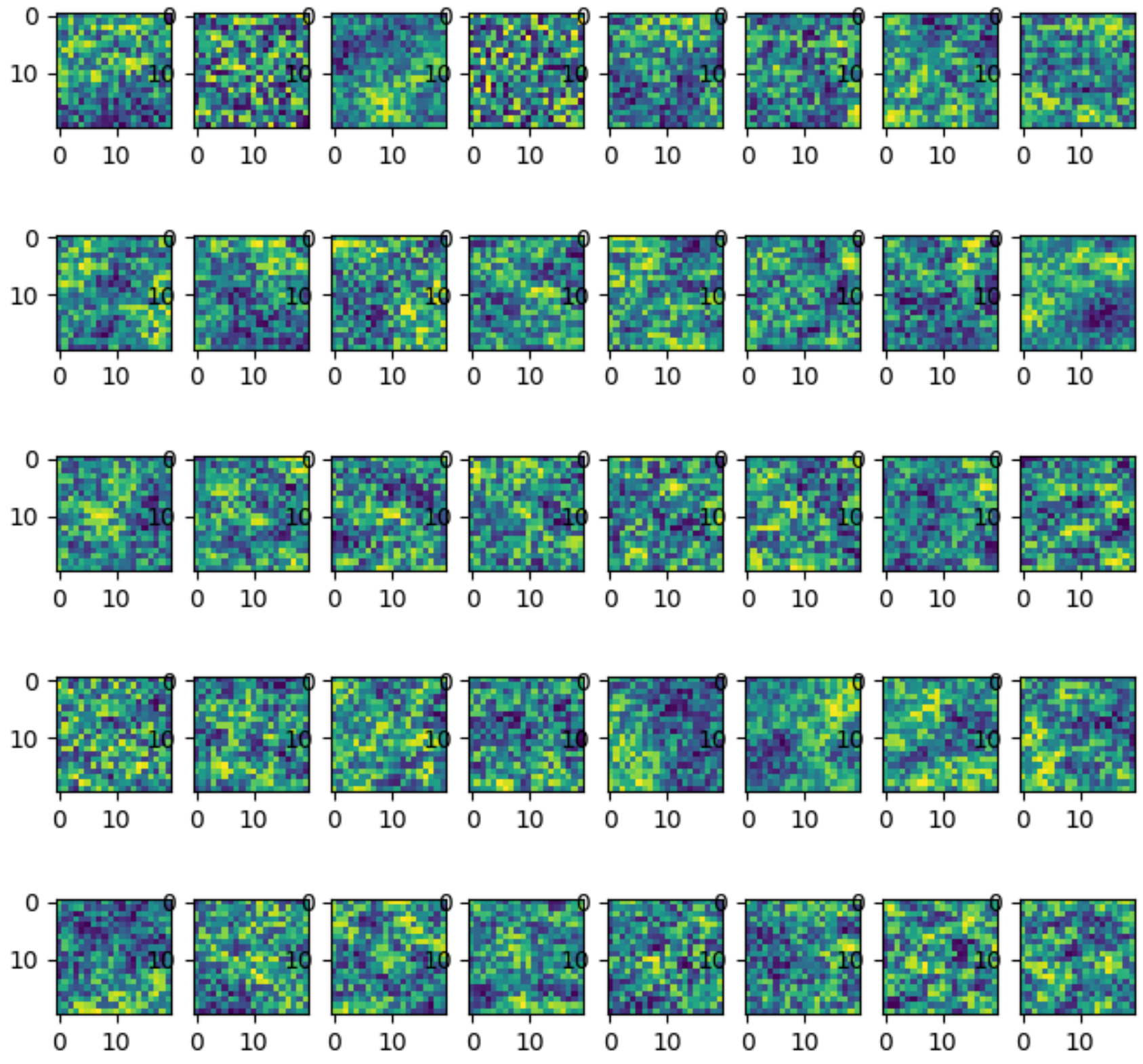

Figure 66: Amino acid feature layer of Soloist dataset

For the Evolver dataset (see section 3.1) we trained a model with 160x160x3 RGB image input sampled by 40 20x20 filters with stride 2 followed by 20 40x40 filters with stride 4. The final hidden layer with 128 fully-connected rectifier units is connected to the output layer of 15 actions. The behavior policy during training was ε-greedy with annealed linearly from 0.8 to 0.2 over the first 8000 frames, and fixed at 0.2 thereafter. We trained for a total of 39200 frames with a learning rate of 2e−4, Nestrov's update rule, a momentum of 0.9 and a replay memory containing the 15000 most recent frames. All layers use a RELU activation function except for the output layer where the softmax activation function is used. In Figure 65 and 66, the trained weights of the amino-acid feature layer are visualized for deepFoldIt-Evolver and Soloist, respectively. As the features of the Evolver net are directly connected to the RGB image, is evident that the net has developed feature detectors specific to groups of amino acids that exploit the amino acid color code (see Figure 18) used by Foldit.

# 5. CONCLUSIONS & FUTURE WORK

We were able to successfully implement a deep reinforcement learning framework DeepFoldit to start playing the protein game Foldit. Though the results did not yet show superhuman performance, initial results are promising. The novel test cases show an improvement of the structure within the limit of 200 moves.

There were some cases where the combination of moves placed the molecule in a position, that occupied the entire window and no actions could be applied. In some cases, this ended up in a locked situation with unlikely recovery moves. We tackled this problem with continuous zoom out, until the agent can perform actions during the training. If this "Zoom out" sequence fails, we abort on this protein and load the next from the training set. The folding process could be made more robust.

The manipulation of the protein could also be made more flexible using a reinforcement learning algorithm that supports continuous actions, avoiding the limitations of the fixed discrete action set defined by us.

Our hyperparameter optimization provides useful settings for training the next architectures. We implemented the first parallelization structure shown in Figure 23. The second structure (Figure 24) would boost the number of training iterations per second. The number of training steps that are required for other similar architectures, in order to produce interesting results was reported with 10 million timesteps [152]. As the longest run in our current implementation performed 1.3 million timesteps this is an obvious next step.

Overall, our current results show that given a set of small unfolded training proteins, DeepFoldit learns action sequences that improve the score both on the training-set and on novel test proteins. Our approach combines the intuitive user interface of Foldit with the power of deep reinforcement learning.

## AKNOWLEDGMENTS

This work was supported by computational time granted from the National Infrastructures for Research and Technology S.A. (GRNET S.A.) in the National HPC facility - ARIS - under project ID pa181001.

We would like to thank to Elias Manolakos, National and Kapodistrian University of Athens Manolakos and Stavros Perantonis, National Center for Scientific Research "Demokritos" for comments on the manuscript.

## REFERENCES

1. Cooper, S., et al., *Predicting protein structures with a multiplayer online game.* Nature, 2010. **466**(7307): p. 756.
2. Horowitz, S., et al., *Determining crystal structures through crowdsourcing and coursework.* Nature communications, 2016. **7**: p. 12549.
3. Khoury, G.A., et al., *WeFold: a coopetition for protein structure prediction.* Proteins: Structure, Function, and Bioinformatics, 2014. **82**(9): p. 1850-1868.
4. Kleffner, R., et al., *Foldit Standalone: a video game-derived protein structure manipulation interface using Rosetta.* Bioinformatics, 2017. **33**(17): p. 2765-2767.
5. Silver, D., et al., *Mastering the game of Go with deep neural networks and tree search.* nature, 2016. **529**(7587): p. 484.
6. Yuan, G.-X., C.-H. Ho, and C.-J. Lin, *Recent advances of large-scale linear classification.* Proceedings of the IEEE, 2012. **100**(9): p. 2584-2603.
7. Heckerman, D., *A tutorial on learning with Bayesian networks*, in *Innovations in Bayesian networks*. 2008, Springer. p. 33-82.
8. Cover, T. and P. Hart, *Nearest neighbor pattern classification.* IEEE transactions on information theory, 1967. **13**(1): p. 21-27.
9. Rabiner, L.R. and B.-H. Juang, *An introduction to hidden Markov models.* ieee assp magazine, 1986. **3**(1): p. 4-16.
10. Kohavi, R. and J.R. Quinlan. *Data mining tasks and methods: Classification: decision-tree discovery*. in *Handbook of data mining and knowledge discovery*. 2002. Oxford University Press, Inc.
11. Hinton, G.E., *Connectionist learning procedures*, in *Machine learning*. 1990, Elsevier. p. 555-610.
12. Dempster, A.P., N.M. Laird, and D.B. Rubin, *Maximum likelihood from incomplete data via the EM algorithm.* Journal of the Royal Statistical Society: Series B (Methodological), 1977. **39**(1): p. 1-22.
13. Kohonen, T., *Self-organized formation of topologically correct feature maps.* Biological cybernetics, 1982. **43**(1): p. 59-69.
14. Ball, G.H. and D.J. Hall, *ISODATA, a novel method of data analysis and pattern classification*. 1965, Stanford research inst Menlo Park CA.
15. Dunn, J.C., *A fuzzy relative of the ISODATA process and its use in detecting compact well-separated clusters.* 1973.
16. Hartigan, J.A., *Clustering algorithms.* 1975.
17. Sutton, R.S. and A.G. Barto, *Reinforcement learning: An introduction*. 2018: MIT press.
18. Mikolov, T., et al. *Recurrent neural network based language model*. in *Eleventh annual conference of the international speech communication association*. 2010.

19. Krizhevsky, A., I. Sutskever, and G.E. Hinton. *Imagenet classification with deep convolutional neural networks*. in *Advances in neural information processing systems*. 2012.

20. Hochreiter, S. and J. Schmidhuber, *Long short-term memory.* Neural computation, 1997. **9**(8): p. 1735-1780.

21. Chung, J., et al., *Empirical evaluation of gated recurrent neural networks on sequence modeling.* arXiv preprint arXiv:1412.3555, 2014.

22. Vincent, P., et al., *Stacked denoising autoencoders: Learning useful representations in a deep network with a local denoising criterion.* Journal of machine learning research, 2010. **11**(Dec): p. 3371-3408.

23. Lee, H., et al. *Convolutional deep belief networks for scalable unsupervised learning of hierarchical representations*. in *Proceedings of the 26th annual international conference on machine learning*. 2009. ACM.

24. Snoek, J., H. Larochelle, and R.P. Adams. *Practical bayesian optimization of machine learning algorithms*. in *Advances in neural information processing systems*. 2012.

25. Bergstra, J.S., et al. *Algorithms for hyper-parameter optimization*. in *Advances in neural information processing systems*. 2011.

26. Franceschi, L., et al. *Forward and reverse gradient-based hyperparameter optimization*. in *Proceedings of the 34th International Conference on Machine Learning-Volume 70*. 2017. JMLR. org.

27. Li, L., et al., *Hyperband: A novel bandit-based approach to hyperparameter optimization.* arXiv preprint arXiv:1603.06560, 2016.

28. McDonald, C. *Machine learning fundamentals (II): Neural networks*. Medium 2017; Available from: https://towardsdatascience.com/machine-learning-fundamentals-ii-neural-networks-f1e7b2cb3eef.

29. Herculano-Houzel, S., *The human brain in numbers: a linearly scaled-up primate brain.* Frontiers in human neuroscience, 2009. **3**: p. 31.

30. Course231. *Convolutional Neural Networks (CNNs / ConvNets)*. Available from: http://cs231n.github.io/neural-networks-1/.

31. Rosenblatt, F., *The perceptron: a probabilistic model for information storage and organization in the brain.* Psychological review, 1958. **65**(6): p. 386.

32. Sibi, P., S.A. Jones, and P. Siddarth, *Analysis of different activation functions using back propagation neural networks.* Journal of Theoretical and Applied Information Technology, 2013. **47**(3): p. 1264-1268.

33. Lau, M.M. and K.H. Lim. *Investigation of activation functions in deep belief network*. in *2017 2nd international conference on control and robotics engineering (ICCRE)*. 2017. IEEE.

34. Geva, *7 Types of Neural Network Activation Functions: How to Choose?* MissingLink.ai: MissingLink.ai.

35. Zhu, X. and A.B. Goldberg, *Introduction to semi-supervised learning.* Synthesis lectures on artificial intelligence and machine learning, 2009. **3**(1): p. 1-130.

36. Masters, D. and C. Luschi, *Revisiting small batch training for deep neural networks.* arXiv preprint arXiv:1804.07612, 2018.

37. Ge, R., et al. *Escaping from saddle points—online stochastic gradient for tensor decomposition*. in *Conference on Learning Theory*. 2015.

38. Srivastava, N., et al., *Dropout: a simple way to prevent neural networks from overfitting.* The journal of machine learning research, 2014. **15**(1): p. 1929-1958.

39. Hubel, D.H. and T.N. Wiesel, *Receptive fields of single neurones in the cat's striate cortex.* The Journal of physiology, 1959. **148**(3): p. 574-591.

40. Hubel, D.H. and T.N. Wiesel, *Receptive fields and functional architecture of monkey striate cortex.* The Journal of physiology, 1968. **195**(1): p. 215-243.

41. Hubel, D.H. and T. Wiesel, *Shape and arrangement of columns in cat's striate cortex.* The Journal of physiology, 1963. **165**(3): p. 559-568.

42. Medium, M.P., *A Brief Guide to Convolutional Neural Network(CNN).* 2019.

43. *Convolutional Neural Networks (CNNs / ConvNets)*. Available from: http://cs231n.github.io/convolutional-networks/.

44. Medium, P. *Understanding of Convolutional Neural Network (CNN) — Deep Learning*. Medium 2018; Available from: https://medium.com/@RaghavPrabhu/understanding-of-convolutional-neural-network-cnn-deep-learning-99760835f148.

45. Hochreiter, S., et al., *Gradient flow in recurrent nets: the difficulty of learning long-term dependencies*. 2001, A field guide to dynamical recurrent neural networks. IEEE Press.

46. Metropolis, N. and S. Ulam, *The monte carlo method.* Journal of the American statistical association, 1949. **44**(247): p. 335-341.

47. Silver, D., et al., *Mastering the game of go without human knowledge.* Nature, 2017. **550**(7676): p. 354.

48. Hassabis, D. and D. Silver, *Alphago zero: Learning from scratch.* deepMind official website, 2017. **18**.

49. Watkins, C.J.C.H., *Learning from delayed rewards.* 1989.

50. Melo, F.S., *Convergence of Q-learning: A simple proof.* Institute Of Systems and Robotics, Tech. Rep, 2001: p. 1-4.

51. Matiisen, T. *Demystifying Deep Reinforcement Learning*. 2015; Available from: https://www.intel.ai/demystifying-deep-reinforcement-learning/#_ftn1.

52. Lin, L.J. *Programming Robots Using Reinforcement Learning and Teaching*. in *AAAI*. 1991.

53. Joseph, R.D., *Contributions to perceptron theory*. 1961: Cornell Univ.

54. Ivakhnenko, A.G.e. and V.G. Lapa, *Cybernetic predicting devices*. 1966, PURDUE UNIV LAFAYETTE IND SCHOOL OF ELECTRICAL ENGINEERING.

55. Ivakhnenko, A.G., *The Group Method of Data of Handling; A rival of the method of stochastic approximation.* Soviet Automatic Control, 1968. **13**: p. 43-55.

56. Ma, J., et al., *Deep neural nets as a method for quantitative structure–activity relationships.* Journal of chemical information and modeling, 2015. **55**(2): p. 263-274.

57. Kadurin, A., et al., *The cornucopia of meaningful leads: Applying deep adversarial autoencoders for new molecule development in oncology.* Oncotarget, 2017. **8**(7): p. 10883.

58. Leung, M.K., et al., *Deep learning of the tissue-regulated splicing code.* Bioinformatics, 2014. **30**(12): p. i121-i129.

59. Xiong, H.Y., et al., *The human splicing code reveals new insights into the genetic determinants of disease.* Science, 2015. **347**(6218): p. 1254806.

60. Zhavoronkov, A., et al., *Deep learning enables rapid identification of potent DDR1 kinase inhibitors.* Nature biotechnology, 2019. **37**(9): p. 1038-1040.

61. Gupta, A., H. Wang, and M. Ganapathiraju. *Learning structure in gene expression data using deep architectures, with an application to gene clustering*. in *2015 IEEE International Conference on Bioinformatics and Biomedicine (BIBM)*. 2015. IEEE.

62. Chen, L., et al. *Learning a hierarchical representation of the yeast transcriptomic machinery using an autoencoder model*. in *BMC bioinformatics*. 2016. BioMed Central.

63. Lee, B., et al., *DNA-level splice junction prediction using deep recurrent neural networks.* arXiv preprint arXiv:1512.05135, 2015.

64. Rives, A., et al., *Biological structure and function emerge from scaling unsupervised learning to 250 million protein sequences.* bioRxiv, 2019: p. 622803.

65. Zhou, J. and O.G. Troyanskaya, *Predicting effects of noncoding variants with deep learning–based sequence model.* Nature methods, 2015. **12**(10): p. 931.

66. Alipanahi, B., et al., *Predicting the sequence specificities of DNA-and RNA-binding proteins by deep learning.* Nature biotechnology, 2015. **33**(8): p. 831.

67. Hassanzadeh, H.R. and M.D. Wang. *DeeperBind: Enhancing prediction of sequence specificities of DNA binding proteins*. in *2016 IEEE International Conference on Bioinformatics and Biomedicine (BIBM)*. 2016. IEEE.

68. Lanchantin, J., et al., *Deep motif: Visualizing genomic sequence classifications.* arXiv preprint arXiv:1605.01133, 2016.

69. Tripathi, R., et al., *DeepLNC, a long non-coding RNA prediction tool using deep neural network.* Network Modeling Analysis in Health Informatics and Bioinformatics, 2016. **5**(1): p. 21.

70. Hill, S.T., et al., *A deep recurrent neural network discovers complex biological rules to decipher RNA protein-coding potential.* Nucleic acids research, 2018. **46**(16): p. 8105-8113.

71. Wang, Y., et al., *Predicting DNA methylation state of CpG dinucleotide using genome topological features and deep networks.* Scientific reports, 2016. **6**: p. 19598.

72. Angermueller, C., et al., *DeepCpG: accurate prediction of single-cell DNA methylation states using deep learning.* Genome biology, 2017. **18**(1): p. 67.

73. Lin, C., et al., *Using Neural Networks To Improve Single-Cell RNA-Seq Data Analysis.* bioRxiv, 2017: p. 129759.

74. Arvaniti, E. and M. Claassen, *Sensitive detection of rare disease-associated cell subsets via representation learning.* Nature communications, 2017. **8**: p. 14825.

75. Shaham, U., et al., *Removal of batch effects using distribution-matching residual networks.* Bioinformatics, 2017. **33**(16): p. 2539-2546.

76. Putin, E., et al., *Deep biomarkers of human aging: application of deep neural networks to biomarker development.* Aging (Albany NY), 2016. **8**(5): p. 1021.

77. Zheng, J., et al., *Deep-RBPPred: Predicting RNA binding proteins in the proteome scale based on deep learning.* Scientific reports, 2018. **8**(1): p. 15264.

78. Stadie, B.C., S. Levine, and P. Abbeel, *Incentivizing exploration in reinforcement learning with deep predictive models.* arXiv preprint arXiv:1507.00814, 2015.

79. He, Z.-L. and P.-K. Wong, *Exploration vs. exploitation: An empirical test of the ambidexterity hypothesis.* Organization science, 2004. **15**(4): p. 481-494.

80. Mnih, V., et al., *Human-level control through deep reinforcement learning.* Nature, 2015. **518**(7540): p. 529.

81. Bohg, J., et al., *Interactive perception: Leveraging action in perception and perception in action.* IEEE Transactions on Robotics, 2017. **33**(6): p. 1273-1291.

82. Liu, F., et al. *3DCNN-DQN-RNN: A deep reinforcement learning framework for semantic parsing of large-scale 3D point clouds*. in *Proceedings of the IEEE International Conference on Computer Vision*. 2017.

83. Brunner, G., et al. *Teaching a machine to read maps with deep reinforcement learning*. in *Thirty-Second AAAI Conference on Artificial Intelligence*. 2018.

84. Mestres, A., et al., *Knowledge-defined networking.* ACM SIGCOMM Computer Communication Review, 2017. **47**(3): p. 2-10.

85. Gavrilovska, L., et al., *Learning and reasoning in cognitive radio networks.* IEEE Communications Surveys & Tutorials, 2013. **15**(4): p. 1761-1777.

86. Haykin, S., *Cognitive radio: brain-empowered wireless communications.* IEEE journal on selected areas in communications, 2005. **23**(2): p. 201-220.

87. Kober, J., J.A. Bagnell, and J. Peters, *Reinforcement learning in robotics: A survey.* The International Journal of Robotics Research, 2013. **32**(11): p. 1238-1274.

88. Deisenroth, M.P., G. Neumann, and J. Peters, *A survey on policy search for robotics.* Foundations and Trends® in Robotics, 2013. **2**(1–2): p. 1-142.

89. Abbeel, P. *Deep Learning for Robotics*. [Keynote Speech slides] 2017; 105]. Available from: https://www.dropbox.com/s/fdw7q8mx3x4wr0c/2017_12_xx_NIPS-keynote-final.pdf.

90. Powles, J. and H. Hodson, *Google DeepMind and healthcare in an age of algorithms.* Health and technology, 2017. **7**(4): p. 351-367.

91. Zhou, Z., et al., *Optimization of molecules via deep reinforcement learning.* Scientific reports, 2019. **9**(1): p. 1-10.

92. Esteva, A., et al., *Dermatologist-level classification of skin cancer with deep neural networks.* Nature, 2017. **542**(7639): p. 115.

93. Raghu, A., et al., *Deep reinforcement learning for sepsis treatment.* arXiv preprint arXiv:1711.09602, 2017.

94. Weng, W.-H., et al., *Representation and reinforcement learning for personalized glycemic control in septic patients.* arXiv preprint arXiv:1712.00654, 2017.

95. Yauney, G. and P. Shah. *Reinforcement learning with action-derived rewards for chemotherapy and clinical trial dosing regimen selection*. in *Machine Learning for Healthcare Conference*. 2018.

96. Wang, L., et al. *Supervised reinforcement learning with recurrent neural network for dynamic treatment recommendation*. in *Proceedings of the 24th ACM SIGKDD International Conference on Knowledge Discovery & Data Mining*. 2018. ACM.

97. Liu, Y., et al. *Deep reinforcement learning for dynamic treatment regimes on medical registry data*. in *2017 IEEE International Conference on Healthcare Informatics (ICHI)*. 2017. IEEE.

98. Schmidhuber, J., *Deep learning in neural networks: An overview.* Neural networks, 2015. **61**: p. 85-117.

99. Mnih, V., et al., *Playing atari with deep reinforcement learning.* arXiv preprint arXiv:1312.5602, 2013.

100. Mnih, V., et al. *Asynchronous methods for deep reinforcement learning*. in *International conference on machine learning*. 2016.

101. Campbell, M., A.J. Hoane Jr, and F.-h. Hsu, *Deep blue.* Artificial intelligence, 2002. **134**(1-2): p. 57-83.

102. OpenAI. *OpenAI. Openai five*. 2018; Available from: https://blog.openai.com/openai-five.

103. Lewis, M., et al., *Deal or no deal? end-to-end learning for negotiation dialogues.* arXiv preprint arXiv:1706.05125, 2017.

104. Semenov, A., et al. *Applications of Machine Learning in Dota 2: Literature Review and Practical Knowledge Sharing*. in *MLSA@ PKDD/ECML*. 2016.

105. Schulman, J., et al., *Proximal policy optimization algorithms.* arXiv preprint arXiv:1707.06347, 2017.

106. Brown, N. and T. Sandholm, *Superhuman AI for heads-up no-limit poker: Libratus beats top professionals.* Science, 2018. **359**(6374): p. 418-424.

107. Pac-Man, M., *Ms. Pac-Man Competition* Ms. Pac-Man Competition **Ms. Pac-Man Competition**

108. Williams, P.R., D. Perez-Liebana, and S.M. Lucas. *Ms. pac-man versus ghost team CIG 2016 competition*. in *2016 IEEE Conference on Computational Intelligence and Games (CIG)*. 2016. IEEE.

109. Microsoft, *Project Malmo.*

110. Johnson, M., et al. *The Malmo Platform for Artificial Intelligence Experimentation*. in *IJCAI*. 2016.

111. Brood-War, A.T.G.R. *bwapi*. 2016; Available from: https://github.com/bwapi/bwapi.

112. Vinyals, O., et al., *Starcraft ii: A new challenge for reinforcement learning.* arXiv preprint arXiv:1708.04782, 2017.

113. Juliani, A. *On "solving" Montezuma's Revenge*. 2018; Available from: https://medium.com/@awjuliani/on-solving-montezumas-revenge-2146d83f0bc3.

114. Brockman, G., et al., *Openai gym.* arXiv preprint arXiv:1606.01540, 2016.

115. Pygame. *Pygame*. 2000; Available from: https://www.pygame.org/news.

116. ALE. *Arcade Learning Environment (ALE) Github Repository*. 2013; Available from: https://github.com/mgbellemare/Arcade-Learning-%20Environment.

117. Bellemare, M.G., et al., *The arcade learning environment: An evaluation platform for general agents.* Journal of Artificial Intelligence Research, 2013. **47**: p. 253-279.

118. Oh, J., et al. *Action-conditional video prediction using deep networks in atari games*. in *Advances in neural information processing systems*. 2015.

119. Kempka, M., et al. *Vizdoom: A doom-based ai research platform for visual reinforcement learning*. in *2016 IEEE Conference on Computational Intelligence and Games (CIG)*. 2016. IEEE.

120. Koutník, J., et al. *Evolving large-scale neural networks for vision-based reinforcement learning*. in *Proceedings of the 15th annual conference on Genetic and evolutionary computation*. 2013. ACM.

121. Twitter. *Twitter torch-twrl: an open-sourced framework for RL development*. 2016 [cited 2019; Available from: https://github.com/twitter/torch-twrl.

122. Crescenzi, P., et al., *On the complexity of protein folding.* Journal of computational biology, 1998. **5**(3): p. 423-465.

123. Gupta, C.L., S. Akhtar, and P. Bajpai, *In silico protein modeling: possibilities and limitations.* EXCLI journal, 2014. **13**: p. 513-515.

124. Andersen, N.H., *Protein Structure, Stability, and Folding. Methods in Molecular Biology. Volume 168 Edited by Kenneth P. Murphy (University of Iowa College of Medicine). Humana Press: Totowa, New Jersey. 2001. ix+ 252 pp. $89.50. ISBN 0-89603-682-0*. 2001, ACS Publications.

125. Berman, H., et al., *The protein data Bank nucleic acids research, 28: 235-242.* URL: www. rcsb. org Citation, 2000.

126. Bairoch, A., et al., *The universal protein resource (UniProt).* Nucleic acids research, 2005. **33**(suppl_1): p. D154-D159.

127. Flock, T., et al., *Deciphering membrane protein structures from protein sequences.* Genome biology, 2012. **13**(6): p. 160.

128. Greer, J., *Comparative model-building of the mammalian serine proteases.* Journal of molecular biology, 1981. **153**(4): p. 1027-1042.

129. Šali, A. and T.L. Blundell, *Comparative protein modelling by satisfaction of spatial restraints.* Journal of molecular biology, 1993. **234**(3): p. 779-815.

130. Lee, J., P.L. Freddolino, and Y. Zhang, *Ab initio protein structure prediction*, in *From protein structure to function with bioinformatics*. 2017, Springer. p. 3-35.

131. Fleishman, S.J., et al., *RosettaScripts: a scripting language interface to the Rosetta macromolecular modeling suite.* PloS one, 2011. **6**(6): p. e20161.

132. Brooks, B.R., et al., *CHARMM: a program for macromolecular energy, minimization, and dynamics calculations.* Journal of computational chemistry, 1983. **4**(2): p. 187-217.

133. Salomon-Ferrer, R., D.A. Case, and R.C. Walker, *An overview of the Amber biomolecular simulation package.* Wiley Interdisciplinary Reviews: Computational Molecular Science, 2013. **3**(2): p. 198-210.

134. Van Gunsteren, W., et al., *Groningen molecular simulation (GROMOS) system.* Groningen, The Netherlands: University of Groningen, 1996.

135. Minecraft, *MineCraft.* https://www.minecraft.net/en-us/.

136. Eiben, C.B., et al., *Increased Diels-Alderase activity through backbone remodeling guided by Foldit players.* Nature biotechnology, 2012. **30**(2): p. 190.

137. Khatib, F., et al., *Algorithm discovery by protein folding game players.* Proceedings of the National Academy of Sciences, 2011. **108**(47): p. 18949-18953.

138. Khatib, F., et al., *Crystal structure of a monomeric retroviral protease solved by protein folding game players.* Nature structural & molecular biology, 2011. **18**(10): p. 1175.

139. Foldit. *Foldit Recipes*. 2009; Available from: http://fold.it/portal/recipes.

140. Cookbook. *Cookbook 5.1*. 2016; Available from: http://foldit.wikia.com/wiki/101_-_Cookbook.

141. Lerusalimschy, R., L.H. De Figueiredo, and W.C. Filho, *Lua—an extensible extension language.* Software: Practice and Experience, 1996. **26**(6): p. 635-652.

142. Rohl, C.A., et al., *Protein structure prediction using Rosetta*, in *Methods in enzymology*. 2004, Elsevier. p. 66-93.

143. Colowick, S.P., et al., *Methods in enzymology*. Vol. 1. 1955: Academic press New York.

144. Chen, K., *Deep reinforcement learning for flappy bird*. 2015, Stanford University.

145. Schaul, T., et al., *Prioritized experience replay.* arXiv preprint arXiv:1511.05952, 2015.

146. Sarvaiya, J.N., S. Patnaik, and K. Kothari, *Image registration using log polar transform and phase correlation to recover higher scale.* Journal of pattern recognition research, 2012. **7**(1): p. 90-105.

147. Thoduka, S. *Log-polar transform*. 2017; Available from: https://sthoduka.github.io/imreg_fmt/docs/log-polar-transform/.

148. LeNail, A. *Alexander LeNail*. 2017; Available from: http://alexlenail.me/NN-SVG/AlexNet.html.

149. GRNET. *National Infrastructures for Research and Technology S.A.*; Available from: https://hpc.grnet.gr/supercomputer/.

150. Chollet, F., *Keras*. 2015.

151. O'Brien, P.C. and T.R. Fleming, *A paired Prentice-Wilcoxon test for censored paired data.* Biometrics, 1987: p. 169-180.

152. Bootcamp, D.R. *Deep RL Bootcamp* 2017; Available from: https://sites.google.com/view/deep-rl-bootcamp/lectures

153. Szegedy, C., et al. *Going deeper with convolutions*. in *Proceedings of the IEEE conference on computer vision and pattern recognition*. 2015.

154. Salimans, T. and D.P. Kingma. *Weight normalization: A simple reparameterization to accelerate training of deep neural networks*. in *Advances in Neural Information Processing Systems*. 2016.

# ANNEX I: Hyperparameters

**Table 8**

| Parameter | Description | Default Value | Tested Values | Best |
|---|---|---|---|---|
| Minibatch | Number of training cases over which each stochastic gradient decent update is computed | 32 | We used 2 Minibatches with size [40,50,70] and the other [80,100,170] | 50 |
| Replay Memory | Stochastic gradient decent updates are sampled from this number of recent frames | 10000 | 800, 2000, 6000, 8000, 9000, 10000, 50000 | 2000, 6000 |
| Discount factor or Gamma | Gamma that is used in the Q-learning update | 0.99 | 0.5, 0.75, 0.95, 0.99 | 0.95 |
| Learning Rate | The learning rate used by SGD activation function.<br>This parameter will define how much the weights are updated after each epoch | 0.00025 | 0.0001, 0.0002, 0.0003, 0.00001, 0.00007, 0.000003, 0.000005, 0.000008, | 0.0001 |
| Gradient Momentum | The gradient momentum used by activation function | 0.95 | 0.75, 0.9, 0.95, 0.99 | 0.95 |
| Initial Exploration (Initial epsilon) | Initial value for e-greedy exploration. Associated with how random you take an action | 1.00 | 0.08, 0.1, 0.25, 0.3, 0.5, 0.8, 0.95 | 0.5 |
| Final Exploration (Final epsilon) | Final value for e-greedy exploration. Associated with how random you take an action | 0.1 | 0.005, 0.05, 0.08, 0.2, 0. 3 | 0.2 |
| Final Exploration Frame | Number of Frames over which the initial value of e is linearly annealed to its final value | 8000 | 100, 1000, 4000, 8000, 16000, 3000000 | 16000 |
| Replay Start or Observation | Timesteps to observe before training | 100 | 100, 320 | 100 |
| Protein Iterations | Number of actions for each protein | - | 200, random | N/A |
| Image Size | Input image Dimensions | 80x80 | 120, 150, 160, 200 | 160 |
| Maximum running steps | Total steps to terminate the program | - | 20000, 30000, 40000 | 30000 |
| Actions per Training | Number of actions performed before model is adapted | 10 | 3, 6, 10, 15, 20 | 6 |
| Decay | It is the weight decay which is an additional term in the weight update rule that causes the weights to exponentially decay to zero, if no other update is scheduled | 0.000002 | 0.000002 | 0.000002 |
| Score Scaling | Scaling factor, 0.1 in (1) in section 3.2.3 | | 0.05, 0.1, 0.125, 0.15, 0.175, 0.2, 0.25 | 0.175 |
| Dropout Rate | Specifies the probability at which outputs of the layer are dropped out, or inversely, the probability at which outputs of the layer are retained. | 0 | 0.05, 0.1, 0.175 | 0 |

# ANNEX II: Parameter values for each run

Run 1 (R1): Same as basic model parameters, input image size is 160x160 and maximum running steps are 20000. Variations in parameters of Convolutional Layers are:

Dense: 128, 256

First convolutional layer depth: 40, 60

First convolutional layer receptive field: 15, 30

First & Second convolutional layer filters: 1,2

Second convolutional layer depth: 10,20

Second convolutional layer receptive field: 20,40

Run 2 (R2): Same as basic model parameters, input image size is 160x160 and maximum running steps are 40000. Variations in parameters of Convolutional Layers are:

Dense: 64,128, 256

Second convolutional layer depth: 10,20,30

Run 3 (R3): Same as basic model parameters, Image size is tested for 150 and 200 and maximum running steps are 40000. Variations in parameters of Convolutional Layers are:

Dense: 64,128, 256

Second convolutional layer depth: 10,20,30

Run 4 (R4): Same as basic model parameters, image size is tested for 150 and 200 and maximum running steps are 30000. Variations in parameters of Convolutional Layers are:

Dense: 32,64,100

Run 5 (R5): Same as basic model parameters with an extra layer as last layer and image size is 150. Variations in parameters of Convolutional Layers are:

Dense: 32,64,100
Third convolutional layer depth: 10,20,30
Third convolutional layer receptive field: 20,30,40

Run 6 (R6): Same as basic model parameters, image size is tested for 120x120 size, learning rate is tested for 2e-4, number of actions per training for 5 actions, no use of log polar and random seed for a specific number or current daytime.

Run 7 (R7): Same as basic model parameters with an extra layer as first layer and image size 150. Variations in parameters of Convolutional Layers are:

First convolutional layer depth: 30, 40, 50 ,60 with a 3x3 receptive field.

Run 8 (R8): Basic model parameters with an extra layer as first layer and image size 150.

Tested for dropout in each layer

Run 9 (R9): Same model as R7, tested for learning rate 2e-4.

Run 10 (R10): Same model as R9 with first convolutional layer depth 40 and learning rate 2e-4. Tested with and without dropout rate for 3x3 and 5x5 receptive field.

Run 11 (R11): Same as basic model parameters with image size 150. Variations in parameters of Convolutional Layers are:

Second convolutional layer depth: 15,20,25
Second convolutional layer receptive field: 15,20,25

Run 12 (R12): Same as basic model parameters, with image size 150 and some variations in parameters of Convolutional Layers, which are:

Second convolutional layer depth: 5,10,15
Second convolutional layer receptive field: 5,10,15

Run 13 (R13): Single Cases. Same as basic model parameters with image size 150. Variations in batch size, protein iterations, score scaling and activation function.

Run 14 (R14): Same as basic model parameters, tested for random number of protein iterations.

Run 15 (R15) & Run 16 (R16): Same as basic model parameters with some variations in parameters of Convolutional Layers, which are:

First convolutional layer depth: 55,60,65
Second convolutional layer receptive field: 10,15,20

Run 17 (R17) & Run 18 (R18): Same as basic model parameters, random seed is the running daytime and mini batch size is tested for 70. Total replicates are 16.

Run 19 (R19): Same as basic model parameters with 70 mini batch size and random seed is the running daytime. Added dropout rate (0.05, 0.1, 0.2), 3 replicates per dropout value.

Run 20 (R20): Same as basic model parameters with 70 mini batch size, random seed the current daytime and some variations in:

Number of Actions per training: 3, 6 (Basic model has 10 actions)

Dropout rate: 0.1, 0.2

Run 21 (R21): Same as basic model parameters with 70 mini batch size, random seed the current daytime and variations in the number of Actions per training (15, 20)

Run 22 (R22): Same as basic model parameters with 70 mini batch size, number of actions per training are 6 and the random seed is the current daytime. Depth of the second layer is tested for values 10 and 20.

Run 23 (R23): Same as model of Run 22. Depth of the second layer is tested for values 15 and 25.

Run 24 (R24): Same as model of Run 22. Depth of the second layer is tested for values 10 and 15.

Run 25 (R25): Same as model of Run 24. Add batch normalization before activation function RELU.

Run 26 (R26): Same as model of Run 24 with depth 15. Add batch normalization after activation function RELU.

Run 27 (R27): Same as model of Run 24 with depth 15. Add batch maxPooling after activation function RELU.

Run 28 (R28): Same as model of Run 24, tested for less memory

Run 29 (R29): Same as model of Run 24, tested for less memory

Run 30 (R30): Same as basic model parameters with image size 160 and 10 second layer depth. Tested for first layer depth (40, 60)

Run 31 (R31):

Tested for second layer depth (10, 20)

Model Parameters

| | | | |
|---|---|---|---|
| Dense | 100 | 1st & 2nd layer filters | 1 |
| 1st layer depth | 40 | 2nd layer receptive field | 20 |
| 1st layer receptive field | 15 | Image size | 160 |

The rest of parameters are the same as the basic models.

Run 32 (R32): ReLU vs LeakyReLU

Run 33 (R33): Tested for second layer depth (15,20)

Model Parameters

| | | | |
|---|---|---|---|
| Dense | 100 | 1st & 2nd layer filters | 1 |
| 1st layer depth | 40 | 2nd layer receptive field | 20 |
| 1st layer receptive field | 15 | Image size | 160 |

The rest of parameters are the same as the basic model.

Run 34 (R34): Same as basic model's parameters with image size 160 and 15 second layer depth. Tested for first layer depth (30,40)

Run 35 (R35): Same as basic model's parameters with image size 160 and 15 second layer depth. Tested for Dense size (64,100)

Run 36 (R36): Same as basic model's parameters with image size 160 and 15 second layer depth. Tested for Dense size (100,128)

Run 37 (R37): Same as basic model's parameters with image size 160 and 15 second layer depth. Tested for Dense size (128,200)

Run 38 (R38): Same as basic model's parameters.Tested for image size (150,160)

Run 39 (R39): Same as basic model's parameters.Tested for logpolar functions.

Run 40 (R40): Model of Run 39 in test set.

Run 41 (R41): Same as basic model's parameters with 30 as first layer convolutional, 15 as second layer depth and 200 as dense.Tested for dropout rate (0, 0.2)

Run 42 (R42): Same as basic model's parameters with 30 as first layer convolutional, 15 as second layer depth.Tested for different actions per training (5, 15)

Run 43 (R43), Run 44 (R44) & Run 45 (R45): Same as basic model's parameters with 30 as first layer convolutional, 15 as second layer depth and 200 as dense.Tested for score scaling (0.05,0.125, 0.150, 0.175, 0.200, 0.250)

Run 46 (R46): Same as basic model's parameters with 30 first layer depth, 15 second layer depth, 0.175 score scaling and 200 as dense.Tested for learning rate (7e-5,3e-4)

Run 47 (R47) & Run 48 (R48): Same as basic model's parameters with 30 first layer depth, 15 second layer depth, 0.175 score scaling and 200 as dense.Tested for momentum (0.75, 0.9, 0.95, 0.99)

Run 49 (R49): Same as basic model's parameters with 30 first layer depth, 15 second layer depth, 0.175 as score scaling and 200 as dense.Tested for gamma (0.75, 0.99)

Run 50 (R50): Same as basic model's parameters with 30 first layer depth, 15 second layer depth, 0.175 as score scaling and 200 as dense.Tested for gamma (0.75, 0.99)

Run 52 (R52): Same as basic model's parameters with 30 first layer depth, 15 second layer depth, 0.175 as score scaling and 200 as dense. Different moving functions. Tested for gamma (0.5, 0.75)

Run 53 (R53): Same as basic model's parameters with 30 first layer depth, 15 second layer depth, 0.175 as score scaling, 200 as dense and 0.75 as gamma. Different moving functions. Tested for final epsilon (0.005, 0.2)

Run 54 (R54) & Run 55 (R55): Same as basic model's parameters with 30 first layer depth, 15 second layer depth, 0.175 as score scaling, 200 as dense and 0.75 as gamma. Different moving functions. Tested for initial epsilon (0.1, 0.25, 0.5, 0.95)

Run 56 (R56): Same as basic model’s parameters with 30 first layer depth, 15 second layer depth, 0.175 as score scaling, 200 as dense and 0.75 as gamma. Different moving functions. Tested for exploration (4000, 16000)

Run 57 (R57): Same as basic model’s parameters with 30 first layer depth, 15 second layer depth, 0.175 as score scaling, 200 as dense and 0.75 as gamma. Different moving functions. Tested for replay memory (800, 2000)

Run 58 (R58): Same as basic model’s parameters with 30 first layer depth, 15 second layer depth, 0.175 as score scaling, 200 as dense and 0.75 as gamma. Different moving functions. Added second dense after the original dense and tested for (50, 100)

Run 59 (R59): Same as basic model’s parameters with 30 first layer depth, 15 second layer depth, 0.175 as score scaling, 200 as dense and 0.75 as gamma. Different moving functions. Added second dense before the original dense and tested for (15, 25)

Run 60 (R60): Same as basic model’s parameters with 30 first layer depth, 15 second layer depth, 0.175 as score scaling, 200 as dense and 0.75 as gamma. Tested for replay memory (6000, 10000)